\documentclass[final,5p,times,twocolumn,unknownkeysallowed]{elsarticle}

\usepackage{doi}
\usepackage{hyperref}
\hypersetup{
	colorlinks=true,        
	linkcolor=blue,         
	citecolor=cyan,         
}

\usepackage[utf8x]{inputenc}
\DeclareUnicodeCharacter{2212}{\textendash}
\usepackage{graphicx}
\usepackage{amssymb}
\usepackage{amsmath}
\usepackage{amsthm}
\usepackage{lineno}
\usepackage{hyperref}
\usepackage{multirow}
\usepackage{bigints}

\usepackage{dcolumn}
\usepackage{bm}
\usepackage{color}

\hypersetup{colorlinks=true,linkcolor=cyan,citecolor=cyan,urlcolor=blue,bookmarks=true}\modulolinenumbers[200]

\biboptions{comma,sort&compress}

\def\a{\alpha}
\def\r{\rho}
\def\s{\sigma}
\def\t{\tau}
\def\m{\mu}
\def\n{\nu}
\def\k{\kappa}
\def\th{\theta}
\def\g{\gamma}\def\G{\Gamma}
\def\L{\Lambda}\def\l{\lambda}
\def\D{\Delta}
\def\la{\langle}
\def\ra{\rangle}
\def\o{\omega}\def\O{\Omega}
\def\d{\delta}
\def\p{\partial}

\newcommand{\be}{\begin{equation}}
\newcommand{\ee}{\end{equation}}
\newcommand{\bea}{\begin{eqnarray}}
\newcommand{\eea}{\end{eqnarray}}

\def\half{\textstyle{\frac{1}{2}}}

\def\bdoc{\begin{document}}
\def\edoc{\end{document}}
\def\beq{\begin{equation}}
\def\eeq{\end{equation}}
\def\bea{\begin{eqnarray}}
\def\eea{\end{eqnarray}}
\def\ben{\begin{enumerate}}
\def\een{\end{enumerate}}
\def\la{\langle}
\def\ra{\rangle}
\def\a{\alpha}
\def\b{\beta}
\def\g{\gamma}
\def\G{\Gamma}
\def\d{\delta}
\def\D{\Delta}
\def\e{\epsilon}

\def\th{\theta}
\def\k{\kappa}
\def\l{\lambda}
\def\m{\mu}
\def\n{\nu}
\def\o{\omega}
\def\p{\pi}
\def\r{\rho}
\def\s{\sigma}
\def\t{\tau}
\def\L{{\cal L}}
\def\S{\Sigma }
\def\gsim{\; \raisebox{-.8ex}{$\stackrel{\textstyle >}{\sim}$}\;}
\def\lsim{\; \raisebox{-.8ex}{$\stackrel{\textstyle <}{\sim}$}\;}
\def\gtrsim{\gsim}
\def\lessim{\lsim}
\def\loc{{\rm local}}
\def\vm{v_{\rm max}}
\def\bh{\bar{h}}
\def\del{\partial}
\def\nab{\nabla}
\def\half{{\textstyle{\frac{1}{2}}}}
\def\fourth{{\textstyle{\frac{1}{4}}}}

\def\bD{{\bf D}}
\def\bE{{\bf E}}
\def\bF{{\bf F}}
\def\bB{{\bf B}}
\def\bP{{\bf P}}
\def\bV{{\bf v}}
\def\bv{{\bf v}}
\def\bx{{\bf x}}
\def\by{{\bf y}}
\def\bz{{\bf z}}
\def\ba{{\bf a}}
\def\bd{{\bf d}}
\def\bs{{\bf s}}
\def\bn{{\bf n}}
\def\bp{{\bf p}}

\def\O{\Omega}

\def\br{{\bf r}}
\def\bnab{{\bf \nab}}

\def\tE{\tilde{E}}
\def\tL{\tilde{L}}

\journal{Journal Name}

\begin{document}

\begin{frontmatter}

\title{Effect of perfect fluid dark matter on particle motion \\ around a static black hole immersed in an external magnetic field 
 }

\author[mainaddress2,mainaddress3,mainaddress4,mainaddress5,mainaddress6]
{Sanjar Shaymatov\cortext[cor2]{Corresponding author}\corref{cor2}}
\ead{sanjar@astrin.uz}
\author[mainaddress7]
{Daniele Malafarina}
\ead{daniele.malafarina@nu.edu.kz}
\author[mainaddress2,mainaddress4,mainaddress5]
{Bobomurat Ahmedov}
\ead{ahmedov@astrin.uz}

\address[mainaddress2]{Ulugh Beg Astronomical Institute, Astronomy St. 33, Tashkent 100052, Uzbekistan}
\address[mainaddress3]{Akfa University, Kichik Halqa Yuli Street 17, Tashkent 100095, Uzbekistan}
\address[mainaddress4]{National University of Uzbekistan, Tashkent 100174, Uzbekistan}
\address[mainaddress5]{Tashkent Institute of Irrigation and Agricultural Mechanization Engineers,\\ Kori Niyoziy 39, Tashkent 100000, Uzbekistan}
\address[mainaddress6]{Power Engineering Faculty, Tashkent State Technical University, Tashkent 100095, Uzbekistan}
\address[mainaddress7]{Department of Physics, Nazarbayev University, Kabanbay Batyr 53, 010000 Nur-Sultan, Kazakhstan}

\date{Received: date / Accepted: date}

\begin{abstract} 
We investigate particle and photon motion in the vicinity of a static and spherically symmetric black hole surrounded by perfect fluid dark matter in the presence of an external asymptotically uniform magnetic field. 
We determine the radius of the innermost stable circular orbit (ISCO) for charged test particles and the radius for unstable circular photon orbits 
and show that the effect of the presence of dark matter shrinks the values of ISCO and photon sphere radii. Based on the analysis of the ISCO radius we further show that the combined effects of dark matter and magnetic field can mimic the spin of a Kerr black hole up to $ a/M \approx 0.75-0.8$. 
Finally, we consider the effect of the presence of dark matter on the center of mass energy of colliding particles in the black hole vicinity. We show that the center of mass energy grows as the value of the dark matter parameter increases. This result, in conjunction with the fact that, in the presence of an external magnetic field, the ISCO radius can become arbitrarily close to the horizon leads to arbitrarily high energy that can be extracted by the collision process, similarly to what is observed in the super-spinning Kerr case.
 
\end{abstract}


\end{frontmatter}

\linenumbers

\section{Introduction}
\label{introduction}

The existence of black holes is a direct consequence of
Einstein's general theory of relativity. 
However, until recently, black holes had been considered as potential explanation for some observed phenomena, like some X-ray sources, but not directly detected.
This changed with the discovery of gravitational waves produced by the merger of pairs of stellar mass black holes by the LIGO and Virgo collaborations~\cite{Abbott16a,Abbott16b} and as well as with the first direct imaging of the supermassive black hole in the galaxy M87 by the Event Horizon Telescope (EHT) collaboration~\cite{Akiyama19L1,Akiyama19L6}. 
These observations allow for the first time to test the nature of the geometry in the vicinity of the black hole's event horizon.
The possibility that the geometry of such astrophysical compact objects may exhibit departures from the Kerr line element has been investigated thoroughly in the literature (see for example \cite{Johannsen11,Bambi-Malafarina13}).

In this context, particle motion around black holes and exotic compact objects has been a productive field of study for several years. 
It is well known that the motion of test particles and photons is strongly affected by the geometry in the strongly relativistic regime, i.e. in the vicinity of the black hole candidate. 
However, other elements may affect the geometry and the particles' geodesics. For example deformations of the source (see for example \cite{Herrera00,Herrera05,Bini12,Toshmatov19d}), the presence of external magnetic fields (see for example \cite{Wald74,Benavides-Gallego19}) and the presence of other matter fields (see for example \cite{Tsukamoto13,Joshi19} ) 
may all contribute to altering the motion of test particles.

A large amount of work has been devoted to the study of charged particle motion in the geometry of a black hole immersed in an external magnetic field~\cite{Prasanna80,Kovar08,Kovar10,
Shaymatov15,Dadhich18,Narzilloev19,
Pavlovic19,Shaymatov19b,Duztas-Jamil20,Stuchlik20,Shaymatov20egb,Shaymatov21c,Shaymatov21b}. 
Similarly, one can consider the motion of test particles in the spacetime describing a black hole immersed in an external matter field. 
{In a realistic astrophysical scenario it is important to take into account the effect of matter fields in the environment surrounding the black hole since in a variety of situations black holes are not in vacuum.}
Among the matter fields that may surround a supermassive black hole candidate the presence of dark matter is particularly important. 
{The first exidence of the existence of dark matter came from the observation of the flattening of rotation curves of giant elliptical and spiral galaxies at large distances~\cite{Rubin80}. 
Recent estimates suggest that dark matter may contribute for up to 90\% of the mass of a galaxy~\cite{Persic96}. Although there is still exist no direct experimental detection of dark matter, astrophysical observations suggest that supermassive black holes in the galactic center are embedded in a giant dark matter halos~\cite{Akiyama19L1,Akiyama19L6}.} 
Thus, from the extrapolation of known dark matter profiles for the outer regions of a galaxy \cite{Sofue13-book} to the inner regions, one can expect the dark matter contribution to be relevant near the galactic center (see for example \cite{Boshkayev19}). In order to model such dark matter envelopes at the center of galaxies we may use some non-vacuum solution of Einstein's equations. For example, Kiselev derived a static and spherically symmetric black hole solution which contains a dark matter profile represented by particular anisotropic fluid~\cite{Kiselev03-dm}. 
Later, based on the standard assumption that the dark matter halo consists of weakly interacting massive particles (WIMPs), together with a linear equation of state with $\omega\simeq 0$, 
Li and Yang~\cite{Li-Yang12} derived another black hole solution surrounded by dark matter, where the dark matter envelope is described by a bounded phantom scalar field.    
There are several other ways that can be considered to include dark matter fields in the background of a black hole geometry
(see for example ~\cite{Xu18,Haroon19,Konoplya19plb,Hendi20,Jusufi19,
Narzilloev20b,Shaymatov21d,Rayimbaev-Shaymatov21a}). 

In the present article we consider a static black hole surrounded by perfect fluid dark matter, as described by the line element proposed in \cite{Li-Yang12}. For this geometry, we investigate photon orbits and the motion of charged test particles in the presence of an external asymptotically uniform magnetic field.

The relevance of the motion of charged particles stems from recent astronomical observations of particle outflows from
active galactic nuclei (AGN) such as winds and
jets. These outflows have been observed in $X$-ray, $\gamma$-ray and Very Long Baseline Interferometry (VLBI) observations and can have energies of the order of $E\approx 10^{42}-10^{47}\:\rm{erg/s}$~\cite{Fender04mnrs,Auchettl17ApJ,IceCube17b}.
A proposed explanation for these observations comes from considering high energy collisions of particles near the black hole horizon. The process was first theoretically described by Banados, Silk and West (BSW) in \cite{banados09} 
and it has since been considered in a large variety of contexts
~\cite{Grib11,Jacobson10,Harada11b,Wei10,Zaslavskii10,
Zaslavskii11b,Zaslavskii11c,Kimura11,Banados11,Frolov12,
Abdujabbarov13a,Liu11,Atamurotov13a,Stuchlik11a,Stuchlik12a,
Igata12,Shaymatov13,Tursunov13,Shaymatov18a,Atamurotov21JCAP}.
Theoretical investigation of high energy phenomena around compact objects has been considered also in alternate theories of gravity
~\cite{Stuchlik14a}, 
in the vicinity of naked singularities
~\cite{Patil10,Patil11,Patil11b}, 
and near regular black hole solutions~\cite{Stuchlik15}.  
The well-known Penrose process was addressed in \cite{Abdujabbarov11} while the effect of spinning test particles on the Penrose collision process was studied in~\cite{Okabayashi20}. 
The presence of a magnetic field 
plays an important role also in understanding the mechanisms of energy extraction from black holes~\cite{
Blandford1977,Wagh89,Morozova14,Alic12ApJ,Moesta12ApJ} and coupling between accretion disks and jets in AGNs~\cite{McKinney07}. 

Since black holes are not endowed with their own magnetic fields~\cite{Ginzburg1964,Anderson70} 
one needs to consider the presence of an external magnetic field induced by nearby objects such the accretion discs around rotating black holes~\cite{Wald74} or magnetars \cite{Morozova10,Morozova12} and neutron stars~\cite{Ginzburg1964,Rezzolla01,deFelice03,deFelice04}. 
Thus, the magnetic field can be regarded as a test field, which does not modify the background geometry~\cite{Frolov10,Aliev02,Abdujabbarov10,Shaymatov14,
Kolos15,Jamil15,Stuchlik16,Tursunov16,Hussain15}.

The paper is organized as follows: In Sec.~\ref{Sec:metic} we briefly discuss the metric for a black hole immersed in a static anisotropic perfect fluid, which can be used to model a dark matter distribution. In Sec.~\ref{Sec:motion} we study the motion of photons and charged particles in the spacetime when an external magnetic field is present. 
The discussion of the effects of dark matter on the collision energy of particles in the spacetime is presented in Sec.~\ref{Sec:energy} and concluding remarks with the relevance of the work for astrophysical black holes are in the Sec.~\ref{Sec:conclusion}.

Throughout the paper we use a system of units in which $G=c=1$. Greek
indices are taken to run from 0 to 3, Latin indices from 1 to 3.

\section{\label{Sec:metic}
The metric}

{ Here we will briefly review the formalism used to model the motion of particles in a solution describing the spacetime of a black immersed in a dark matter fluid.}
The metric describing a static and spherically symmetric black hole immersed in perfect fluid dark matter in Schwarzschild coordinates $(t, r, \theta, \varphi)$ is given by \cite{Li-Yang12} 
\\
\begin{eqnarray}\label{Eq:metric1}
ds^2&=&-F(r)dt^2+F(r)^{-1}dr^2 + r^2 d\Omega^2\, ,  
\end{eqnarray}
where $d\Omega^2$ is the line element on the unit 2-sphere and where we have defined
\be\label{Eq:F} 
F(r)=\left(1-\frac{2M}{r}+\frac{\lambda}{r} \log\frac{r}{\vert\lambda \vert}\right)\, ,
\ee
with $M$ being black hole mass and $\lambda$ related to the dark matter density and pressure. In the case of vanishing $\lambda$, the spacetime metric (\ref{Eq:metric1}) reduces to the Schwarzschild metric, while for $\lambda\neq 0$ the stress energy-momentum tensor of the dark matter distribution is that of an anisotropic perfect fluid $T^\mu_\nu={\rm diag}(-\rho,p_{r},p_{\theta},p_{\phi})$ where density, radial and tangential pressures are given by
\begin{eqnarray}\label{rho}
\rho=-p_{r}= \frac{\lambda}{8\pi r^3}\,  \mbox{~~~and~~~} p_{\theta}=p_{\phi}=\frac{\lambda}{16\pi r^3}\, .
\end{eqnarray}
In order to model a dark matter distribution we shall restrict ourselves to the case $\lambda>0$ which gives positive energy density. The horizon is located at the root of the following equation 
\begin{eqnarray}\label{Eq:h}
r^2-2Mr+\lambda\, r \log\frac{r}{\vert\lambda \vert}=0\, .
\end{eqnarray}
{For small values of parameter $\lambda \ll M$ the above equation gives $r_0\simeq 2M$, and thus we can approximate it with $r^2-2Mr+\lambda\, r_0 \log{r_0}/{\vert\lambda \vert}=0$ which has the following approximate root}
\begin{eqnarray}
r_{h}=M+\sqrt{M^2-2 M \lambda \log\frac{2M}{\vert\lambda \vert}}\, .
\end{eqnarray}

We notice that the radius of the horizon is $r_h=2M$ for two values of $\lambda$, i.e. $\lambda=0$ and $\lambda=2M$. Therefore $r_h$ decreases as the dark matter profile is introduced and it reaches a minimum for $\lambda=\lambda_0<2M$.

\begin{figure}
\centering
 \includegraphics[width=0.45\textwidth]{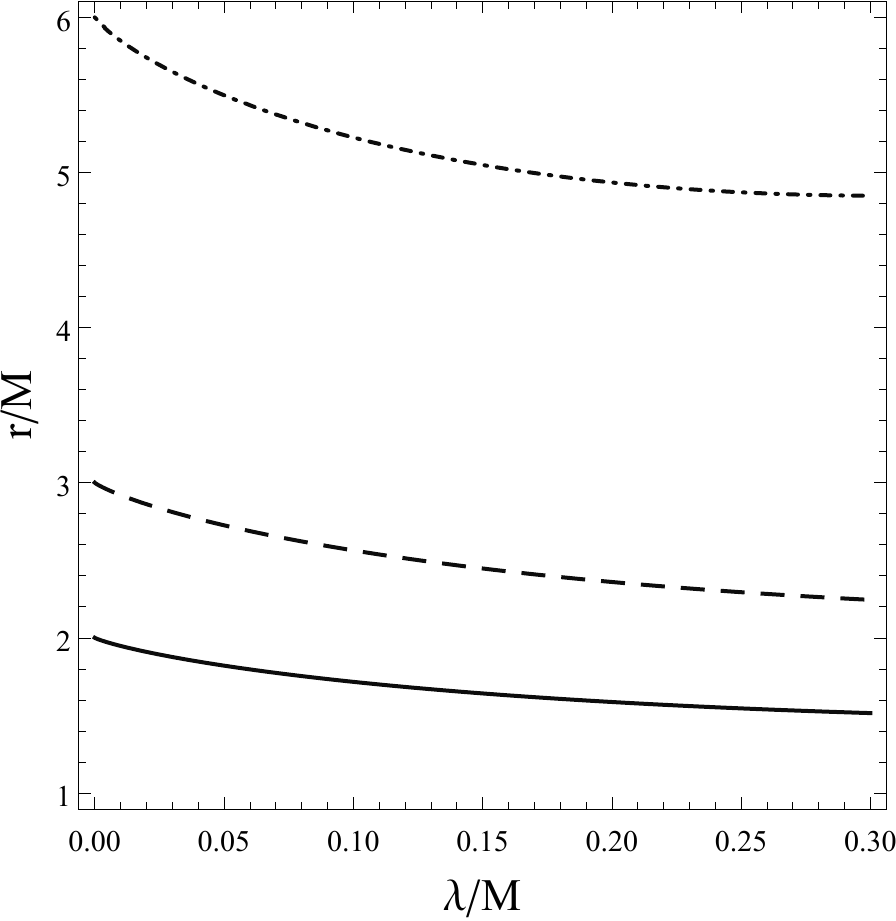}%
\caption{\label{fig:hor} The dependence of the horizon radius $r_{h}$ (thick line), photon sphere $r_{ph}$ (dashed line) and ISCO (dot-dashed line) on the perfect fluid dark matter parameter $\lambda$. For small values of $\lambda$ all three radii have maximum value in the absence of dark matter.}
\end{figure}

We move now to the study of particle and photon motion in the above spacetime. From the usual relation for the 4-momentum $p_\mu p^\mu=k$ we have that for massive particles one has to set $k=-m^2$ (with $m$ the mass of the test particle) while for photons one has to set $k=0$. From the Hamilton-Jacobi formalism we obtain the action $S$ in the form 
\begin{eqnarray}\label{separation}
S= -Et+L\varphi+S_{r}(r)+S_{\theta}(\theta)\, ,
\end{eqnarray}
where $E$ and $L$ are the usual conserved quantities associated with the time translations and spatial rotations and describe the energy $E$ and angular momentum $L$ of the particle or photon, respectively and  $S_{r}$ and $S_{\theta}$ are functions of only $r$ and $\theta$, respectively. Now it is straightforward to obtain the Hamilton-Jacobi equation in the following form
\begin{eqnarray}
\label{Eq:Sep1}
k&=&-\frac{E^2}{F(r)}+F(r)
\left(\frac{\partial S_{r}}{\partial r}\right)^2
+\frac{1}{r^2}\left(\frac{\partial S_{\theta}}{\partial \theta}\right)^2+ \frac{L^2}{r^2\sin^2\theta} \, .\nonumber\\
\end{eqnarray}
Due to the spherical symmetry of the spacetime we can restrict the analysis to the equatorial plane $\theta=\pi/2$.  Then from the separability of the action given in Eq.~(\ref{Eq:Sep1}) we obtain the radial equations of motion for particles and photons in the form
\bea \label{Veff3}
\dot{r}^2=E^2+kF(r)-\frac{L^2}{r^2}F(r)
\, ,
\eea
where, for simplicity, we shall now set $k=-1$ for massive particles so that $E$ and $L$ describe energy and angular momentum per unit mass. 
We can use the effective potential to find the radii of circular orbits for given values of $E$ and $L$ by solving simultaneously $\dot{r}=\ddot{r}=0$. This is equivalent to solving simultaneously
\begin{eqnarray}
V_{eff}(r,E,L)=0, \mbox{~~~} \frac{\partial V_{eff}(r,E,L)}{\partial r}=0\, ,
\end{eqnarray}
for the function $V_{eff}(r,E,L)$ defined by
\begin{eqnarray}
V_{eff}(r,E,L)&=& E^2r^2+kr^2F(r)-L^2F(r)\, .
\end{eqnarray}

In the case of massive particles the radius of the ISCO $r_i$ is obtained from the minimum value of the angular momentum for which circular orbits are allowed. Therefore one determines $L^2$ from $V_{eff}'=0$ and then $r_i$ from $V_{eff}''=0$. In our case this gives $r_i$ implicitly from the condition
\begin{eqnarray} \label{L}
0&=&2 \lambda \left(r^2-6 L(r)^2\right) \log \frac{r}{\lambda}-(4M+3 \lambda) r^2\nonumber+\\ 
&+&\left(24M-6 r+7 \lambda\right)L(r)^2  \, ,
\end{eqnarray} 
with $L=L(r)$ determined by $V_{eff}'=0$.
On the other hand, for photons it is sufficient to use $V_{eff}'=0$ to find the following condition for the radius of the photon sphere
\be 
6M-2r+\lambda\left(1-3\log\frac{r}{\lambda}\right)=0\, .
\ee 
 
In the limit of $\lambda \ll 1$ we can write the approximate expressions for the ISCO radius $r_i$ and the photon orbit $r_{ph}$ as
\begin{eqnarray}\label{Eq:rph}
r_i&\approx& 6 M+\left[4-3 \log \left(\frac{6 M}{\lambda}\right)\right]\lambda +O(\lambda^2)\, ,\\
r_{ph}&\approx& 3M+\frac{1}{2} \left[1- \log \left(\frac{27}{8}\right)\right]\lambda +O(\lambda^2) \, .
\end{eqnarray}
This clearly shows $r_i=6M$ and $r_{ph}=3M$ in the limit $\lambda\rightarrow 0$, which corresponds to the ISCO and photon orbit in the case of the Schwarzschild black hole. 

The behaviour of the black hole horizon $r_h$, photon orbit $r_{ph}$ and ISCO $r_i$ is shown in Fig.~\ref{fig:hor}. It is clear that each radius decreases as a consequence of the presence of $\lambda$ in the limit of small $\lambda$. However, due to the repulsive nature of the radial pressures the effect of the dark matter profile ${\rho}$ can turn repulsive for larger values of $\lambda$ thus resulting in increasing values for the three radii here considered as $\lambda$ grows. In the following we will restrict our attention to more realistic case where the black hole mass dominates over the dark matter distribution, therefore considering $\lambda<M$.

\section{\label{Sec:motion}
Charged particle motion}

We now consider the motion of charged particles in the geometry described by the line element \eqref{Eq:metric1} once an external magnetic field is present. For simplicity we will consider the magnetic field to be uniform at large distances. Also we assume that the presence of the electromagnetic field does not affect the background geometry {due to the fact that the magnetic field strength is respectively of order $B_1\sim 10^{8}~\rm{Gauss}$ for stellar mass black holes and $B_2\sim 10^{4}~\rm{Gauss}$ for the supermassive black holes (i.e.,  $B_{1,2}\ll B_{max}\sim 10^{19}M_\odot/M~ \rm{Gauss}$, {with $B_{max}$ corresponding to the upper limit for the magnetic field for which the black hole is referred to as \textit{strongly magnetized}})~\cite{Piotrovich10,Baczko16,Dallilar2018}.} However, the presence of the magnetic field alters the motion of charged particles depending on its strength. We can introduce the following dimensionless parameter to characterize the magnetic field strength~\cite{Frolov10}:
\begin{eqnarray}\label{b}
b\equiv\frac{qBMG}{mc^4}\, ,
\end{eqnarray}
where $q$ is the test particle's charge, $B$ is the modulus of the uniform external magnetic field and we have reintroduced $G$ and $c$ in order to be able to make quantitative estimates.
Notice that depending on the signs of $q$ and $B$ we may have $b$ positive, when charge and magnetic field are `aligned', or negative otherwise.
This parameter is of the order of $b\sim 10^{7}$ for a proton around a stellar mass black hole of mass $M\sim 10M_{\odot}$ and of the order of $b\sim10^{11}$ for a proton around a supermassive black hole of mass $M\sim10^9 M_{\odot}$, where $M_{\odot}$ is the mass of the Sun. This quantity can be larger for an electron due to the fact of its mass is smaller as compared to the mass of proton. Thus, these estimates clearly show that the effect of the magnetic field on a charged
particle motion can dominate over gravitational field for particles on orbits near the black hole~\cite{Frolov10,Frolov12,Stuchlik16,Tursunov16}.

{ The formalism to describe the motion of charged particles in a spacetime endowed with an external magnetic field has been widely used since the pioneering work \cite{Wald74}.}
In the following, we consider an external asymptotically uniform magnetic field surrounding the black hole immersed in the dark matter fluid. 
Following Wald~\cite{Wald74} we consider the timelike and spacelike Killing vectors $\xi^{\alpha}_{(t)}=(\partial/\partial
t)^{\alpha}$ and $\xi^{\alpha}_{(\varphi)}=(\partial/\partial \phi)^{\alpha}$ associated to the time translational and rotational symmetries of the spcetime (\ref{Eq:metric1}). We can write the Killing equations as  
\begin{eqnarray}\label{Eq:Killing}
\label{kelling} \xi_{\alpha; \beta}+\xi_{\beta; \alpha}= 0\, ,
\end{eqnarray}
from which we obtain the following equation
\begin{eqnarray}\label{Eq:Killing1}
\label{weq} \Box \xi^\alpha = \xi^{\alpha ;\beta}_{\ \ \ \ \
;\beta}=R^{\mu}_{\ \delta}\xi^{\delta}\,,
\end{eqnarray}
where $R_{\mu\nu}$ is the Ricci tensor.
It is obvious that the right-hand side of expression
~(\ref{weq}) vanishes in the vacuum case, so that the equation takes the simpler form 
$ \Box \xi^\alpha=0$.
Then, in the vacuum case, the Maxwell equations for the vector potential
$A^\alpha$ take the same form as in the Lorentz
gauge, i.e. $ \Box A^\alpha=0$ . 
However, the line element \eqref{Eq:metric1} is not Ricci flat ($R_{\alpha\beta}\neq0$), thus implying that the vector potential for the magnetic field surrounding the source must be modified. 
The right-hand side of Eq.~(\ref{Eq:Killing1}) can be defined as
$R^{\alpha}_{\ \delta}\xi^{\delta}=\eta^{\alpha}$, so that
Maxwell's equations take the form
\begin{eqnarray}\label{Eq:mxw}
\mathcal{F}^{\alpha \beta}_{\ \ \ ;\beta}=-2C_{0}\left(\xi^{\alpha ;\beta}_{\ \ \ \ \
;\beta}- \eta^{\alpha}\right)=0 \, .
\end{eqnarray}
{Then the electromagnetic field tensor (Faraday tensor) $\mathcal{F}_{\alpha \beta}$ can be written as}
\begin{eqnarray}
\mathcal{F}_{\alpha \beta}&=&C_0\left(\xi_{ \beta;\alpha}-\xi_{ \alpha;\beta}\right)+
2\left(a_{
\beta,\alpha}-a_{ \alpha,\beta}\right)
=\nonumber\\
&=& -2C_0\left(\xi_{ \alpha;\beta}+a_{
\beta,\alpha}-a_{ \alpha,\beta}\right) \, ,
\end{eqnarray}
where $C_0$ is a constant and the 4-potential $a^{\alpha}$, due to the non vanishing Ricci tensor, is defined by $\Box a^{\alpha}=\eta^{\alpha}$. As said, the Ricci tensor does not vanish in the case considered here where we have a dark matter distribution surrounding the black hole, i.e. $R_{\alpha\beta}\neq 0$. It is a well-known fact that if the Ricci tensor vanishes the Killing vector can be used as the four potential for magnetic field surrounding the black hole. In our case the vector potential must be modified accordingly since the background spacetime is not vacuum. From the above we see that, in this case, the vector potential of the electromagnetic filed can be defined in the following way~\cite{Shaymatov18a,Abdujabbarov10} 
\begin{eqnarray}\label{Vec-pot}
A^\alpha=C_1 \xi^\alpha_{(t)} +C_2 \xi^\alpha_{(\phi)}+a^\alpha\, ,
\end{eqnarray}
with $C_1$ and $C_2$ being integration constants.
The vector $a^\alpha$ can be obtained from the following equation
\begin{eqnarray}\label{Eq:21}
\Box a^{\alpha}=a^{\alpha;\beta}_{\ \ \ ;\beta}&=& \left(C_{1} \xi^{\gamma}_{(t)}+C_{2} \xi^{\gamma}_{(\phi)}\right) R^{\alpha}_{\ \gamma} \ .
\end{eqnarray}
Taking Eqs.~(\ref{Vec-pot}) and (\ref{Eq:21}) into account we obtain the following equation    
\begin{eqnarray}\label{Eq:22}
&&r \left(r-2M+\lambda \log \frac{r}{\vert\lambda\vert}\right) f^{\prime \prime}(r)+\nonumber\\&+&\left(2M+\lambda-\lambda \log \frac{r}{\vert\lambda\vert}\right) f^{\prime}(r)-2 f(r)=0\, ,
\end{eqnarray}
{where $f(r)$ is the radial part of the vector potential of the electromagnetic field $A_{\phi}(r,\theta)$.
It is not possible to solve the above equation analytically, but it can be approximated for small values of $\lambda$. Thus Eq.~\ref{Eq:22} solves to give the following analytical expression }
\begin{eqnarray}
f(r)=r^2+r \left(1+\log\frac{M}{r}\right)\lambda+O(\lambda^2)\, .
\end{eqnarray}
From the condition of the spacetime being static we immediately obtain one of the integration constants as $C_1=0$. Thus the vector potential can be written as $A_{\phi}(r,\theta)=C_2f(r)\sin^2\theta$.  Taking the second integration constant to be $C_2=B/2$ leads to 
the covariant components of the 4-potential of the electromagnetic field to take the following form
\begin{eqnarray}
\label{4pot} A_t&=&A_{r}= A_{\theta}=0\, , \nonumber\\
A_{\varphi}&=&\frac{B}{2}r^2\left[1+\frac{\lambda}{r}\left(1+\log\frac{M}{r}\right)+O(\lambda^2)\right]\sin^2\theta 
\, .\nonumber\\ 
\end{eqnarray}
The field in the above satisfies the source-free Maxwell equation (\ref{Eq:mxw}) for $\lambda \ll M$, i.e. \begin{eqnarray}
\mathcal{F}^{\alpha \beta}_{\ \ \ ;\beta} =A^{\alpha; \beta}_{\ \ \ ;\beta}-R^{\alpha}_{\gamma}A^{\gamma}=0\, .
\end{eqnarray}

The four-velocity of the zero angular
momentum observers (ZAMOs) in the spacetime under consideration is
\begin{eqnarray}
\label{zamo_con} &(u^{\alpha})_{_{\textrm{ZAMO}}}
=\left\{\frac{1}{\sqrt{F}},0,0,0\right\}\, ,
\end{eqnarray}
{with $F$ given by Eq.~(\ref{Eq:F}) and} from which we obtain the non vanishing components of the Faraday tensor $\mathcal{F}$ measured by the ZAMOs as
\begin{eqnarray}\label{Eq:faraday1}
\mathcal{F}_{r\phi}&=& B r \left(1+\frac{\lambda}{2r}\log \frac{M}{r}\right)\sin^2\theta\, ,\\
\mathcal{F}_{\theta
\phi}&=& {B}r^2\left[1+\frac{\lambda}{r}\left(1+\log\frac{M}{r}\right)\right] \sin\theta  \cos\theta\, .
\label{Eq:faraday2}
\end{eqnarray}
From the above one easily obtains the expressions for the
orthonormal components of the electromagnetic field measured by
the ZAMO observers as
\begin{eqnarray}
\label{M1}  B^{\hat r}
&=&-B~\left[1+\frac{\lambda}{r}\left(1+\log\frac{M}{r}\right)\right]\cos\theta \, , \\
\label{M2}
B^{\hat\theta} 
&=&B~\sqrt{F}\left[1+\frac{\lambda}{2r}\log\frac{M}{r}\right]\sin\theta \, . 
\end{eqnarray}

\begin{figure*}
\centering
  \includegraphics[width=0.3\textwidth]{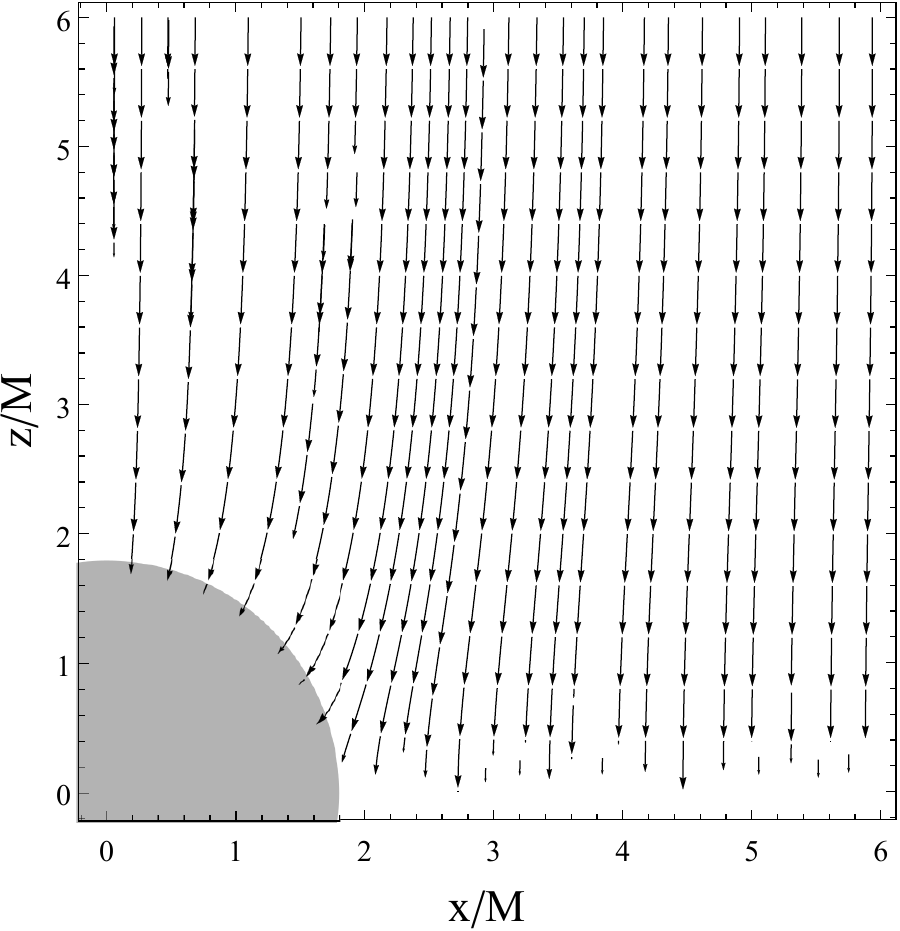}
  \includegraphics[width=0.3\textwidth]{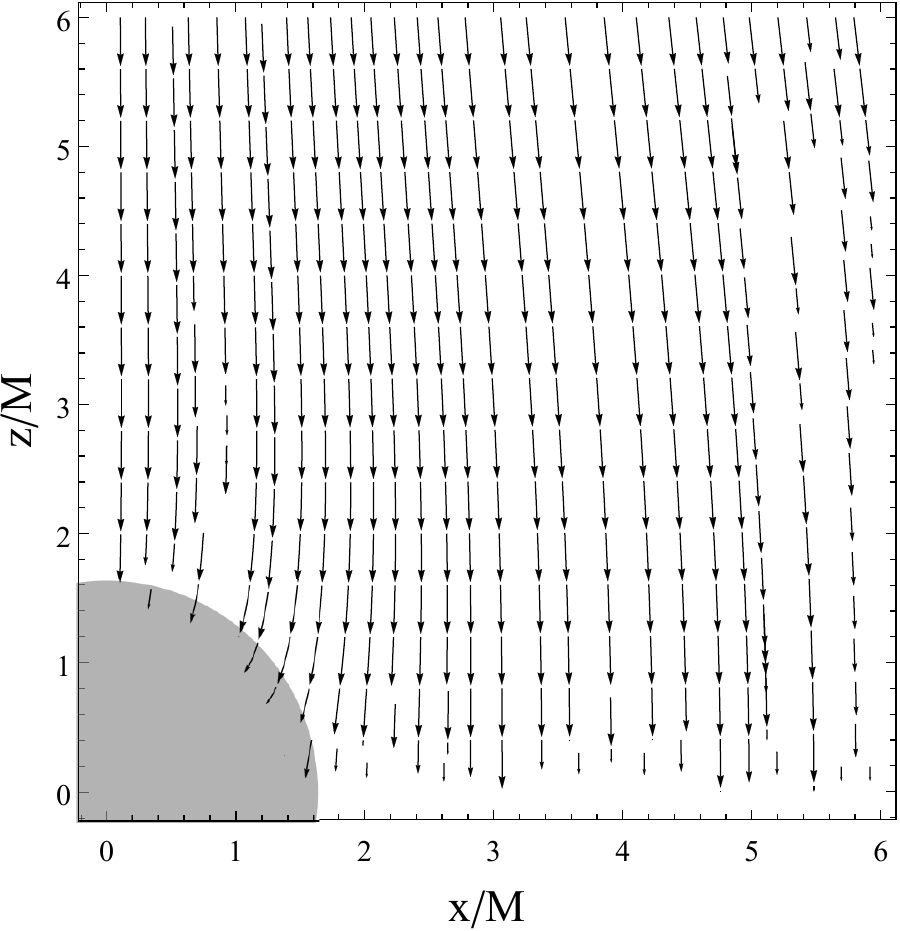}
  \includegraphics[width=0.3\textwidth]{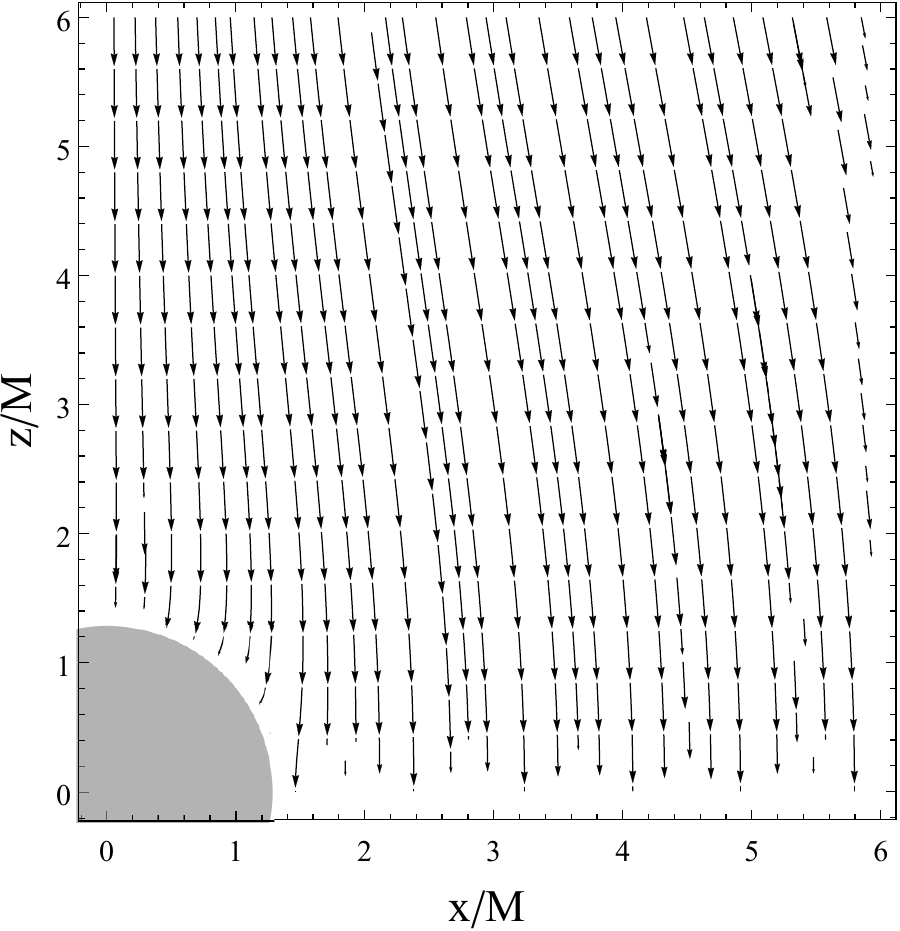} %
 
\caption{\label{config} The configuration of magnetic field
lines in the vicinity of a black hole surrounded by perfect fluid dark matter for different values of $\lambda$ and a given magnetic field with strength $B=0.5$ (Note that we consider $B$ as a dimensionless quantity on the basis of Eq.~\eqref{b} having set $G=c=1$). The values of the dark matter parameter used in the figures are $\lambda=0.05$ (left panel), $\lambda=0.1$ (middle panel) and $\lambda=0.2$ (right panel). The vertical axis $z$ corresponds to the symmetry axis $\theta=0$ along which the magnetic field is oriented, the horizontal axis $x$ corresponds to an arbitrary radial direction orthogonal to $z$. {Note that the boundary of the gray-shaded area represents the black hole horizon.} }
\end{figure*}

As expected, the radial and polar components of the electric field do not appear because of the vanishing of dragging of inertial frames due to the spin parameter of the black hole being zero.
In the limit of flat spacetime or in the limit of large distances, i.e. for \hbox{$M/r\rightarrow 0$}, 
Eqs. (\ref{M1}) and (\ref{M2}) reduce to
\begin{eqnarray}
\label{Eq:limit} && 
B^{\hat r} =-B\cos\theta\, , \; \; 
B^{\hat\theta}=B\sin\theta\, , 
\end{eqnarray}
which describe an homogeneous magnetic field in flat spacetime. The magnetic field lines are shown in Fig.~\ref{config}.

The Hamiltonian for the system of a charged particle around a static and spherically symmetric black hole immersed in an external magnetic field is
\begin{eqnarray}
 H  \equiv \frac{1}{2}g^{\alpha\beta}(\pi_\alpha - q A_\alpha)(\pi_\beta - q A_\beta)\, ,
\label{Eq:H}
\end{eqnarray}
with $\pi_\alpha$ being the canonical momentum of the charged particle and the
four-vector potential of the electromagnetic field $A_\alpha$ given by Eq.~(\ref{4pot}). 
The Hamiltonian is a constant $H=\tilde{k}/2$, where $\tilde{k}=-m^2$ (with $m$ the mass of the charged particle)~\cite{Misner73}.

Accordingly Hamilton's equations of motion in terms of $x^\alpha$ and $\pi_\alpha$ are
\begin{eqnarray}
  \frac{dx^\alpha}{d\tau} &=& \frac{\partial H}{\partial \pi_\alpha}   \, ,
\label{Eq:H-1}  \\
  \frac{d\pi_\alpha}{d\tau} &=& - \frac{\partial H}{\partial x^\alpha} \, ,
\label{Eq:H-2}
\end{eqnarray}
where $ {\tau}=\varsigma/m$ is the affine parameter with proper time $\varsigma$. 
As usual, the first equation is a constraint equation providing the definition of the four-momentum of the charged particle.
Note that
the action $S$ corresponding to the Hamilton-Jacobi equation can be separated in the following form
\begin{eqnarray}\label{Eq:separation1}
S= -\frac{1}{2}\tilde{k}\tau-Et+L\varphi+S_{r}(r)+S_{\theta}(\theta)\ ,
\end{eqnarray}
where the quantities $E \equiv -\pi_t$ and $L \equiv \pi_{\varphi}$, together with the rest energy of the test particle $m$, are the constants of motion, namely, energy and axial angular momentum of the charged particle. 
Using Eqs (\ref{Eq:H}) and (\ref{Eq:separation1}), one can easily obtain the Hamilton-Jacobi equation in the form
\begin{eqnarray}\label{Eq:separable}
\tilde{k}&=&-F(r)^{-1}\left(E+qA_{t}\right)^2
+ F(r)
\left(\frac{\partial S_{r}}{\partial r}\right)^2
\nonumber\\&+&\frac{1}{r^2}\left(\frac{\partial S_{\theta}}{\partial \theta}\right)^2+
\frac{\left(L-qA_{\varphi}\right)^2}{r^2\sin^2\theta} \, . 
\end{eqnarray}
A fourth constant of motion can be obtained due to separability of the action. However, since the fourth constant of motion is related to the latitudinal motion of test particles, we won't need to specify it when considering motion in the equatorial plane.
\begin{figure*}
\centering
  \includegraphics[width=0.3\textwidth]{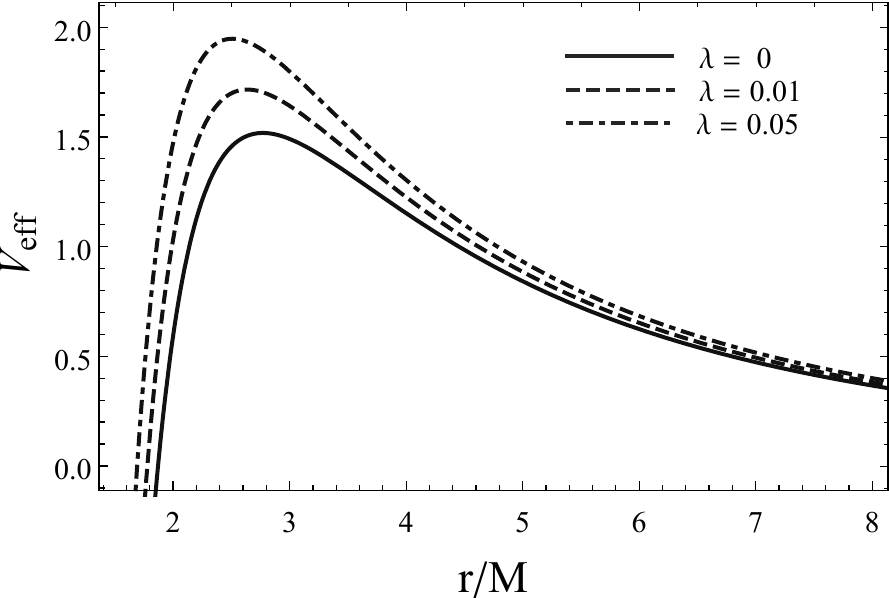}
  \includegraphics[width=0.3\textwidth]{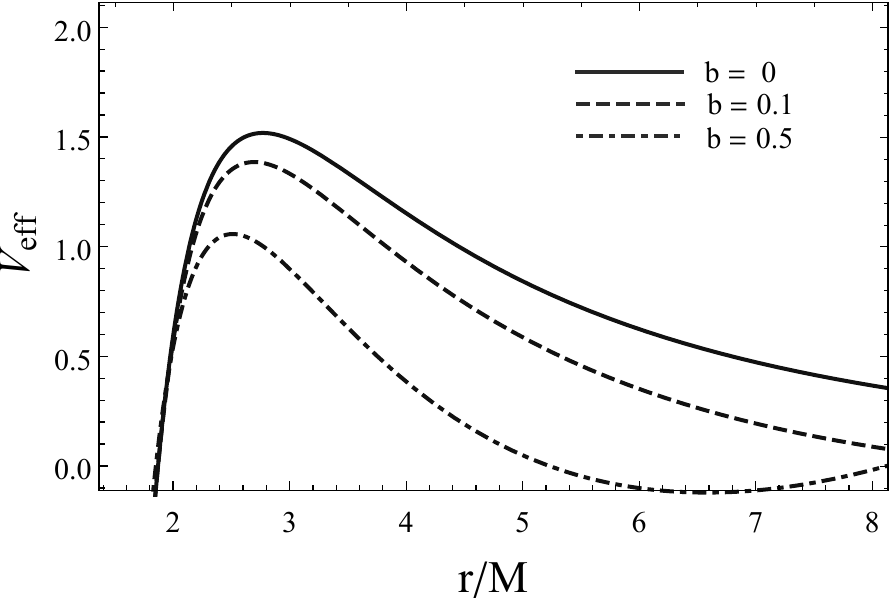}
  \includegraphics[width=0.3\textwidth]{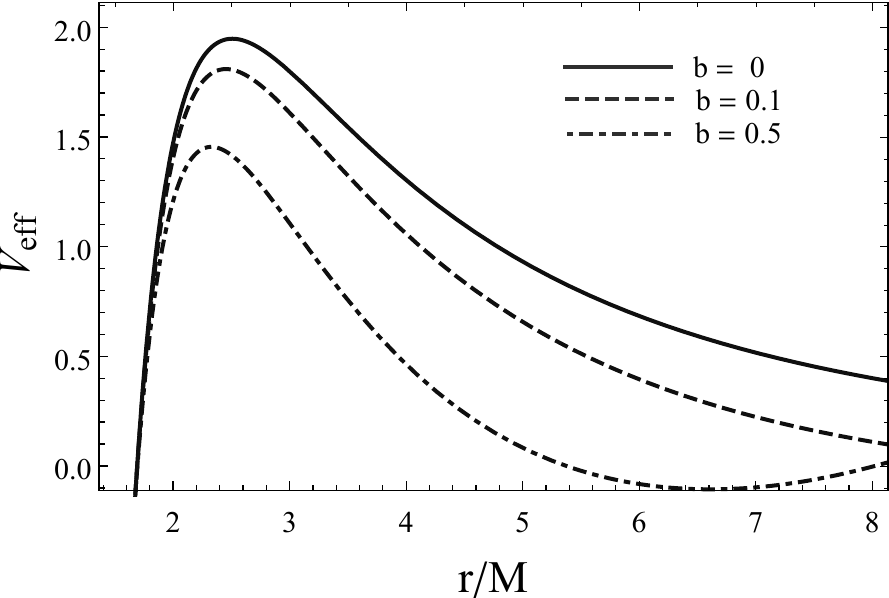}
  
  \includegraphics[width=0.35\textwidth]{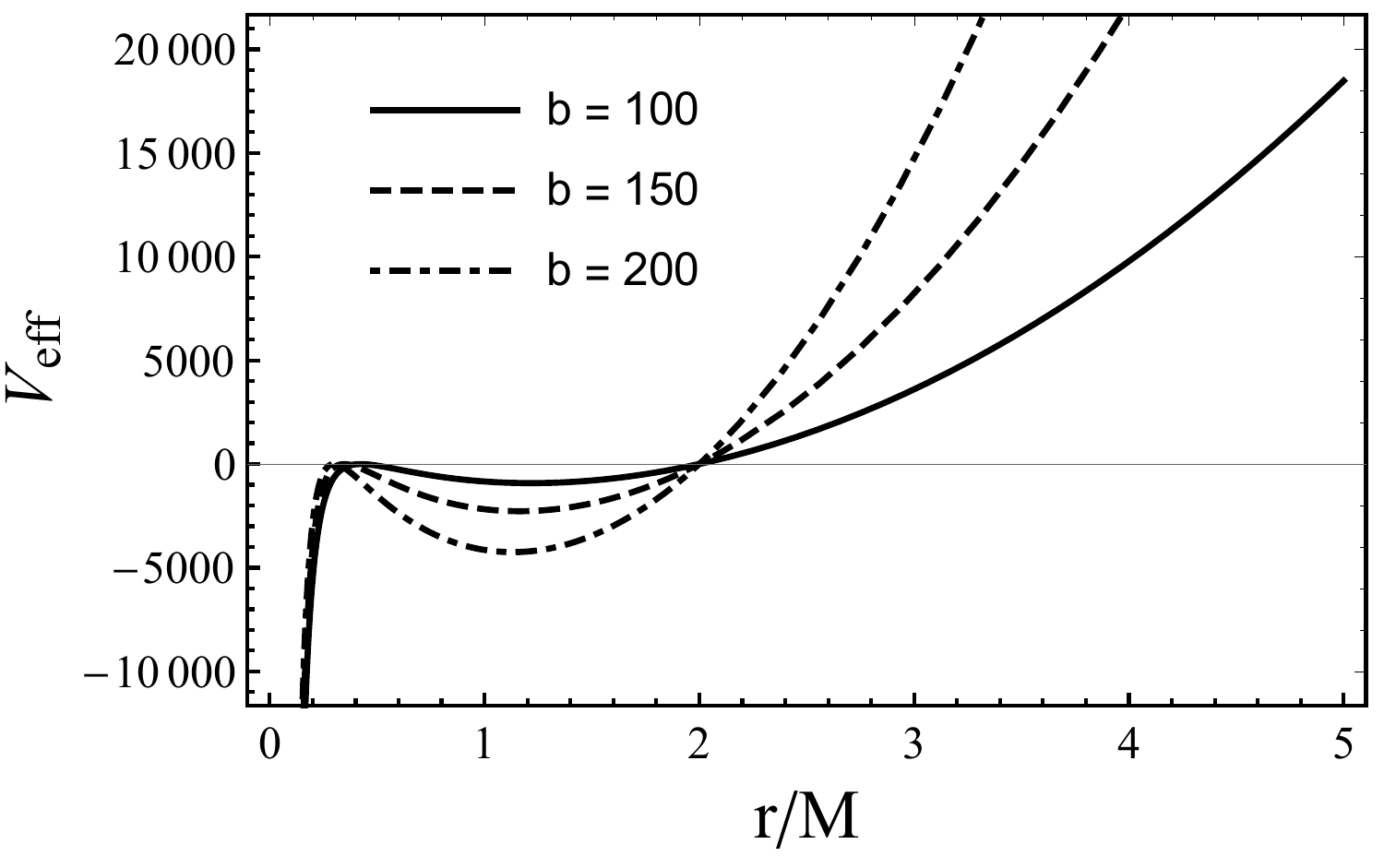}
  \includegraphics[width=0.35\textwidth]{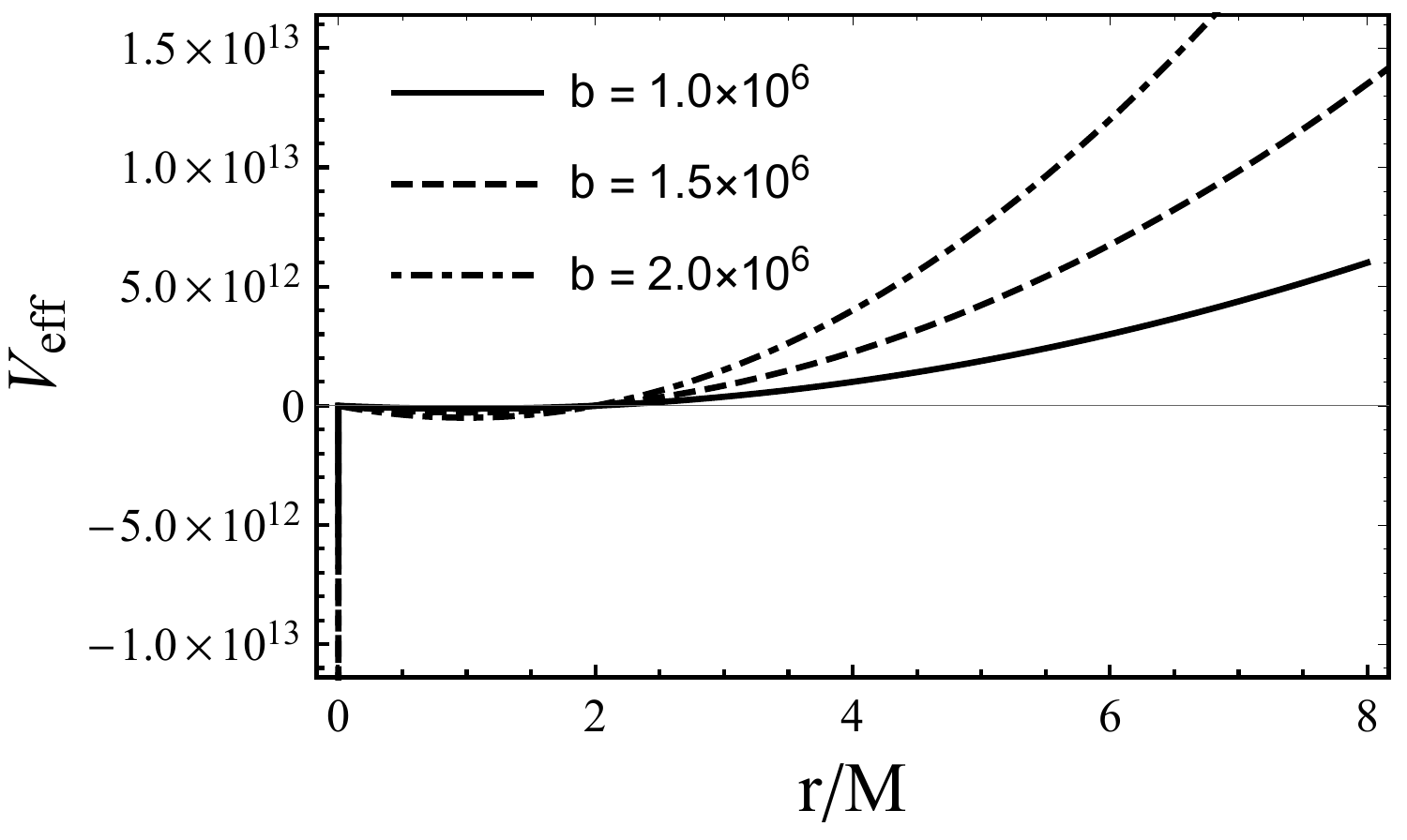}

\caption{\label{eff-pot} {Radial dependence of the effective potential for massive particles around a black hole in perfect fluid dark matter immersed in an external asymptotically uniform magnetic field. Top row, left panel: $V_{eff}$ is plotted for different values of $\lambda$ in the case without magnetic field, i.e. $b=0$. Top row, middle panel:
$V_{eff}$ is plotted for different values of $b$ in the case without dark matter, i.e. $\lambda=0$.
Top row, right panel: $V_{eff}$ is plotted for different values of $b$ in the case of fixed $\lambda=0.05$. The two panels in the bottom row show how $V_{eff}$ is affected by considering more realistic values of $b\gg1$ while keeping fixed $\lambda=0.05$.  }}
\end{figure*}

By virtue of Eq.~(\ref{Eq:separable}), the radial equation of motion  for the charged test particle can be written in the usual form
\begin{equation}
\frac{1}{2}\dot{r}^{2} + V_{eff}(r;\mathcal{L},\lambda,b)=\mathcal{E}^2\, ,
\end{equation}
where the effective potential $V_{eff}(r;\mathcal{L},\lambda,b)$
which determines the motion of the particle is given by
\begin{equation} \label{Veff2}
V_{eff}= F(r)  \left(
1+\frac{\left[\mathcal{L}-\frac{b}{M}\left(1+\frac{\lambda}{r}\left(1+\log\frac{M}{r}\right)\right)~r^{2}\right]^2}{r^2}\right)\, , 
\end{equation}
and where we have used the specific constants of motion per unit mass, namely $\mathcal{E}=E/m$,
$\mathcal{L}=L/m$. The magnetic parameter $b$ measuring the effect of the magnetic field on the charged particle motion is given in Eq.~\eqref{b}. In the case of vanishing $\lambda$ and $b$ parameters, Eq.~(\ref{Veff2}) recovers the effective potential for the Schwarzschild spacetime. 

\begin{table}[h]
\caption{\label{1tab} The values of the ISCO radius $r_i$ for charged particles moving around the black hole surrounded by perfect fluid dark matter for different values of $\lambda$ and $b$. Notice that $r_i$ has maximum value in the Schwarzschild case.}
\small
\centering
\begin{tabular}{c|cccccc}
 &  &  &  
& $b$ &  &  \\ \hline
{$\rm \lambda$} & 0.000 & 0.001 &  0.005
& 0.010 & 0.050 & 0.100 \\
    &  & -0.001 &  -0.005 & -0.010 & -0.050 & -0.100 \\
\hline
 0.000 & 6.00000 &5.99826 & 5.95709 & 5.84140 & 4.69667 & 3.98268 \\
    &          &5.99829 &5.95986 &5.85955 &5.02571 &4.63195 \\\\
0.001  &5.97790  &5.97619  &5.93543  &5.82080 &4.68252 &3.97087 \\
      &        &5.97620  &5.93811  &5.83864  &5.00876 &4.61541 \\\\
0.005  &5.91367  &5.91199 &5.87244 &5.76075 &4.64058 &3.93563 \\
&             &5.91203 &5.87481 &5.77759 &4.95849 &4.56610 \\\\
0.010 &5.84820   &5.84653 &5.80819 &5.69935 &4.59709 &3.89889 \\
      &       &5.84662 &5.81020 &5.71512 &4.90642 &4.51480 \\ \\
0.050 &5.48472   &5.48305 &5.45034 &5.35596 &4.34514 &3.68334 \\
      &      &5.48356 &5.45085 &5.36506 &4.60582 &4.21597 \\\\
0.100 &5.18427   &5.18244 &5.15126 &5.06887 &4.12166 &3.48825 \\
      &      &5.18352 &5.15476 &5.07115 &4.34040 & 3.94815 \\\hline
\end{tabular}
\end{table}

\begin{table}
\begin{center}
\caption{\label{2tab} {The values of the ISCO radius $r_i$ for charged particles moving around the black hole for different values of $b\gg 1$ while keeping fixed dark matter parameter $\lambda=0$. }}
\begin{tabular}{c c c}
\\ \hline
$b$  & $r_{i}\,(b>0)$   & $r_{i}\,(b<0)$\\[1.0ex]\hline
 $\pm 1.0$      &$2.4399176$    &$4.3078984492$\\[1ex]
 $\pm 10^2$      &$2.0057514$   &$4.3027761691$\\[1ex]
 $\pm 10^3$      &$2.0005771$   &$4.3027756428$\\[1ex]
 $\pm 10^4$      &$2.0000577$   &$4.3027756377$\\[1ex]
 $\pm 10^6$      &$2.0000006$   &$4.3027756363$\\ 
 \hline
\end{tabular}
\end{center}
\end{table}

\begin{figure*}
\centering
  \includegraphics[width=0.4\textwidth]{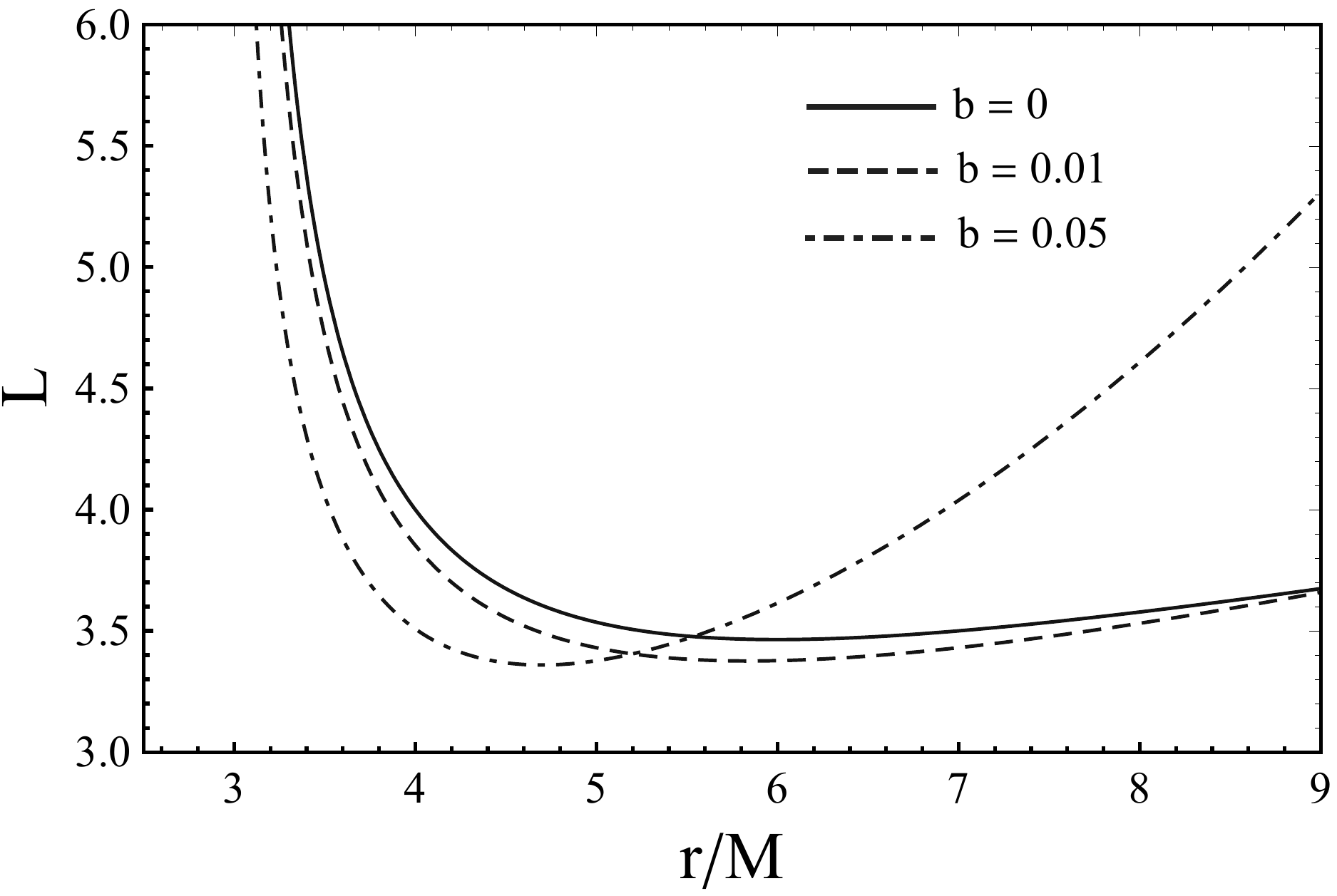}
     \includegraphics[width=0.4\textwidth]{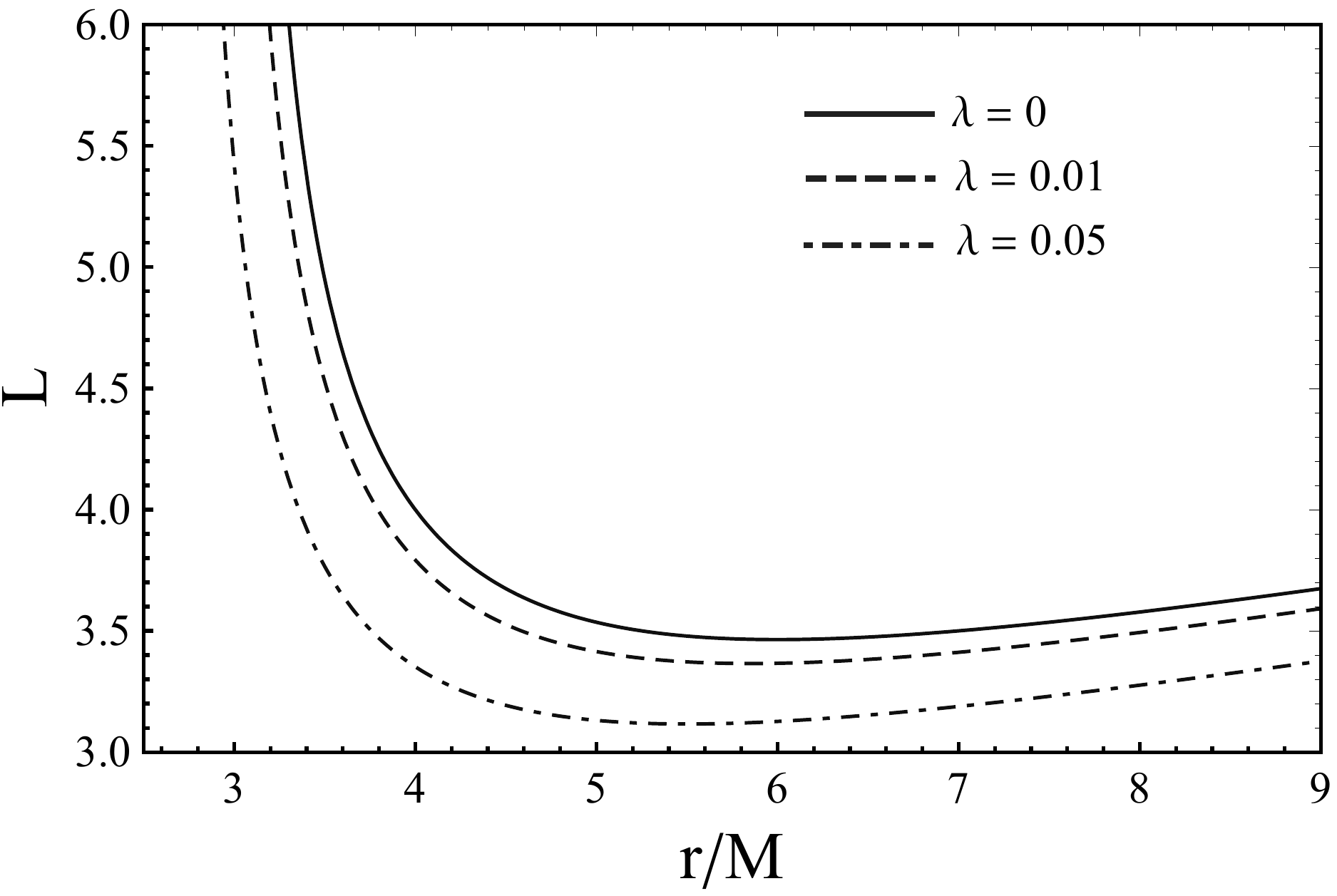}
     
     \includegraphics[width=0.4\textwidth]{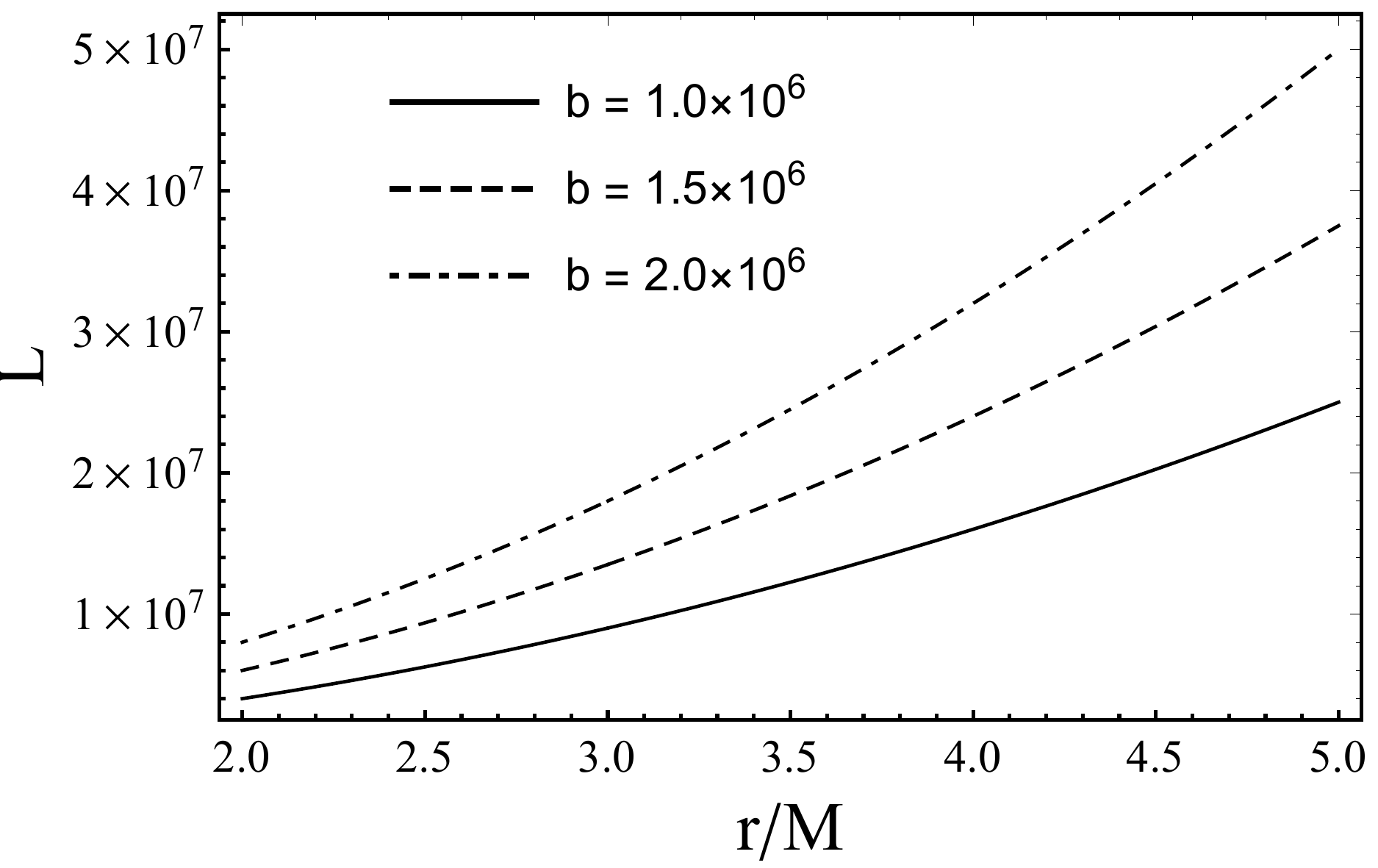}
     \includegraphics[width=0.4\textwidth]{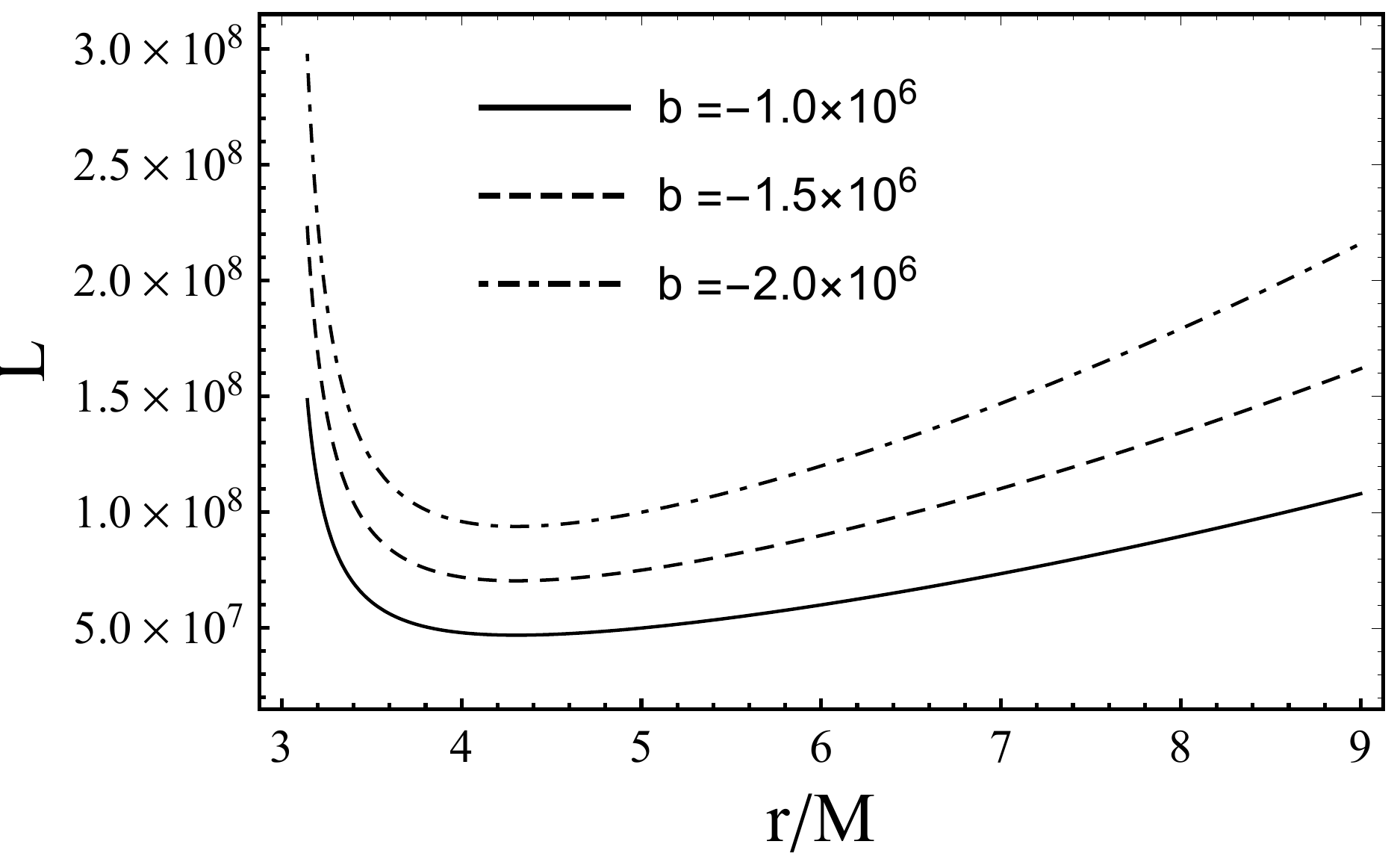}
\caption{\label{Fig:L} 
{The dependence of the specific angular
momentum on the radial motion of charged particles
moving in the vicinity of a black hole immersed in perfect fluid dark matter. Top row, left panel: $L$ is plotted for different illustrative values of magnetic parameter
${b}$ without the dark matter parameter, $\lambda=0$.  Top row, right panel: $L$ is plotted for different values of dark matter parameter $\lambda$ without magnetic field, $b=0$. The two panels in the bottom row show more realistic values of the magnetic parameter ${\pm b\gg1}$ while keeping fixed dark matter parameter $\lambda=0.05$. The case $\lambda=0$, $b=0$ corresponds to Schwarzschild.} }
\end{figure*}

In Fig.~\ref{eff-pot} we show the radial dependence of the effective potential (\ref{Veff2}) for different values of $\lambda$ and $b$.
We see that the presence of dark matter, i.e. $\lambda>0$ has the opposite effect with respect to the magnetic field when $b>0$, in terms of the strength of the potential, therefore suggesting the possibility that these two effects may cancel each other at some radius for certain values of $\lambda$ and $b$.
On the other hand we notice that regardless of the sign of the magnetic field parameter, the ISCO radius is always smaller with respect to the Schwarzschild case, thus suggesting that the geometry could be distinguished from the Schwarzschild geometry, provided that one is able to have an independent measurement of $M$. 
{Notice that the values of $b$ and $\lambda$ in the first row of Fig.~\ref{eff-pot} are purely illustrative. Regarding $b$ the upper limit discussed at the beginning of this section may be greatly reduced when the test particles are atoms or molecules, which may have the same charge but much larger mass than elementary particles such as electrons. The effective potential for more realistic values of $b$ is illustrated in the second row of Fig.~\ref{eff-pot}. Regarding $\lambda$, since we do not know the characteristic dark matter densities at the center of galaxies (namely within a few Schwarzschild radii from the central object) we can not estimate a realistic value. For this reason, within the assumption that $\lambda<<M$ (i.e. the dark matter mass is much smaller than the black hole mass), we have chosen values of $\lambda$ that help illustrate the features of the model.  
However we can be somewhat more quantitatively considering dark matter estimates within few parsecs from the central object. There are several studies that discuss the well-known disagreement between the results that stems from numerical simulations and observations of low-mass galaxies, known as the core-cusp problem, according to which dark matter densities as inferred from observations lie between $\rho\sim\left(10^{-2}-10^{-1}\right) M_{\odot}/pc^3$ (see for example [99]). Following this estimate we may obtain the corresponding value of $\lambda$ as per the model considered here. For example: the dark matter parameter would be of the order $\lambda\sim \left(10^{-21}-10^{-20}\right)$ for the Sgr A$^{\star}$ and $\lambda\sim \left(10^{-12}-10^{-11}\right)$ for the galaxy M87. }

We shall now consider circular orbits and in particular study the innermost stable circular orbit (ISCO) for charged particles moving in the black hole spacetime with asymptotically uniform magnetic field and dark matter.
The condition for circular orbits $\dot{r}=\ddot{r}=0$ is obtained by taking the following conditions for the effective potential and its first derivative  
\begin{eqnarray}\label{isco1}
V_{eff}(r,\mathcal{L},\lambda,b)=\mathcal{E}^2, \mbox{~~~~} \frac{\partial V_{eff}(r,\mathcal{L},\lambda,b)}{\partial r}=0\, .
\end{eqnarray}
We can then solve the above equations to find the corresponding values of the specific energy $\mathcal{E}$ and angular
momentum $\mathcal{L}$ at the circular orbits. 
In Fig.~\ref{Fig:L} we show the dependence on $\lambda$ and $b$ of the specific angular momentum $\mathcal{L}$ for particles in circular orbit. For small radii both $b$ and $\lambda$ have similar effect, thus reducing the value of $\mathcal{L}$ for the particle to be on circular orbit.
Finally, to find the ISCO we solve the equation for the second derivative of the effective potential to vanish
\begin{eqnarray}\label{isco2}
 \frac{\partial^2 V_{eff}(r,\mathcal{L},\lambda,b)}{\partial r^2}=0\, .
\end{eqnarray}
{In table~\ref{1tab} and \ref{2tab}}, we show the ISCO radius obtained by solving Eqs. (\ref{isco1}) and (\ref{isco2}) numerically for different values of the magnetic and dark matter parameters. 
{ In table~\ref{1tab} we use illustrative values of $b$, while in table~\ref{2tab} we consider more realistic values which have a stronger effect on the radius of the ISCO. We see then the location of $r_i$ would be strongly dominated by large values of $b$.}
It can be seen that the radius of the ISCO decreases due to the combined effects of dark matter and magnetic field.

We can compare the behaviour of test particles with the Schwarzschild case by considering the difference between the effective potential $V_{eff}^{\rm Schw}$ for Schwarzschild spacetime and the effective potential $V_{eff}(r,\mathcal{L},\lambda,b)$ for the black hole surrounded by dark matter and magnetic field. Their difference is given by 
\begin{eqnarray}
&& R(r;\mathcal{L},\lambda,b)=V_{eff}^{\rm Schw}(r)-V_{eff}(r;\lambda,b)=
\nonumber\\&& \left(1-\frac{2M}{r}\right) \left(1+\frac{\mathcal{L}^2}{r^2}\right)-\left(1-\frac{2M}{r}+\frac{\lambda}{r} \log\frac{r}{\vert\lambda \vert}\right) \nonumber\\ &&\times \left\{1+\frac{\left[\mathcal{L}-\frac{b}{M}\Big(1+\frac{\lambda}{r}\left(1+\log\frac{M}{r}\right)\Big)~r^{2}\right]^2}{r^2}\right\}\, .
\end{eqnarray}
For a given radius $r$ there may exist non zero values of both $\lambda$ and $b$ for which the effective potential is the same as Schwarzschild, i.e. for which $R(r;\mathcal{L},\lambda,b)=0$. 

\begin{figure}

\includegraphics[width=0.45\textwidth]{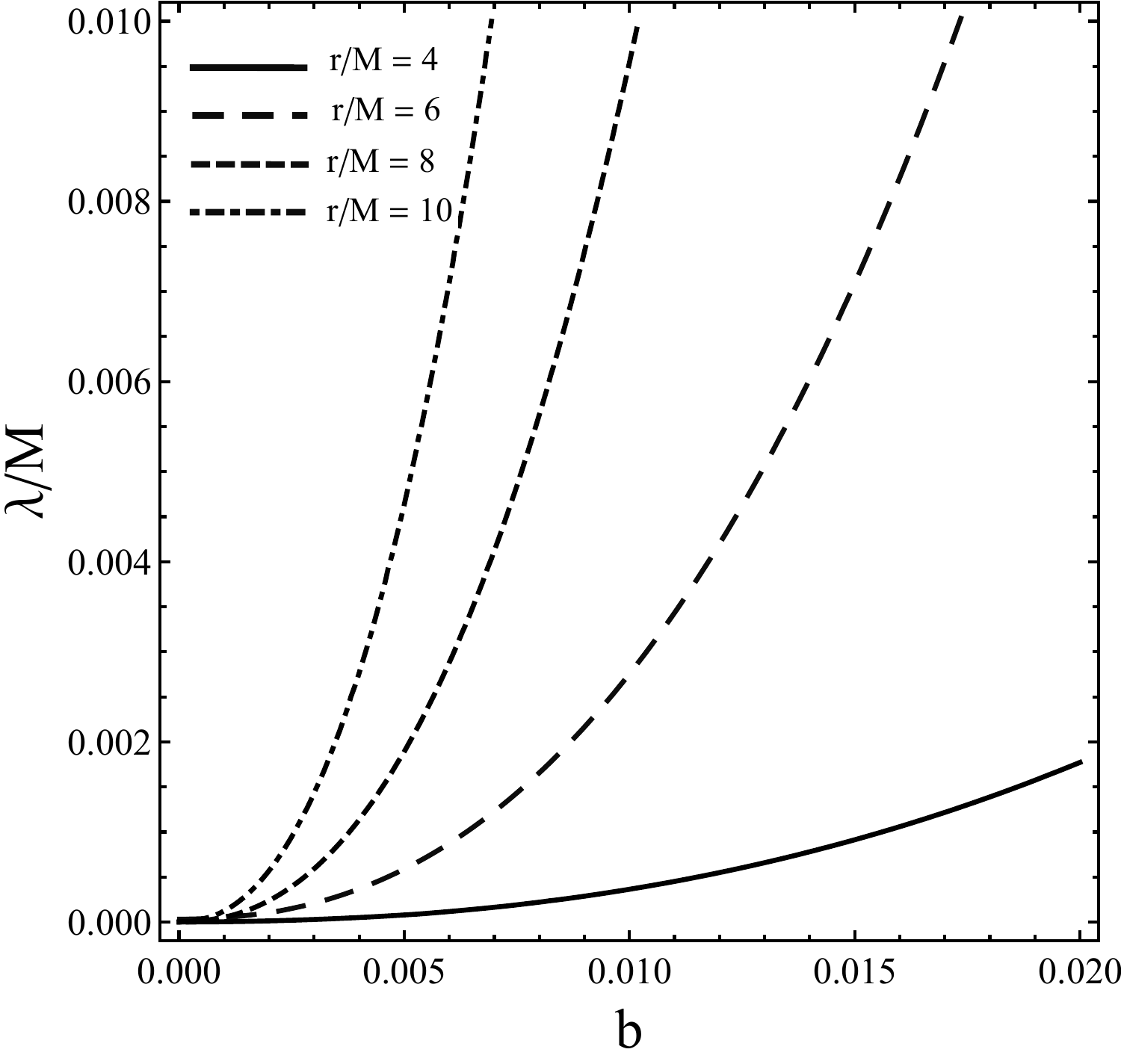} %

\caption{\label{Fig5} \textcolor{black}{The dependence of the dark matter parameter $\lambda$ on the magnetic field parameter $b$ for a charged test particle on circular orbit at a fixed radius $r$ such that the orbit mimics a circular orbit in the Schwarzschild spacetime.} 
}
\end{figure}

We can evaluate the implicit function $\lambda(b)$ for which $R(r,\mathcal{L},\lambda,b)=0$ at a given $r$ of circular orbit by imposing the following conditions 
\begin{eqnarray}\label{func1}
R(r;\mathcal{L},\lambda,b)=0, \mbox{~~~~} \frac{\partial R(r;\mathcal{L},\lambda,b)}{\partial r}=0\, .
\end{eqnarray}
In Fig.~\ref{Fig5}, we show the relation between the dark matter parameter $\lambda$ and the magnetic field parameter $b$ for which the effective potential mimics the Schwarzschild case.
Additionally from Fig.~\ref{Fig5} one can see the combined effect of the magnetic field and dark matter around a black hole. Namely, as one moves at larger radii, for any given value of $b$ it is necessary a larger value of $\lambda$ in order to mimic the Schwarzschild behaviour.

\subsection{Astrophysical applications}\label{Sec:apllications}

The recent detection of gravitational waves by the LIGO and Virgo scientific collaborations \cite{Abbott16a,Abbott16b} and the first image of supermassive black hole candidate at the center of the galaxy M87~\cite{Akiyama19L1,Akiyama19L6}, have opened the door to precise measurements of the parameter associated with the geometry of astrophysical black hole candidates. However, such objects are still regarded as black holes candidates due to the fact that the parameters associated to the geometry have not been precisely measured. In order to confirm that the observations are indeed due to relativistic black holes it is important to determine whether there exist degeneracies in the determination of crucial parameters such as mass, angular momentum or quadrupole moment between different geometries. In other words it is important to exclude possible `black hole mimickers' as viable sources.
 
In this respect the attractive effects of the perfect fluid dark matter considered here would play an important role in altering the geodesics of test particles thus affecting observable properties such as the ISCO. This behavior is similar to what one obtains for a Kerr black hole with co-rotating accretion disk and therefore the presence of dark matter could mimic the effects of the black hole's angular momentum (see \cite{Bambi-Malafarina13,Boshkayev20} for similar examples).
For illustrative purposes, let us focus on stable circular orbits and compare how the ISCO depends on the rotation parameter $a$ in the Kerr geometry with its dependence on $\lambda$ in the geometry considered here. 
Following Bardeen \textit{et al.}~\cite{Bardeen72} the very well known expression for the ISCO $r_{\rm i}$ of test particle around Kerr black hole is given by 
\begin{eqnarray}
r_{\rm i}= 3 + Z_2 \pm \sqrt{(3- Z_1)(3+ Z_1 +2 Z_2 )} \ ,
\end{eqnarray}
where
\begin{eqnarray} 
Z_1 &  = & 
1+\left( \sqrt[3]{1+a}+ \sqrt[3]{1-a} \right) 
\sqrt[3]{1-a^2} \, ,
\\ 
Z_2 & = & \sqrt{3 a^2 + Z_1^2} \ .
\end{eqnarray}

\begin{figure}
  \includegraphics[width=0.45\textwidth]{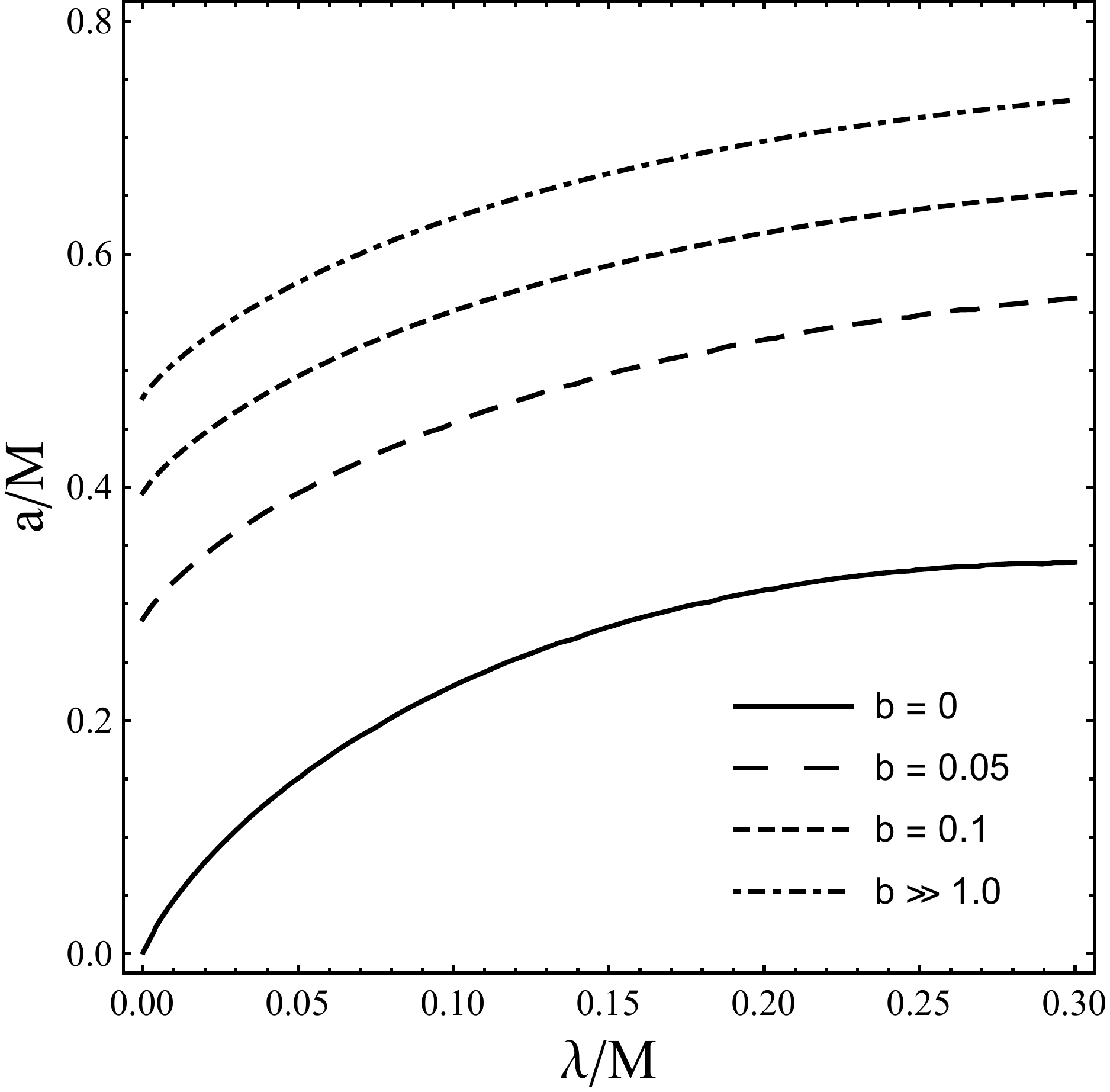}%
\caption{\label{Fig:a_vs_lambda} The degeneracy for the location of the ISCO between the Kerr case and the perfect fluid dark matter case. The plot shows the values of $a$ as a function of $\lambda$ for which the radius of the ISCO in the Kerr geometry is the same as the ISCO in the perfect fluid dark matter geometry. The degeneracy is illustrated for different values of $b$, showing that increase in the external magnetic field allows to mimic higher values of the angular momentum for a given value of $\lambda$. {Note that the degeneracy becomes unchanged for large values of $b\gg1$, indicating that combined effects of dark matter and magnetic field can mimic the black hole rotation parameter up to $ a/M \approx 0.75-0.8$ only. }}
\end{figure}

As can be seen from Fig.~\ref{Fig:a_vs_lambda}, there is a degeneracy for the value of the ISCO radius between perfect fluid dark matter and Kerr space-times. For a given value of $\lambda$ there exist a Kerr geometry with a given value of $a$ that has the same ISCO. The degeneracy is preserved once one introduces the external magnetic field and the effect of the combined dark matter and magnetic field can be mimicked by a Kerr geometry with a larger value of $a$. Or, in other words, test particles under the combined effects of dark matter and magnetic field on circular orbits around a static and spherically symmetric black hole may have the same orbits as particles around a Kerr black hole.

From an observational point of view, far away observers would not be able to distinguish between the two geometries by analyzing electromagnetic radiations emitted by gas on the accretion disk orbiting around the central object. Notice that the combined effects of dark matter and magnetic field can mimic the black hole rotation parameter up to $ a/M \approx 0.75-0.8$, whereas dark matter alone can mimic only up to $a/M\approx 0.35$. 
This suggests that a measurement of the angular momentum of a black hole candidate may be affected up to 30\% by the presence of dark matter in its surroundings and even more if external magnetic fields are present. When applied to astrophysical black holes, the above qualitative argument, suggests that the measurement of highly spinning black hole candidates may be also regarded as due to black holes with lower spin immersed in a dark matter envelope (see for example~\cite{Bambi17-BHs,Walton13,Patrick11b-Seyfert,Patrick11a-Seyfert,Tan12,Gallo05,Gallo11}).

\section{\label{Sec:energy}
Particle collisions}

In this section, we investigate the effect of perfect fluid dark matter on the center of mass energy for collisions of two particles in the vicinity of a static black hole immersed in an external magnetic field. As it is usually done, we consider two particles  
with energies at infinity equal to the rest masses $m_i$ ($i=1,2$). 
Let us then write the four-momentum and the total momentum of the two particles as  
\begin{eqnarray}
\pi_{i}^{\alpha}&=&m_{i}u_{i}^{\alpha}, \\
\pi_{tot}^{\alpha}&=&\pi_{1}^{\alpha}+\pi_{2}^{\alpha}\, ,
\end{eqnarray}
with $u_{i}^{\alpha}$ being the four velocity of the $i$-th particle ($i=1,2$). Following BSW \cite{banados09} we define the general form of the center of mass energy $E_{\rm cm}$ as
\begin{eqnarray}\label{c.m.}
\frac{E_{\rm cm}^{2}}{2m_1 m_2}=\frac{m_{1}^2+m_{2}^2}{2m_1
m_2}-g_{\alpha\beta}u^{\alpha}_{1}u^{\beta}_{2}\, .
\end{eqnarray}
Our aim is to understand the effect of dark matter and external magnetic field on the amount of energy extracted from the collision of two particles. 
In the following, we will focus on a specific collision scenario that was originally developed by Frolov~\cite{Frolov12}. More specifically, we consider the collision between a neutral particle in free fall from spatial infinity with a charged particle revolving at a circular orbit, and more specifically we will consider the ISCO orbit. In~\cite{Frolov12} it was found that for a static and spherically symmetric black hole surrounded by an external magnetic field the scenario leads to arbitrary high center of mass energy.
Of course, there exist several factors that would prevent particle collisions around astrophysical black holes from achieving arbitrary high energies (see for example \cite{Berti09}), however the scenario is interesting as it allows to explore a regime where high energy phenomena may occur.

For convenience, let us denote with  the subscript $n$ the freely falling neutral particle with mass $m_1=m_{n}$ and with the subscript $q$ the charged particle on circular orbit with mass $m_2=m_{q}$.
Then their four momenta shall be $\pi_{n}$ and $\pi_{q}$ and the total momentum $\pi^{\alpha}_{tot}=\pi^{\alpha}_{n}+\pi^{\alpha}_{q}$ for which the center of mass energy of these two particles is written as
\begin{eqnarray}
 \label{Eq:energy1} E^{2}_{\rm cm} &=&-\pi^{\alpha}_{tot}\pi_{tot\alpha }=m^2_{n}+m^2_{q}-2g_{\alpha\beta}\pi^{\alpha}_{n}\pi^{\beta}_{q}\, .
 \end{eqnarray}

The four momentum of a charged particle moving at an arbitrary circular orbit with radius $r$ is 
\begin{eqnarray} \label{pi}
\label{angular} \pi^{\alpha}_q & = & m_q \gamma\left[
\left(\frac{r}{r-2M+\lambda  \log\frac{r}{\vert\lambda \vert}}\right)^{1/2}\delta^{\alpha}_{t}+\frac{1}{r}\upsilon\delta^{\alpha}_{\varphi}\right]\, , \nonumber\\
\end{eqnarray}
with $\upsilon$ being the particle's velocity for the rest frame with Lorentz factor $\gamma=1/\sqrt{1-\upsilon^{2}}$. 
For the sake of clarity, we can rewrite our quantities in terms of $\tilde{L}_q=L_{q}/(2m_qM)$, $\tilde{r}=r/2M$ and $\tilde{\lambda}=\lambda/2M$.
Then the expression for the angular momentum together with $d\varphi/d\varsigma=\upsilon\g/r$ leads to
\begin{eqnarray}
\upsilon\g=\frac{\tilde{L}_q}{\tilde{r}}-b\left[1+\frac{\tilde{\lambda}}{\tilde{r}}\left(1+\log\frac{1}{2\tilde{r}}\right)\right]\tilde{r}\, ,
\end{eqnarray}
with the Lorentz factor $\gamma$ given by
\begin{eqnarray} \label{gamma}
\gamma^{2}=1+\left(\frac{\tilde{L}_q}{\tilde{r}}-b\left[1+\frac{\tilde{\lambda}}{\tilde{r}}\left(1+\log\frac{1}{2\tilde{r}}\right)\right]\tilde{r} \right)^2\, ,
\end{eqnarray}
so that we find the velocity as 
\begin{eqnarray} 
\upsilon=\frac{\tilde{L}_q-b\left[\tilde{r}^2+\tilde{\lambda}\tilde{r}\left(1+\log\frac{1}{2\tilde{r}}\right)\right]}{\sqrt{\tilde{r}^2+\left(\tilde{L}_q-b\left[\tilde{r}^2+\tilde{\lambda}\tilde{r}\left(1+\log\frac{1}{2\tilde{r}}\right)\right] \right)^2}}\, .
\end{eqnarray}

Hence, recalling Eq. \eqref{Eq:energy1}
and employing Eq.~(\ref{pi}) we derive the center of mass energy for the collision of a neutral particle in free fall from spatial infinity with a charged particle orbiting at a circular orbit with radius $r$ as
\begin{eqnarray}
 \label{energy2} E^{2}_{\rm cm} &=&
m^2_{n}+m^2_{q}+2m_{q}\gamma
\nonumber\\ &\times& \left[{E}_{n}\left(\frac{r}{r-2M+\lambda  \log\frac{r}{\vert\lambda \vert}}\right)^{1/2}-\frac{ \upsilon}{r}{L}_{n}\right] \, .
\end{eqnarray}
Notice that the above expression diverges for $r=r_h$. Of course, stable circular orbits in the vicinity of the horizon are generally not allowed as $r_i>r_h$. However, we have seen that the value of the ISCO decreases as we increase the absolute value of the magnetic field (i.e. $b$) and the contribution due to the dark matter part (i.e. $\lambda$). This allows for the extraction of larger collision energies from a black hole immersed in an external magnetic field and surrounded by a dark matter distribution as compared to the Schwarzschild case.
Considering almost radial fall from infinity for the neutral particle one can neglect the second term in the bracket of Eq.~(\ref{energy2}) due to the fact that $L_n$ is small enough as compared to the first term. Hence, Eq.~(\ref{energy2}) yields 
\beq 
E^{2}_{\rm cm} \label{c.m.limit1}\approx \frac{2m_{q} \gamma
{E}_{n}\sqrt{r} }{\sqrt{r-2M+\lambda  \log\frac{r}{\vert\lambda \vert}}}\,. 
\eeq
To evaluate the value of the center of mass energy for the collision between a freely falling neutral particle with ${E}_{n}=m_n$ and a charged particle on circular orbit we need to
determine the angular momentum $\tilde{L}_q$ of the charged particle as a function of the radius of the circular orbit
from $V_{eff}'(\tilde{r},\tilde{L},\tilde{\lambda},b)=0$ similarly to what was done in Eq.\eqref{L} with the addition of the external magnetic field. From the effective potential given in Eq.~(\ref{Veff2}) we get 
\begin{eqnarray} \label{Eq:L:isco}
\tilde{{L}}_q&=& \frac{b~\tilde{r}^2 \left(3 \tilde{r}-1\right)^{1/2} }{\left(3-\tilde{r}\right)^{1/2}}\left\{ 1-\frac{\tilde{\lambda}}{4 b^2  (3 \tilde{r}-1)\tilde{r}^2 }\right.+
\nonumber\\
&-&\frac{\tilde{\lambda} \left(7  +\big((3-\tilde{r})(3 \tilde{r}-1)\big)^{1/2}\right)}{2 (\tilde{r}-3) (3 \tilde{r}-1)}+
\nonumber\\
&+& \left.\frac{ 9\tilde{\lambda} \tilde{r}-8\tilde{\lambda} \tilde{r} \log (\tilde{r}/\vert\tilde{\lambda} \vert)}{2 (\tilde{r}-3) (3 \tilde{r}-1)}\right\} \, .
\end{eqnarray} 

\begin{figure*}

  \includegraphics[width=0.45\textwidth]{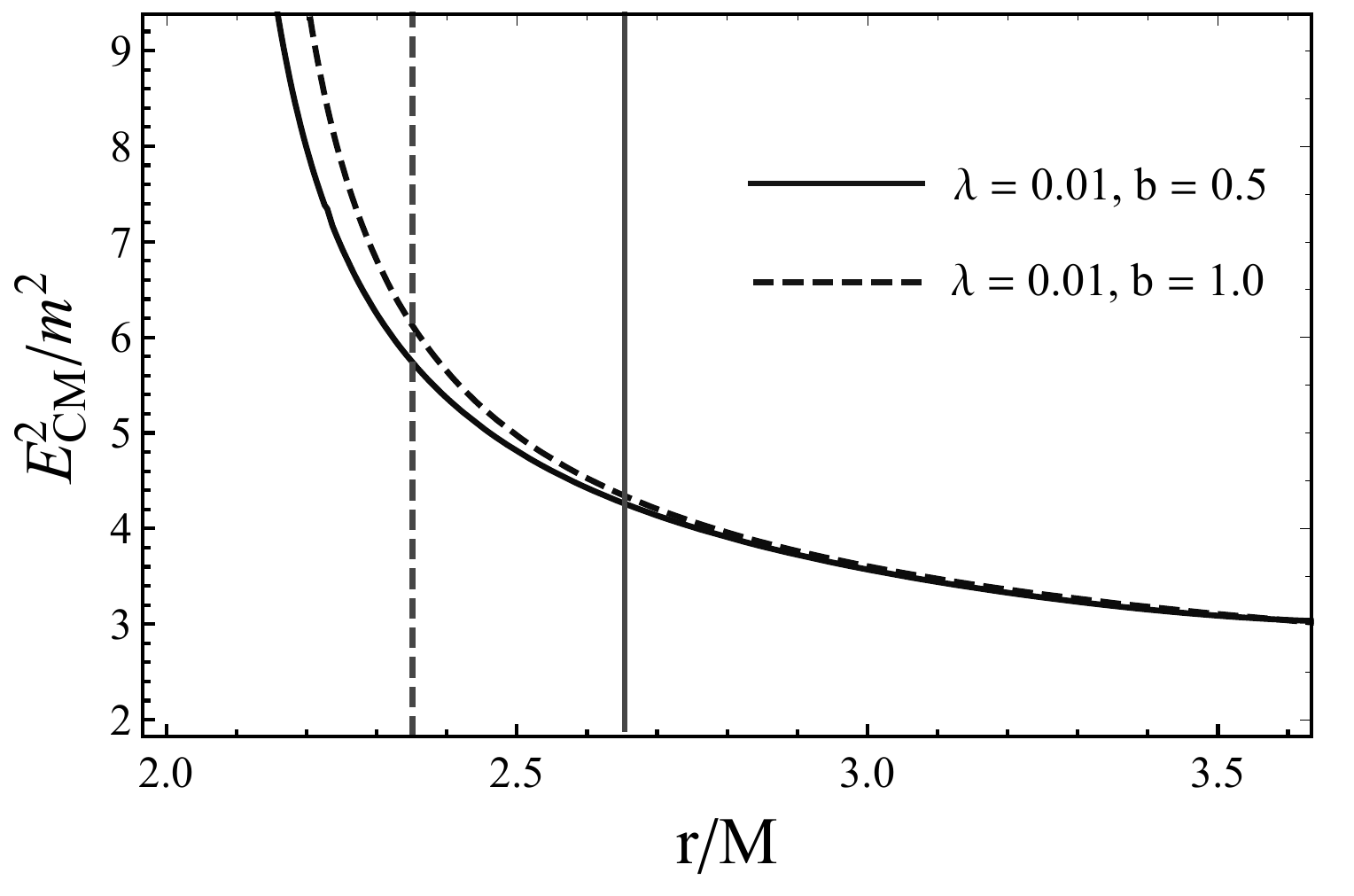}
  \includegraphics[width=0.45\textwidth]{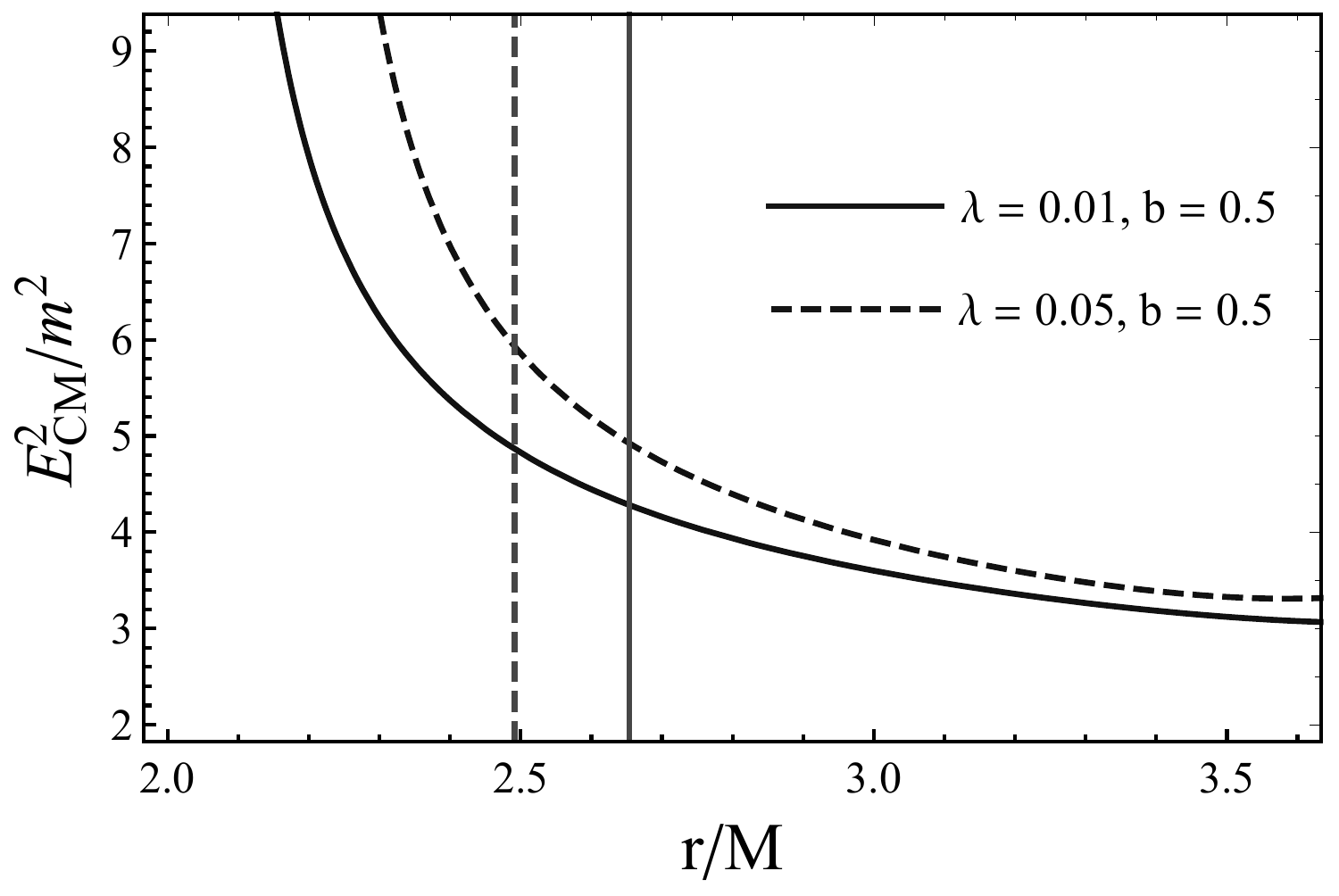}
\caption{\label{Fig:CM} {The center of mass energy $E_{\rm cm}$ for the collision between a neutral particle in free fall from spatial infinity with a charged particle on a circular orbit in the geometry of a black hole immersed in perfect fluid dark matter and an external magnetic field as a function of the radius of the circular orbit $r$. One must keep in mind that $E_{\rm cm}(r)$ here is defined only when stable circular orbits are allowed and therefore, for any given values of $\lambda$ and $b$ the plot will be valid only for $r>r_i$, indicated in the figure by vertical lines. It is easy to notice that at any given radius $E_{\rm cm}$ increases when $\lambda$ and/or $b$ increase.}}
\end{figure*}

We can then express the center of mass energy $E_{\rm cm}(r)$ for the collision as a function of the radius of circular orbit $r$ for any given set of values of $\tilde{\lambda}$ and $b$.
In Fig.~\ref{Fig:CM} one can see the effect of perfect fluid dark matter and the external magnetic field on the extracted energy from the collision. 
The center of mass energy increases when increasing the values of $\lambda$ and $b$.
However, one should keep in mind that $E_{\rm cm}$ is evaluated for charged particles on circular orbits and therefore it is valid only in the regime where stable circular orbits are allowed, that is for $r>r_i$.
It is then useful to evaluate the center of mass energy at the ISCO orbit, which is the maximum possible extracted energy for any given $\lambda$ and $b$. By substituting Eq.~(\ref{Eq:L:isco}) into $V_{eff}''(\tilde{r},\tilde{L},\tilde{\lambda},b)=0$, we derive the following condition that implicitly determines the value of the ISCO 
\begin{eqnarray}\label{Eq:isco}
0&=&\left(4 \tilde{r}^2-9 \tilde{r}+3\right)+\sqrt{(3 \tilde{r}-1) (3-\tilde{r})}+
\nonumber\\
&+&\frac{\tilde{\lambda} \left(2+3 (\tilde{r}-1)^2 \log \frac{\tilde{r}}{\vert\tilde{\lambda} \vert}\right)}{\big((3-\tilde{r}) (3 \tilde{r}-1)\big)^{1/2}}+
\nonumber\\
&-&\frac{\tilde{\lambda} \left( 3 \left(\tilde{r}^2+2 \tilde{r}-3\right) \log \frac{\tilde{r}}{\vert\tilde{\lambda} \vert}-2 \left(4 \tilde{r}^2-9 \tilde{r}+6\right)\right)}{\tilde{r}-3}+
\nonumber\\
&-&\frac{\tilde{\lambda} \left(2 \left(3 \tilde{r}^2-10 \tilde{r}+3\right) \log \frac{\tilde{r}}{\vert\tilde{\lambda} \vert}+6 \tilde{r}^2-11 \tilde{r}+3\right)}{4 b^2 (3 \tilde{r}-1)\tilde{r}^2 }+
\nonumber\\
&-&\frac{\tilde{\lambda} \left(3-\tilde{r}\right)^{1/2}}{4 b^2 \left(3 \tilde{r}-1\right)^{1/2}\tilde{r}^2 }-\frac{(\tilde{\lambda}+1) (3-\tilde{r})}{2 b^2 \tilde{r}^2}\, .
\end{eqnarray}
The equation above allows us to determine $r_i$ in the case in which $\lambda \ll 1$ and $b\gg 1$. In this limit we get 
\begin{eqnarray} \label{Eq:isco_q}
\frac{r_i}{M}&\approx& 2+\frac{2}{\sqrt{3}~b}\left[1+\frac{\lambda}{2M}\left(1+\log \frac{\lambda}{M}\right) +O\left(\lambda^2\right)\right]\nonumber\\&&+O\left(b^{-2}\right)\, ,
\end{eqnarray}
which shows how the ISCO radius approaches the black hole horizon for large values of the magnetic parameter $b$. In the vacuum case, i.e. for $\lambda=0$, the approximate expression for the ISCO radius becomes
\begin{eqnarray}
\frac{r_i}{M}&\approx& 2+\frac{2}{\sqrt{3}~b}+O\left(b^{-2}\right)\, ,
\end{eqnarray}
which corresponds to the result obtained by Frolov and Shoom ~\cite{Frolov10}.
Also it is worth noting that the ISCO radius decreases also in case the dark matter parameter $\lambda$ increases.  

\begin{figure}
\includegraphics[width=0.45\textwidth]{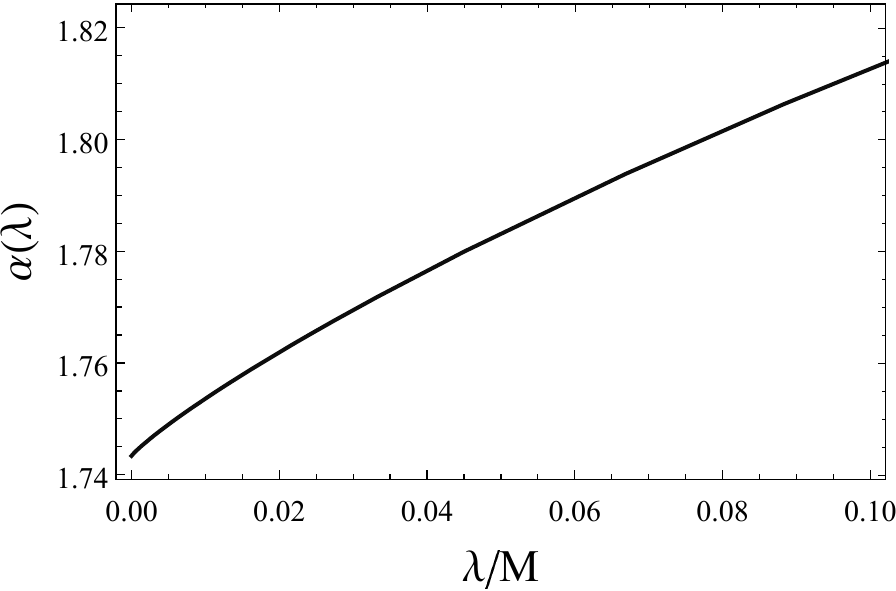} %
\caption{\label{Fig6} \textcolor{black}{The dependence of the function $\alpha(\lambda)$ on the dark matter parameter $\lambda$ for the collision of neutral particles with charged particles orbiting at the ISCO radius. See text for details.} }
\end{figure}

Considering the limiting case of a charged particle orbiting at the ISCO, from Eq.~\eqref{c.m.limit1} we get
\begin{eqnarray}\label{Eq:finalEN}
 \frac{E_{\rm cm}}{m}&\approx& 
 \alpha(\lambda)~b^{1/4}\, ,
 \end{eqnarray}
with $m_q=m_n=m$ and
\be\label{Eq:alpha}
\alpha(\lambda)=(2\gamma)^{1/2}\left[\frac{2\sqrt{3}}{2+\left(1+\log{\lambda}\right){\lambda}}\right]^{1/4}\, .
\ee
In the limiting case of small $\lambda$, with $b\gg1$ and evaluating $\gamma$ at the ISCO we can rewrite Eq.~(\ref{Eq:alpha}) as 
 \begin{eqnarray}
  \alpha(\lambda)&=&\sqrt{2}\left[\frac{2}{\sqrt{3}}+\left(\frac{31 }{24 \sqrt{3}}+\frac{\log {\lambda}}{6 \sqrt{3}}\right)\lambda\right]^{1/2}\nonumber\\&\times&\left(\frac{2\sqrt{3}}{2+\left(1+\log{\lambda}\right){\lambda}}\right)^{1/4}\, .
 \end{eqnarray}
It is not difficult to show that $\alpha$ is an increasing function of $\lambda$ as it can be seen from Fig.~\ref{Fig6}, which helps to understand the dependence of $E_{\rm cm}$ on the dark matter parameter as seen in Fig.~\ref{Fig:CM}. 
 
Also, the above expression becomes $\alpha(0)=2/(3^{1/8})$ in the vacuum case. Thus, the expression (\ref{Eq:finalEN}) represents the center of mass energy for collision of the charged and neutral particles at the ISCO. In the limit $\lambda \rightarrow 0$, one can easily recover the result of Ref.~\cite{Frolov12}
 \begin{eqnarray}
 \frac{E_{\rm cm}}{m}\approx 1.74337~b^{1/4}\, .
 \end{eqnarray}
In table~\ref{4tab}, we show the collision energy $E_{\rm cm}$ and the radius of the ISCO $r_i$ for different values of the dark matter parameter $\lambda$. It is immediately clear that, while the value of $r_i$ decreases with $\lambda$, the center of mass energy extracted via the collision increases with increasing value of the dark matter parameter.
Therefore in the presence of dark matter one always obtains a larger value for $E_{\rm cm}$ with respect to vacuum case. Also, as $b$ grows, $E_{\rm cm}$ increases, diverging in the limit of $r_i\rightarrow r_h$ obtained for $b\rightarrow +\infty$.

\begin{table}
\begin{center}
\caption{\label{4tab} The values of the ISCO radius $r_i$ and the center of mass energy $E_{\rm cm}/m$ for the collision of charged particles orbiting at the ISCO with neutral particles or radial fall for different values of the dark matter parameter $\lambda$ at fixed values of the magnetic field parameter $b$. }
\begin{tabular}{c c  c}
\\ \hline
$\lambda$  & $r_{i}$   & ${E_{\rm cm}}/{m}$\\[1.0ex]\hline
 0.000      &$2+1.15470~b^{-1}$    &$1.74337 ~b^{1/4}$\\[1ex]
 0.001      &$2+1.15129~b^{-1}$    &$1.74472 ~b^{1/4}$\\[1ex]
 0.005      &$2+1.14229~b^{-1}$    &$1.74898 ~b^{1/4}$\\[1ex]
 0.010      &$2+1.13389~b^{-1}$    &$1.75361 ~b^{1/4}$\\[1ex]
 0.050      &$2+1.09709~b^{-1}$    &$1.78322 ~b^{1/4}$\\[1ex]
 0.100      &$2+1.07950~b^{-1}$    &$1.81277~b^{1/4}$\\ \hline
\end{tabular}
\end{center}
\end{table}

\section{Conclusions}
\label{Sec:conclusion}

Accretion disks around super-massive black hole candidates are the primary source of information about gravity in the strong field regime and the geometry surrounding such black hole candidates \cite{Abramowicz13}.  
Energetic particles are produced by collisions in the accretion disk and the disk's luminosity depends on the underlying geometry. 
However, in a realistic scenario the object can not be considered to be in vacuum, as we know that dark matter distributions exist at the center of galaxies. Also magnetic fields play an important role in the dynamics of charged particles around black holes, especially close to the black hole's horizon.
Therefore, in order to have confidence in the conclusions drawn from the observations of accretion disks, it is important to study the effects that the presence of external matter fields and magnetic fields have on the particles in the disks.

In this paper, we studied particle motion in the vicinity of a static, spherically symmetric black hole surrounded by perfect fluid dark matter and immersed in an external asymptotically uniform magnetic field. 
We showed that the radius of the innermost stable circular orbit for charged particles decreases under the effects of dark matter and external magnetic field. 
A similar behaviour occurs for the unstable photon orbits.  

We showed that the combined effects of dark matter and magnetic field can cancel each other out only at a specific radius depending on the values of the dark matter and magnetic parameters. This suggests that the emitted spectrum of an accretion disk in the geometry considered here could be distinguished from the spectrum of the disk around a static black hole.

Further, we showed that the orbits of test particles under the combined effects of dark matter and magnetic field may mimic the same orbits around a Kerr black hole. From an observational point of view, this result suggests that the determination of the ISCO from the observation of electromagnetic radiation emitted by the accretion disk could not suffice to establish the value of the source's angular momentum. In fact it would not be possible for far away observers to distinguish between a Kerr black hole in vacuum from a black hole with smaller angular momentum and immersed in a dark matter envelope.

Also, as a consequence of the decrease in value for the ISCO, we showed that the center of mass energy for the collision of neutral and charged particles increases under the effects of both dark matter and external magnetic field.
In this work, following \cite{Frolov12}, we considered the simple collision scenario of a freely falling neutral particle colliding with a charged particle revolving at a circular orbit. We showed that the center of mass energy becomes maximum for charged particles at the ISCO. And since the ISCO can become arbitrarily close to the horizon as the magnetic field strength increases, we showed that the center of mass energy can become arbitrarily large.

These theoretical studies can help constraint the validity of alternative models to black holes in explaining astrophysical observations.

\section{Acknowledgments}
S.S. and B.A. acknowledge Nazarbayev University, Nur-Sultan, Kazakhstan for warm hospitality. This research is supported in part by the Abdus Salam International Centre for Theoretical Physics under the Grant No. OEA-NT-01. D.M. acknowledges support from Nazarbayev University Faculty Development Competitive Research Grant No. 090118FD5348.

%
\section*{References}

\bibliographystyle{apsrev4-1}
\bibliography{gravreferences}

\begin{thebibliography}{109}%
\makeatletter
\providecommand \@ifxundefined [1]{%
 \@ifx{#1\undefined}
}%
\providecommand \@ifnum [1]{%
 \ifnum #1\expandafter \@firstoftwo
 \else \expandafter \@secondoftwo
 \fi
}%
\providecommand \@ifx [1]{%
 \ifx #1\expandafter \@firstoftwo
 \else \expandafter \@secondoftwo
 \fi
}%
\providecommand \natexlab [1]{#1}%
\providecommand \enquote  [1]{``#1''}%
\providecommand \bibnamefont  [1]{#1}%
\providecommand \bibfnamefont [1]{#1}%
\providecommand \citenamefont [1]{#1}%
\providecommand \href@noop [0]{\@secondoftwo}%
\providecommand \href [0]{\begingroup \@sanitize@url \@href}%
\providecommand \@href[1]{\@@startlink{#1}\@@href}%
\providecommand \@@href[1]{\endgroup#1\@@endlink}%
\providecommand \@sanitize@url [0]{\catcode `\\12\catcode `\$12\catcode
  `\&12\catcode `\#12\catcode `\^12\catcode `\_12\catcode `\%12\relax}%
\providecommand \@@startlink[1]{}%
\providecommand \@@endlink[0]{}%
\providecommand \url  [0]{\begingroup\@sanitize@url \@url }%
\providecommand \@url [1]{\endgroup\@href {#1}{\urlprefix }}%
\providecommand \urlprefix  [0]{URL }%
\providecommand \Eprint [0]{\href }%
\providecommand \doibase [0]{http://dx.doi.org/}%
\providecommand \selectlanguage [0]{\@gobble}%
\providecommand \bibinfo  [0]{\@secondoftwo}%
\providecommand \bibfield  [0]{\@secondoftwo}%
\providecommand \translation [1]{[#1]}%
\providecommand \BibitemOpen [0]{}%
\providecommand \bibitemStop [0]{}%
\providecommand \bibitemNoStop [0]{.\EOS\space}%
\providecommand \EOS [0]{\spacefactor3000\relax}%
\providecommand \BibitemShut  [1]{\csname bibitem#1\endcsname}%
\let\auto@bib@innerbib\@empty
\bibitem [{\citenamefont {{Abbott}}\ and\ \citenamefont {et~al. {(Virgo and
  LIGO Scientific Collaborations)}}(2016{\natexlab{a}})}]{Abbott16a}%
  \BibitemOpen
  \bibfield  {author} {\bibinfo {author} {\bibfnamefont {B.~P.}\ \bibnamefont
  {{Abbott}}}\ and\ \bibinfo {author} {\bibnamefont {et~al. {(Virgo and LIGO
  Scientific Collaborations)}}},\ }\href {\doibase
  10.1103/PhysRevLett.116.061102} {\bibfield  {journal} {\bibinfo  {journal}
  {Phys. Rev. Lett.}\ }\textbf {\bibinfo {volume} {116}},\ \bibinfo {eid}
  {061102} (\bibinfo {year} {2016}{\natexlab{a}})},\ \Eprint
  {http://arxiv.org/abs/1602.03837} {arXiv:1602.03837 [gr-qc]} \BibitemShut
  {NoStop}%
\bibitem [{\citenamefont {{Abbott}}\ and\ \citenamefont {et~al. {(Virgo and
  LIGO Scientific Collaborations)}}(2016{\natexlab{b}})}]{Abbott16b}%
  \BibitemOpen
  \bibfield  {author} {\bibinfo {author} {\bibfnamefont {B.~P.}\ \bibnamefont
  {{Abbott}}}\ and\ \bibinfo {author} {\bibnamefont {et~al. {(Virgo and LIGO
  Scientific Collaborations)}}},\ }\href {\doibase
  10.1103/PhysRevLett.116.241102} {\bibfield  {journal} {\bibinfo  {journal}
  {Phys. Rev. Lett.}\ }\textbf {\bibinfo {volume} {116}},\ \bibinfo {eid}
  {241102} (\bibinfo {year} {2016}{\natexlab{b}})},\ \Eprint
  {http://arxiv.org/abs/1602.03840} {arXiv:1602.03840 [gr-qc]} \BibitemShut
  {NoStop}%
\bibitem [{\citenamefont {{Akiyama}}\ and\ \citenamefont {et~al. {(Event
  Horizon Telescope Collaboration)}}(2019{\natexlab{a}})}]{Akiyama19L1}%
  \BibitemOpen
  \bibfield  {author} {\bibinfo {author} {\bibfnamefont {K.}~\bibnamefont
  {{Akiyama}}}\ and\ \bibinfo {author} {\bibnamefont {et~al. {(Event Horizon
  Telescope Collaboration)}}},\ }\href {\doibase 10.3847/2041-8213/ab0ec7}
  {\bibfield  {journal} {\bibinfo  {journal} {Astrophys. J.}\ }\textbf
  {\bibinfo {volume} {875}},\ \bibinfo {eid} {L1} (\bibinfo {year}
  {2019}{\natexlab{a}})},\ \Eprint {http://arxiv.org/abs/1906.11238}
  {arXiv:1906.11238 [astro-ph.GA]} \BibitemShut {NoStop}%
\bibitem [{\citenamefont {{Akiyama}}\ and\ \citenamefont {et~al. {(Event
  Horizon Telescope Collaboration)}}(2019{\natexlab{b}})}]{Akiyama19L6}%
  \BibitemOpen
  \bibfield  {author} {\bibinfo {author} {\bibfnamefont {K.}~\bibnamefont
  {{Akiyama}}}\ and\ \bibinfo {author} {\bibnamefont {et~al. {(Event Horizon
  Telescope Collaboration)}}},\ }\href {\doibase 10.3847/2041-8213/ab1141}
  {\bibfield  {journal} {\bibinfo  {journal} {Astrophys. J.}\ }\textbf
  {\bibinfo {volume} {875}},\ \bibinfo {eid} {L6} (\bibinfo {year}
  {2019}{\natexlab{b}})},\ \Eprint {http://arxiv.org/abs/1906.11243}
  {arXiv:1906.11243 [astro-ph.GA]} \BibitemShut {NoStop}%
\bibitem [{\citenamefont {{Johannsen}}\ and\ \citenamefont
  {{Psaltis}}(2011)}]{Johannsen11}%
  \BibitemOpen
  \bibfield  {author} {\bibinfo {author} {\bibfnamefont {T.}~\bibnamefont
  {{Johannsen}}}\ and\ \bibinfo {author} {\bibfnamefont {D.}~\bibnamefont
  {{Psaltis}}},\ }\href {\doibase 10.1103/PhysRevD.83.124015} {\bibfield
  {journal} {\bibinfo  {journal} {Phys. Rev. D}\ }\textbf {\bibinfo {volume}
  {83}},\ \bibinfo {eid} {124015} (\bibinfo {year} {2011})},\ \Eprint
  {http://arxiv.org/abs/1105.3191} {arXiv:1105.3191 [gr-qc]} \BibitemShut
  {NoStop}%
\bibitem [{\citenamefont {{Bambi}}\ and\ \citenamefont
  {{Malafarina}}(2013)}]{Bambi-Malafarina13}%
  \BibitemOpen
  \bibfield  {author} {\bibinfo {author} {\bibfnamefont {C.}~\bibnamefont
  {{Bambi}}}\ and\ \bibinfo {author} {\bibfnamefont {D.}~\bibnamefont
  {{Malafarina}}},\ }\href {\doibase 10.1103/PhysRevD.88.064022} {\bibfield
  {journal} {\bibinfo  {journal} {Phys. Rev. D}\ }\textbf {\bibinfo {volume}
  {88}},\ \bibinfo {eid} {064022} (\bibinfo {year} {2013})},\ \Eprint
  {http://arxiv.org/abs/1307.2106} {arXiv:1307.2106 [gr-qc]} \BibitemShut
  {NoStop}%
\bibitem [{\citenamefont {{Herrera}}\ \emph {et~al.}(2000)\citenamefont
  {{Herrera}}, \citenamefont {{Paiva}}, \citenamefont {{Santos}},\ and\
  \citenamefont {{Ferrari}}}]{Herrera00}%
  \BibitemOpen
  \bibfield  {author} {\bibinfo {author} {\bibfnamefont {L.}~\bibnamefont
  {{Herrera}}}, \bibinfo {author} {\bibfnamefont {F.~M.}\ \bibnamefont
  {{Paiva}}}, \bibinfo {author} {\bibfnamefont {N.~O.}\ \bibnamefont
  {{Santos}}}, \ and\ \bibinfo {author} {\bibfnamefont {V.}~\bibnamefont
  {{Ferrari}}},\ }\href {\doibase 10.1142/S021827180000061X} {\bibfield
  {journal} {\bibinfo  {journal} {‎Int. J. Mod. Phys. D}\ }\textbf {\bibinfo
  {volume} {9}},\ \bibinfo {pages} {649} (\bibinfo {year} {2000})},\ \Eprint
  {http://arxiv.org/abs/gr-qc/9812023} {arXiv:gr-qc/9812023 [gr-qc]}
  \BibitemShut {NoStop}%
\bibitem [{\citenamefont {{Herrera}}\ \emph {et~al.}(2005)\citenamefont
  {{Herrera}}, \citenamefont {{Magli}},\ and\ \citenamefont
  {{Malafarina}}}]{Herrera05}%
  \BibitemOpen
  \bibfield  {author} {\bibinfo {author} {\bibfnamefont {L.}~\bibnamefont
  {{Herrera}}}, \bibinfo {author} {\bibfnamefont {G.}~\bibnamefont {{Magli}}},
  \ and\ \bibinfo {author} {\bibfnamefont {D.}~\bibnamefont {{Malafarina}}},\
  }\href {\doibase 10.1007/s10714-005-0120-1} {\bibfield  {journal} {\bibinfo
  {journal} {Gen. Relativ. Gravit.}\ }\textbf {\bibinfo {volume} {37}},\
  \bibinfo {pages} {1371} (\bibinfo {year} {2005})},\ \Eprint
  {http://arxiv.org/abs/gr-qc/0407037} {arXiv:gr-qc/0407037 [gr-qc]}
  \BibitemShut {NoStop}%
\bibitem [{\citenamefont {{Bini}}\ \emph {et~al.}(2012)\citenamefont {{Bini}},
  \citenamefont {{Boshkayev}},\ and\ \citenamefont {{Geralico}}}]{Bini12}%
  \BibitemOpen
  \bibfield  {author} {\bibinfo {author} {\bibfnamefont {D.}~\bibnamefont
  {{Bini}}}, \bibinfo {author} {\bibfnamefont {K.}~\bibnamefont {{Boshkayev}}},
  \ and\ \bibinfo {author} {\bibfnamefont {A.}~\bibnamefont {{Geralico}}},\
  }\href {\doibase 10.1088/0264-9381/29/14/145003} {\bibfield  {journal}
  {\bibinfo  {journal} {Class. Quantum Grav.}\ }\textbf {\bibinfo {volume}
  {29}},\ \bibinfo {eid} {145003} (\bibinfo {year} {2012})},\ \Eprint
  {http://arxiv.org/abs/1306.4803} {arXiv:1306.4803 [gr-qc]} \BibitemShut
  {NoStop}%
\bibitem [{\citenamefont {{Toshmatov}}\ and\ \citenamefont
  {{Malafarina}}(2019)}]{Toshmatov19d}%
  \BibitemOpen
  \bibfield  {author} {\bibinfo {author} {\bibfnamefont {B.}~\bibnamefont
  {{Toshmatov}}}\ and\ \bibinfo {author} {\bibfnamefont {D.}~\bibnamefont
  {{Malafarina}}},\ }\href {\doibase 10.1103/PhysRevD.100.104052} {\bibfield
  {journal} {\bibinfo  {journal} {Phys. Rev. D}\ }\textbf {\bibinfo {volume}
  {100}},\ \bibinfo {eid} {104052} (\bibinfo {year} {2019})},\ \Eprint
  {http://arxiv.org/abs/1910.11565} {arXiv:1910.11565 [gr-qc]} \BibitemShut
  {NoStop}%
\bibitem [{\citenamefont {{Wald}}(1974)}]{Wald74}%
  \BibitemOpen
  \bibfield  {author} {\bibinfo {author} {\bibfnamefont {R.~M.}\ \bibnamefont
  {{Wald}}},\ }\href {\doibase 10.1103/PhysRevD.10.1680} {\bibfield  {journal}
  {\bibinfo  {journal} {Phys. Rev. D}\ }\textbf {\bibinfo {volume} {10}},\
  \bibinfo {pages} {1680} (\bibinfo {year} {1974})}\BibitemShut {NoStop}%
\bibitem [{\citenamefont {{Benavides-Gallego}}\ \emph
  {et~al.}(2019)\citenamefont {{Benavides-Gallego}}, \citenamefont
  {{Abdujabbarov}}, \citenamefont {{Malafarina}}, \citenamefont {{Ahmedov}},\
  and\ \citenamefont {{Bambi}}}]{Benavides-Gallego19}%
  \BibitemOpen
  \bibfield  {author} {\bibinfo {author} {\bibfnamefont {C.~A.}\ \bibnamefont
  {{Benavides-Gallego}}}, \bibinfo {author} {\bibfnamefont {A.}~\bibnamefont
  {{Abdujabbarov}}}, \bibinfo {author} {\bibfnamefont {D.}~\bibnamefont
  {{Malafarina}}}, \bibinfo {author} {\bibfnamefont {B.}~\bibnamefont
  {{Ahmedov}}}, \ and\ \bibinfo {author} {\bibfnamefont {C.}~\bibnamefont
  {{Bambi}}},\ }\href {\doibase 10.1103/PhysRevD.99.044012} {\bibfield
  {journal} {\bibinfo  {journal} {Phys. Rev. D}\ }\textbf {\bibinfo {volume}
  {99}},\ \bibinfo {eid} {044012} (\bibinfo {year} {2019})},\ \Eprint
  {http://arxiv.org/abs/1812.04846} {arXiv:1812.04846 [gr-qc]} \BibitemShut
  {NoStop}%
\bibitem [{\citenamefont {{Tsukamoto}}\ and\ \citenamefont
  {{Harada}}(2013)}]{Tsukamoto13}%
  \BibitemOpen
  \bibfield  {author} {\bibinfo {author} {\bibfnamefont {N.}~\bibnamefont
  {{Tsukamoto}}}\ and\ \bibinfo {author} {\bibfnamefont {T.}~\bibnamefont
  {{Harada}}},\ }\href {\doibase 10.3390/galaxies1030261} {\bibfield  {journal}
  {\bibinfo  {journal} {Galaxies}\ }\textbf {\bibinfo {volume} {1}},\ \bibinfo
  {pages} {261} (\bibinfo {year} {2013})},\ \Eprint
  {http://arxiv.org/abs/1201.1738} {arXiv:1201.1738 [gr-qc]} \BibitemShut
  {NoStop}%
\bibitem [{\citenamefont {{Dey}}\ and\ \citenamefont
  {{Joshi}}(2019)}]{Joshi19}%
  \BibitemOpen
  \bibfield  {author} {\bibinfo {author} {\bibfnamefont {D.}~\bibnamefont
  {{Dey}}}\ and\ \bibinfo {author} {\bibfnamefont {P.~S.}\ \bibnamefont
  {{Joshi}}},\ }\href@noop {} {\bibfield  {journal} {\bibinfo  {journal} {arXiv
  e-prints}\ ,\ \bibinfo {eid} {arXiv:1907.12738}} (\bibinfo {year} {2019})},\
  \Eprint {http://arxiv.org/abs/1907.12738} {arXiv:1907.12738 [gr-qc]}
  \BibitemShut {NoStop}%
\bibitem [{\citenamefont {{Prasanna}}(1980)}]{Prasanna80}%
  \BibitemOpen
  \bibfield  {author} {\bibinfo {author} {\bibfnamefont {A.~R.}\ \bibnamefont
  {{Prasanna}}},\ }\href {\doibase 10.1007/BF02724339} {\bibfield  {journal}
  {\bibinfo  {journal} {Nuovo Cimento Rivista Serie}\ }\textbf {\bibinfo
  {volume} {3}},\ \bibinfo {pages} {1} (\bibinfo {year} {1980})}\BibitemShut
  {NoStop}%
\bibitem [{\citenamefont {{Kov{\'a}{\v r}}}\ \emph {et~al.}(2008)\citenamefont
  {{Kov{\'a}{\v r}}}, \citenamefont {{Stuchl{\'{\i}}k}},\ and\ \citenamefont
  {{Karas}}}]{Kovar08}%
  \BibitemOpen
  \bibfield  {author} {\bibinfo {author} {\bibfnamefont {J.}~\bibnamefont
  {{Kov{\'a}{\v r}}}}, \bibinfo {author} {\bibfnamefont {Z.}~\bibnamefont
  {{Stuchl{\'{\i}}k}}}, \ and\ \bibinfo {author} {\bibfnamefont
  {V.}~\bibnamefont {{Karas}}},\ }\href {\doibase
  10.1088/0264-9381/25/9/095011} {\bibfield  {journal} {\bibinfo  {journal}
  {Class. Quantum Grav.}\ }\textbf {\bibinfo {volume} {25}},\ \bibinfo {eid}
  {095011} (\bibinfo {year} {2008})},\ \Eprint {http://arxiv.org/abs/0803.3155}
  {arXiv:0803.3155} \BibitemShut {NoStop}%
\bibitem [{\citenamefont {{Kov{\'a}{\v r}}}\ \emph {et~al.}(2010)\citenamefont
  {{Kov{\'a}{\v r}}}, \citenamefont {{Kop{\'a}{\v c}ek}}, \citenamefont
  {{Karas}},\ and\ \citenamefont {{Stuchl{\'{\i}}k}}}]{Kovar10}%
  \BibitemOpen
  \bibfield  {author} {\bibinfo {author} {\bibfnamefont {J.}~\bibnamefont
  {{Kov{\'a}{\v r}}}}, \bibinfo {author} {\bibfnamefont {O.}~\bibnamefont
  {{Kop{\'a}{\v c}ek}}}, \bibinfo {author} {\bibfnamefont {V.}~\bibnamefont
  {{Karas}}}, \ and\ \bibinfo {author} {\bibfnamefont {Z.}~\bibnamefont
  {{Stuchl{\'{\i}}k}}},\ }\href {\doibase 10.1088/0264-9381/27/13/135006}
  {\bibfield  {journal} {\bibinfo  {journal} {Class. Quantum Grav.}\ }\textbf
  {\bibinfo {volume} {27}},\ \bibinfo {eid} {135006} (\bibinfo {year}
  {2010})},\ \Eprint {http://arxiv.org/abs/1005.3270} {arXiv:1005.3270
  [astro-ph.HE]} \BibitemShut {NoStop}%
\bibitem [{\citenamefont {{Shaymatov}}\ \emph {et~al.}(2015)\citenamefont
  {{Shaymatov}}, \citenamefont {{Patil}}, \citenamefont {{Ahmedov}},\ and\
  \citenamefont {{Joshi}}}]{Shaymatov15}%
  \BibitemOpen
  \bibfield  {author} {\bibinfo {author} {\bibfnamefont {S.}~\bibnamefont
  {{Shaymatov}}}, \bibinfo {author} {\bibfnamefont {M.}~\bibnamefont
  {{Patil}}}, \bibinfo {author} {\bibfnamefont {B.}~\bibnamefont {{Ahmedov}}},
  \ and\ \bibinfo {author} {\bibfnamefont {P.~S.}\ \bibnamefont {{Joshi}}},\
  }\href {\doibase 10.1103/PhysRevD.91.064025} {\bibfield  {journal} {\bibinfo
  {journal} {Phys. Rev. D}\ }\textbf {\bibinfo {volume} {91}},\ \bibinfo {eid}
  {064025} (\bibinfo {year} {2015})},\ \Eprint {http://arxiv.org/abs/1409.3018}
  {arXiv:1409.3018 [gr-qc]} \BibitemShut {NoStop}%
\bibitem [{\citenamefont {{Dadhich}}\ \emph {et~al.}(2018)\citenamefont
  {{Dadhich}}, \citenamefont {{Tursunov}}, \citenamefont {{Ahmedov}},\ and\
  \citenamefont {{Stuchl{\'\i}k}}}]{Dadhich18}%
  \BibitemOpen
  \bibfield  {author} {\bibinfo {author} {\bibfnamefont {N.}~\bibnamefont
  {{Dadhich}}}, \bibinfo {author} {\bibfnamefont {A.}~\bibnamefont
  {{Tursunov}}}, \bibinfo {author} {\bibfnamefont {B.}~\bibnamefont
  {{Ahmedov}}}, \ and\ \bibinfo {author} {\bibfnamefont {Z.}~\bibnamefont
  {{Stuchl{\'\i}k}}},\ }\href {\doibase 10.1093/mnrasl/sly073} {\bibfield
  {journal} {\bibinfo  {journal} {Mon. Not. Roy. Astron. Soc.}\ }\textbf
  {\bibinfo {volume} {478}},\ \bibinfo {pages} {L89} (\bibinfo {year}
  {2018})},\ \Eprint {http://arxiv.org/abs/1804.09679} {arXiv:1804.09679
  [astro-ph.HE]} \BibitemShut {NoStop}%
\bibitem [{\citenamefont {{Narzilloev}}\ \emph {et~al.}(2019)\citenamefont
  {{Narzilloev}}, \citenamefont {{Abdujabbarov}}, \citenamefont {{Bambi}},\
  and\ \citenamefont {{Ahmedov}}}]{Narzilloev19}%
  \BibitemOpen
  \bibfield  {author} {\bibinfo {author} {\bibfnamefont {B.}~\bibnamefont
  {{Narzilloev}}}, \bibinfo {author} {\bibfnamefont {A.}~\bibnamefont
  {{Abdujabbarov}}}, \bibinfo {author} {\bibfnamefont {C.}~\bibnamefont
  {{Bambi}}}, \ and\ \bibinfo {author} {\bibfnamefont {B.}~\bibnamefont
  {{Ahmedov}}},\ }\href {\doibase 10.1103/PhysRevD.99.104009} {\bibfield
  {journal} {\bibinfo  {journal} {Phys. Rev. D}\ }\textbf {\bibinfo {volume}
  {99}},\ \bibinfo {eid} {104009} (\bibinfo {year} {2019})},\ \Eprint
  {http://arxiv.org/abs/1902.03414} {arXiv:1902.03414 [gr-qc]} \BibitemShut
  {NoStop}%
\bibitem [{\citenamefont {{Pavlovi{\'c}}}\ \emph {et~al.}(2019)\citenamefont
  {{Pavlovi{\'c}}}, \citenamefont {{Saveliev}},\ and\ \citenamefont
  {{Sossich}}}]{Pavlovic19}%
  \BibitemOpen
  \bibfield  {author} {\bibinfo {author} {\bibfnamefont {P.}~\bibnamefont
  {{Pavlovi{\'c}}}}, \bibinfo {author} {\bibfnamefont {A.}~\bibnamefont
  {{Saveliev}}}, \ and\ \bibinfo {author} {\bibfnamefont {M.}~\bibnamefont
  {{Sossich}}},\ }\href {\doibase 10.1103/PhysRevD.100.084033} {\bibfield
  {journal} {\bibinfo  {journal} {Phys. Rev. D}\ }\textbf {\bibinfo {volume}
  {100}},\ \bibinfo {eid} {084033} (\bibinfo {year} {2019})},\ \Eprint
  {http://arxiv.org/abs/1908.01888} {arXiv:1908.01888 [gr-qc]} \BibitemShut
  {NoStop}%
\bibitem [{\citenamefont {{Shaymatov}}(2019)}]{Shaymatov19b}%
  \BibitemOpen
  \bibfield  {author} {\bibinfo {author} {\bibfnamefont {S.}~\bibnamefont
  {{Shaymatov}}},\ }\href {\doibase 10.1142/S2010194519600206} {\bibfield
  {journal} {\bibinfo  {journal} {Int. J. Mod. Phys. Conf. Ser.}\ }\textbf
  {\bibinfo {volume} {49}},\ \bibinfo {eid} {1960020} (\bibinfo {year}
  {2019})}\BibitemShut {NoStop}%
\bibitem [{\citenamefont {{D{\"u}zta{\c{s}}}}\ \emph
  {et~al.}(2020)\citenamefont {{D{\"u}zta{\c{s}}}}, \citenamefont {{Jamil}},
  \citenamefont {{Shaymatov}},\ and\ \citenamefont
  {{Ahmedov}}}]{Duztas-Jamil20}%
  \BibitemOpen
  \bibfield  {author} {\bibinfo {author} {\bibfnamefont {K.}~\bibnamefont
  {{D{\"u}zta{\c{s}}}}}, \bibinfo {author} {\bibfnamefont {M.}~\bibnamefont
  {{Jamil}}}, \bibinfo {author} {\bibfnamefont {S.}~\bibnamefont
  {{Shaymatov}}}, \ and\ \bibinfo {author} {\bibfnamefont {B.}~\bibnamefont
  {{Ahmedov}}},\ }\href {\doibase 10.1088/1361-6382/ab9d96} {\bibfield
  {journal} {\bibinfo  {journal} {Class. Quantum Grav.}\ }\textbf {\bibinfo
  {volume} {37}},\ \bibinfo {pages} {175005} (\bibinfo {year} {2020})},\
  \Eprint {http://arxiv.org/abs/1808.04711} {arXiv:1808.04711 [gr-qc]}
  \BibitemShut {NoStop}%
\bibitem [{\citenamefont {{Stuchl{\'\i}k}}\ \emph {et~al.}(2020)\citenamefont
  {{Stuchl{\'\i}k}}, \citenamefont {{Kolo{\v{s}}}}, \citenamefont
  {{Kov{\'a}{\v{r}}}}, \citenamefont {{Slan{\'y}}},\ and\ \citenamefont
  {{Tursunov}}}]{Stuchlik20}%
  \BibitemOpen
  \bibfield  {author} {\bibinfo {author} {\bibfnamefont {Z.}~\bibnamefont
  {{Stuchl{\'\i}k}}}, \bibinfo {author} {\bibfnamefont {M.}~\bibnamefont
  {{Kolo{\v{s}}}}}, \bibinfo {author} {\bibfnamefont {J.}~\bibnamefont
  {{Kov{\'a}{\v{r}}}}}, \bibinfo {author} {\bibfnamefont {P.}~\bibnamefont
  {{Slan{\'y}}}}, \ and\ \bibinfo {author} {\bibfnamefont {A.}~\bibnamefont
  {{Tursunov}}},\ }\href {\doibase 10.3390/universe6020026} {\bibfield
  {journal} {\bibinfo  {journal} {Universe}\ }\textbf {\bibinfo {volume} {6}},\
  \bibinfo {pages} {26} (\bibinfo {year} {2020})}\BibitemShut {NoStop}%
\bibitem [{\citenamefont {{Shaymatov}}\ \emph {et~al.}(2020)\citenamefont
  {{Shaymatov}}, \citenamefont {{Vrba}}, \citenamefont {{Malafarina}},
  \citenamefont {{Ahmedov}},\ and\ \citenamefont
  {{Stuchl{\'\i}k}}}]{Shaymatov20egb}%
  \BibitemOpen
  \bibfield  {author} {\bibinfo {author} {\bibfnamefont {S.}~\bibnamefont
  {{Shaymatov}}}, \bibinfo {author} {\bibfnamefont {J.}~\bibnamefont {{Vrba}}},
  \bibinfo {author} {\bibfnamefont {D.}~\bibnamefont {{Malafarina}}}, \bibinfo
  {author} {\bibfnamefont {B.}~\bibnamefont {{Ahmedov}}}, \ and\ \bibinfo
  {author} {\bibfnamefont {Z.}~\bibnamefont {{Stuchl{\'\i}k}}},\ }\href
  {\doibase 10.1016/j.dark.2020.100648} {\bibfield  {journal} {\bibinfo
  {journal} {Phys. Dark Universe}\ }\textbf {\bibinfo {volume} {30}},\ \bibinfo
  {eid} {100648} (\bibinfo {year} {2020})},\ \Eprint
  {http://arxiv.org/abs/2005.12410} {arXiv:2005.12410 [gr-qc]} \BibitemShut
  {NoStop}%
\bibitem [{\citenamefont {{Shaymatov}}\ \emph
  {et~al.}(2021{\natexlab{a}})\citenamefont {{Shaymatov}}, \citenamefont
  {{Narzilloev}}, \citenamefont {{Abdujabbarov}},\ and\ \citenamefont
  {{Bambi}}}]{Shaymatov21c}%
  \BibitemOpen
  \bibfield  {author} {\bibinfo {author} {\bibfnamefont {S.}~\bibnamefont
  {{Shaymatov}}}, \bibinfo {author} {\bibfnamefont {B.}~\bibnamefont
  {{Narzilloev}}}, \bibinfo {author} {\bibfnamefont {A.}~\bibnamefont
  {{Abdujabbarov}}}, \ and\ \bibinfo {author} {\bibfnamefont {C.}~\bibnamefont
  {{Bambi}}},\ }\href {\doibase 10.1103/PhysRevD.103.124066} {\bibfield
  {journal} {\bibinfo  {journal} {Phys. Rev. D}\ }\textbf {\bibinfo {volume}
  {103}},\ \bibinfo {eid} {124066} (\bibinfo {year} {2021}{\natexlab{a}})},\
  \Eprint {http://arxiv.org/abs/2105.00342} {arXiv:2105.00342 [gr-qc]}
  \BibitemShut {NoStop}%
\bibitem [{\citenamefont {Shaymatov}\ and\ \citenamefont
  {Atamurotov}(2021)}]{Shaymatov21b}%
  \BibitemOpen
  \bibfield  {author} {\bibinfo {author} {\bibfnamefont {S.}~\bibnamefont
  {Shaymatov}}\ and\ \bibinfo {author} {\bibfnamefont {F.}~\bibnamefont
  {Atamurotov}},\ }\href {\doibase 10.3390/galaxies9020040} {\bibfield
  {journal} {\bibinfo  {journal} {Galaxies}\ }\textbf {\bibinfo {volume} {9}},\
  \bibinfo {eid} {40} (\bibinfo {year} {2021})},\ \Eprint
  {http://arxiv.org/abs/2007.10793} {arXiv:2007.10793 [gr-qc]} \BibitemShut
  {NoStop}%
\bibitem [{\citenamefont {{Rubin}}\ \emph {et~al.}(1980)\citenamefont
  {{Rubin}}, \citenamefont {{Ford}},\ and\ \citenamefont
  {{Thonnard}}}]{Rubin80}%
  \BibitemOpen
  \bibfield  {author} {\bibinfo {author} {\bibfnamefont {V.~C.}\ \bibnamefont
  {{Rubin}}}, \bibinfo {author} {\bibfnamefont {J.}~\bibnamefont {{Ford}},
  \bibfnamefont {W.~K.}}, \ and\ \bibinfo {author} {\bibfnamefont
  {N.}~\bibnamefont {{Thonnard}}},\ }\href {\doibase 10.1086/158003} {\bibfield
   {journal} {\bibinfo  {journal} {Astrophys. J.}\ }\textbf {\bibinfo {volume}
  {238}},\ \bibinfo {pages} {471} (\bibinfo {year} {1980})}\BibitemShut
  {NoStop}%
\bibitem [{\citenamefont {{Persic}}\ \emph {et~al.}(1996)\citenamefont
  {{Persic}}, \citenamefont {{Salucci}},\ and\ \citenamefont
  {{Stel}}}]{Persic96}%
  \BibitemOpen
  \bibfield  {author} {\bibinfo {author} {\bibfnamefont {M.}~\bibnamefont
  {{Persic}}}, \bibinfo {author} {\bibfnamefont {P.}~\bibnamefont {{Salucci}}},
  \ and\ \bibinfo {author} {\bibfnamefont {F.}~\bibnamefont {{Stel}}},\ }\href
  {\doibase 10.1093/mnras/278.1.27} {\bibfield  {journal} {\bibinfo  {journal}
  {Mon. Not. R. Astron. Soc.}\ }\textbf {\bibinfo {volume} {281}},\ \bibinfo
  {pages} {27} (\bibinfo {year} {1996})},\ \Eprint
  {http://arxiv.org/abs/astro-ph/9506004} {arXiv:astro-ph/9506004 [astro-ph]}
  \BibitemShut {NoStop}%
\bibitem [{\citenamefont {{Sofue}}(2013)}]{Sofue13-book}%
  \BibitemOpen
  \bibfield  {author} {\bibinfo {author} {\bibfnamefont {Y.}~\bibnamefont
  {{Sofue}}},\ }\enquote {\bibinfo {title} {{Mass Distribution and Rotation
  Curve in the Galaxy}},}\ in\ \href {\doibase 10.1007/978-94-007-5612-0_19}
  {\emph {\bibinfo {booktitle} {Planets, Stars and Stellar Systems. Volume 5:
  Galactic Structure and Stellar Populations}}},\ Vol.~\bibinfo {volume} {5},\
  \bibinfo {editor} {edited by\ \bibinfo {editor} {\bibfnamefont {T.~D.}\
  \bibnamefont {{Oswalt}}}\ and\ \bibinfo {editor} {\bibfnamefont
  {G.}~\bibnamefont {{Gilmore}}}}\ (\bibinfo {year} {2013})\ p.\ \bibinfo
  {pages} {985}\BibitemShut {NoStop}%
\bibitem [{\citenamefont {{Boshkayev}}\ and\ \citenamefont
  {{Malafarina}}(2019)}]{Boshkayev19}%
  \BibitemOpen
  \bibfield  {author} {\bibinfo {author} {\bibfnamefont {K.}~\bibnamefont
  {{Boshkayev}}}\ and\ \bibinfo {author} {\bibfnamefont {D.}~\bibnamefont
  {{Malafarina}}},\ }\href {\doibase 10.1093/mnras/stz219} {\bibfield
  {journal} {\bibinfo  {journal} {Mon. Not. R. Astron. Soc.}\ }\textbf
  {\bibinfo {volume} {484}},\ \bibinfo {pages} {3325} (\bibinfo {year}
  {2019})},\ \Eprint {http://arxiv.org/abs/1811.04061} {arXiv:1811.04061
  [gr-qc]} \BibitemShut {NoStop}%
\bibitem [{\citenamefont {{Kiselev}}(2003)}]{Kiselev03-dm}%
  \BibitemOpen
  \bibfield  {author} {\bibinfo {author} {\bibfnamefont {V.~V.}\ \bibnamefont
  {{Kiselev}}},\ }\href@noop {} {\bibfield  {journal} {\bibinfo  {journal}
  {arXiv e-prints}\ } (\bibinfo {year} {2003})},\ \Eprint
  {http://arxiv.org/abs/gr-qc/0303031} {arXiv:gr-qc/0303031 [gr-qc]}
  \BibitemShut {NoStop}%
\bibitem [{\citenamefont {{Li}}\ and\ \citenamefont
  {{Yang}}(2012)}]{Li-Yang12}%
  \BibitemOpen
  \bibfield  {author} {\bibinfo {author} {\bibfnamefont {M.-H.}\ \bibnamefont
  {{Li}}}\ and\ \bibinfo {author} {\bibfnamefont {K.-C.}\ \bibnamefont
  {{Yang}}},\ }\href {\doibase 10.1103/PhysRevD.86.123015} {\bibfield
  {journal} {\bibinfo  {journal} {Phys. Rev. D}\ }\textbf {\bibinfo {volume}
  {86}},\ \bibinfo {eid} {123015} (\bibinfo {year} {2012})},\ \Eprint
  {http://arxiv.org/abs/1204.3178} {arXiv:1204.3178 [astro-ph.CO]} \BibitemShut
  {NoStop}%
\bibitem [{\citenamefont {{Xu}}\ \emph {et~al.}(2018)\citenamefont {{Xu}},
  \citenamefont {{Hou}},\ and\ \citenamefont {{Wang}}}]{Xu18}%
  \BibitemOpen
  \bibfield  {author} {\bibinfo {author} {\bibfnamefont {Z.}~\bibnamefont
  {{Xu}}}, \bibinfo {author} {\bibfnamefont {X.}~\bibnamefont {{Hou}}}, \ and\
  \bibinfo {author} {\bibfnamefont {J.}~\bibnamefont {{Wang}}},\ }\href
  {\doibase 10.1088/1361-6382/aabcb6} {\bibfield  {journal} {\bibinfo
  {journal} {Class. Quantum Grav.}\ }\textbf {\bibinfo {volume} {35}},\
  \bibinfo {eid} {115003} (\bibinfo {year} {2018})},\ \Eprint
  {http://arxiv.org/abs/1711.04538} {arXiv:1711.04538 [gr-qc]} \BibitemShut
  {NoStop}%
\bibitem [{\citenamefont {{Haroon}}\ \emph {et~al.}(2019)\citenamefont
  {{Haroon}}, \citenamefont {{Jamil}}, \citenamefont {{Jusufi}}, \citenamefont
  {{Lin}},\ and\ \citenamefont {{Mann}}}]{Haroon19}%
  \BibitemOpen
  \bibfield  {author} {\bibinfo {author} {\bibfnamefont {S.}~\bibnamefont
  {{Haroon}}}, \bibinfo {author} {\bibfnamefont {M.}~\bibnamefont {{Jamil}}},
  \bibinfo {author} {\bibfnamefont {K.}~\bibnamefont {{Jusufi}}}, \bibinfo
  {author} {\bibfnamefont {K.}~\bibnamefont {{Lin}}}, \ and\ \bibinfo {author}
  {\bibfnamefont {R.~B.}\ \bibnamefont {{Mann}}},\ }\href {\doibase
  10.1103/PhysRevD.99.044015} {\bibfield  {journal} {\bibinfo  {journal} {Phys.
  Rev. D}\ }\textbf {\bibinfo {volume} {99}},\ \bibinfo {eid} {044015}
  (\bibinfo {year} {2019})},\ \Eprint {http://arxiv.org/abs/1810.04103}
  {arXiv:1810.04103 [gr-qc]} \BibitemShut {NoStop}%
\bibitem [{\citenamefont {{Konoplya}}(2019)}]{Konoplya19plb}%
  \BibitemOpen
  \bibfield  {author} {\bibinfo {author} {\bibfnamefont {R.~A.}\ \bibnamefont
  {{Konoplya}}},\ }\href {\doibase 10.1016/j.physletb.2019.05.043} {\bibfield
  {journal} {\bibinfo  {journal} {Phys. Lett. B}\ }\textbf {\bibinfo {volume}
  {795}},\ \bibinfo {pages} {1} (\bibinfo {year} {2019})},\ \Eprint
  {http://arxiv.org/abs/1905.00064} {arXiv:1905.00064 [gr-qc]} \BibitemShut
  {NoStop}%
\bibitem [{\citenamefont {{Hendi}}\ \emph {et~al.}(2020)\citenamefont
  {{Hendi}}, \citenamefont {{Nemati}}, \citenamefont {{Lin}},\ and\
  \citenamefont {{Jamil}}}]{Hendi20}%
  \BibitemOpen
  \bibfield  {author} {\bibinfo {author} {\bibfnamefont {S.~H.}\ \bibnamefont
  {{Hendi}}}, \bibinfo {author} {\bibfnamefont {A.}~\bibnamefont {{Nemati}}},
  \bibinfo {author} {\bibfnamefont {K.}~\bibnamefont {{Lin}}}, \ and\ \bibinfo
  {author} {\bibfnamefont {M.}~\bibnamefont {{Jamil}}},\ }\href {\doibase
  10.1140/epjc/s10052-020-7829-6} {\bibfield  {journal} {\bibinfo  {journal}
  {Eur. Phys. J. C}\ }\textbf {\bibinfo {volume} {80}},\ \bibinfo {eid} {296}
  (\bibinfo {year} {2020})},\ \Eprint {http://arxiv.org/abs/2001.01591}
  {arXiv:2001.01591 [gr-qc]} \BibitemShut {NoStop}%
\bibitem [{\citenamefont {Jusufi}\ \emph {et~al.}(2019)\citenamefont {Jusufi},
  \citenamefont {Jamil}, \citenamefont {Salucci}, \citenamefont {Zhu},\ and\
  \citenamefont {Haroon}}]{Jusufi19}%
  \BibitemOpen
  \bibfield  {author} {\bibinfo {author} {\bibfnamefont {K.}~\bibnamefont
  {Jusufi}}, \bibinfo {author} {\bibfnamefont {M.}~\bibnamefont {Jamil}},
  \bibinfo {author} {\bibfnamefont {P.}~\bibnamefont {Salucci}}, \bibinfo
  {author} {\bibfnamefont {T.}~\bibnamefont {Zhu}}, \ and\ \bibinfo {author}
  {\bibfnamefont {S.}~\bibnamefont {Haroon}},\ }\href {\doibase
  10.1103/PhysRevD.100.044012} {\bibfield  {journal} {\bibinfo  {journal}
  {Phys. Rev. D}\ }\textbf {\bibinfo {volume} {100}},\ \bibinfo {pages}
  {044012} (\bibinfo {year} {2019})}\BibitemShut {NoStop}%
\bibitem [{\citenamefont {{Narzilloev}}\ \emph {et~al.}(2020)\citenamefont
  {{Narzilloev}}, \citenamefont {{Rayimbaev}}, \citenamefont {{Shaymatov}},
  \citenamefont {{Abdujabbarov}}, \citenamefont {{Ahmedov}},\ and\
  \citenamefont {{Bambi}}}]{Narzilloev20b}%
  \BibitemOpen
  \bibfield  {author} {\bibinfo {author} {\bibfnamefont {B.}~\bibnamefont
  {{Narzilloev}}}, \bibinfo {author} {\bibfnamefont {J.}~\bibnamefont
  {{Rayimbaev}}}, \bibinfo {author} {\bibfnamefont {S.}~\bibnamefont
  {{Shaymatov}}}, \bibinfo {author} {\bibfnamefont {A.}~\bibnamefont
  {{Abdujabbarov}}}, \bibinfo {author} {\bibfnamefont {B.}~\bibnamefont
  {{Ahmedov}}}, \ and\ \bibinfo {author} {\bibfnamefont {C.}~\bibnamefont
  {{Bambi}}},\ }\href {\doibase 10.1103/PhysRevD.102.104062} {\bibfield
  {journal} {\bibinfo  {journal} {Phys. Rev. D}\ }\textbf {\bibinfo {volume}
  {102}},\ \bibinfo {eid} {104062} (\bibinfo {year} {2020})},\ \Eprint
  {http://arxiv.org/abs/2011.06148} {arXiv:2011.06148 [gr-qc]} \BibitemShut
  {NoStop}%
\bibitem [{\citenamefont {{Shaymatov}}\ \emph
  {et~al.}(2021{\natexlab{b}})\citenamefont {{Shaymatov}}, \citenamefont
  {{Ahmedov}},\ and\ \citenamefont {{Jamil}}}]{Shaymatov21d}%
  \BibitemOpen
  \bibfield  {author} {\bibinfo {author} {\bibfnamefont {S.}~\bibnamefont
  {{Shaymatov}}}, \bibinfo {author} {\bibfnamefont {B.}~\bibnamefont
  {{Ahmedov}}}, \ and\ \bibinfo {author} {\bibfnamefont {M.}~\bibnamefont
  {{Jamil}}},\ }\href {\doibase 10.1140/epjc/s10052-021-09398-w} {\bibfield
  {journal} {\bibinfo  {journal} {Eur. Phys. J. C}\ }\textbf {\bibinfo {volume}
  {81}},\ \bibinfo {eid} {588} (\bibinfo {year}
  {2021}{\natexlab{b}})}\BibitemShut {NoStop}%
\bibitem [{\citenamefont {{Rayimbaev}}\ \emph {et~al.}(2021)\citenamefont
  {{Rayimbaev}}, \citenamefont {{Shaymatov}},\ and\ \citenamefont
  {{Jamil}}}]{Rayimbaev-Shaymatov21a}%
  \BibitemOpen
  \bibfield  {author} {\bibinfo {author} {\bibfnamefont {J.}~\bibnamefont
  {{Rayimbaev}}}, \bibinfo {author} {\bibfnamefont {S.}~\bibnamefont
  {{Shaymatov}}}, \ and\ \bibinfo {author} {\bibfnamefont {M.}~\bibnamefont
  {{Jamil}}},\ }\href {\doibase 10.1140/epjc/s10052-021-09488-9} {\bibfield
  {journal} {\bibinfo  {journal} {Eur. Phys. J. C}\ }\textbf {\bibinfo {volume}
  {81}},\ \bibinfo {eid} {699} (\bibinfo {year} {2021})},\ \Eprint
  {http://arxiv.org/abs/2107.13436} {arXiv:2107.13436 [gr-qc]} \BibitemShut
  {NoStop}%
\bibitem [{\citenamefont {{Fender}}\ \emph {et~al.}(2004)\citenamefont
  {{Fender}}, \citenamefont {{Belloni}},\ and\ \citenamefont
  {{Gallo}}}]{Fender04mnrs}%
  \BibitemOpen
  \bibfield  {author} {\bibinfo {author} {\bibfnamefont {R.~P.}\ \bibnamefont
  {{Fender}}}, \bibinfo {author} {\bibfnamefont {T.~M.}\ \bibnamefont
  {{Belloni}}}, \ and\ \bibinfo {author} {\bibfnamefont {E.}~\bibnamefont
  {{Gallo}}},\ }\href {\doibase 10.1111/j.1365-2966.2004.08384.x} {\bibfield
  {journal} {\bibinfo  {journal} {Mon. Not. R. Astron. Soc.}\ }\textbf
  {\bibinfo {volume} {355}},\ \bibinfo {pages} {1105} (\bibinfo {year}
  {2004})},\ \Eprint {http://arxiv.org/abs/astro-ph/0409360}
  {arXiv:astro-ph/0409360 [astro-ph]} \BibitemShut {NoStop}%
\bibitem [{\citenamefont {{Auchettl}}\ \emph {et~al.}(2017)\citenamefont
  {{Auchettl}}, \citenamefont {{Guillochon}},\ and\ \citenamefont
  {{Ramirez-Ruiz}}}]{Auchettl17ApJ}%
  \BibitemOpen
  \bibfield  {author} {\bibinfo {author} {\bibfnamefont {K.}~\bibnamefont
  {{Auchettl}}}, \bibinfo {author} {\bibfnamefont {J.}~\bibnamefont
  {{Guillochon}}}, \ and\ \bibinfo {author} {\bibfnamefont {E.}~\bibnamefont
  {{Ramirez-Ruiz}}},\ }\href {\doibase 10.3847/1538-4357/aa633b} {\bibfield
  {journal} {\bibinfo  {journal} {Astrophys. J.}\ }\textbf {\bibinfo {volume}
  {838}},\ \bibinfo {eid} {149} (\bibinfo {year} {2017})},\ \Eprint
  {http://arxiv.org/abs/1611.02291} {arXiv:1611.02291 [astro-ph.HE]}
  \BibitemShut {NoStop}%
\bibitem [{\citenamefont {{The IceCube Collaboration}}\ and\ \citenamefont
  {et~al.}(2018)}]{IceCube17b}%
  \BibitemOpen
  \bibfield  {author} {\bibinfo {author} {\bibnamefont {{The IceCube
  Collaboration}}}\ and\ \bibinfo {author} {\bibnamefont {et~al.}},\ }\href
  {\doibase 10.1126/science.aat1378} {\bibfield  {journal} {\bibinfo  {journal}
  {Science}\ }\textbf {\bibinfo {volume} {361}},\ \bibinfo {eid} {eaat1378}
  (\bibinfo {year} {2018})},\ \Eprint {http://arxiv.org/abs/1807.08816}
  {arXiv:1807.08816 [gr-qc]} \BibitemShut {NoStop}%
\bibitem [{\citenamefont {{Ba{\~n}ados}}\ \emph {et~al.}(2009)\citenamefont
  {{Ba{\~n}ados}}, \citenamefont {{Silk}},\ and\ \citenamefont
  {{West}}}]{banados09}%
  \BibitemOpen
  \bibfield  {author} {\bibinfo {author} {\bibfnamefont {M.}~\bibnamefont
  {{Ba{\~n}ados}}}, \bibinfo {author} {\bibfnamefont {J.}~\bibnamefont
  {{Silk}}}, \ and\ \bibinfo {author} {\bibfnamefont {S.~M.}\ \bibnamefont
  {{West}}},\ }\href {\doibase 10.1103/PhysRevLett.103.111102} {\bibfield
  {journal} {\bibinfo  {journal} {Phys. Rev. Lett.}\ }\textbf {\bibinfo
  {volume} {103}},\ \bibinfo {eid} {111102} (\bibinfo {year}
  {2009})}\BibitemShut {NoStop}%
\bibitem [{\citenamefont {{Grib}}\ and\ \citenamefont
  {{Pavlov}}(2011)}]{Grib11}%
  \BibitemOpen
  \bibfield  {author} {\bibinfo {author} {\bibfnamefont {A.~A.}\ \bibnamefont
  {{Grib}}}\ and\ \bibinfo {author} {\bibfnamefont {Y.~V.}\ \bibnamefont
  {{Pavlov}}},\ }\href {\doibase 10.1134/S0202289311010099} {\bibfield
  {journal} {\bibinfo  {journal} {Gravitation and Cosmology}\ }\textbf
  {\bibinfo {volume} {17}},\ \bibinfo {pages} {42} (\bibinfo {year} {2011})},\
  \Eprint {http://arxiv.org/abs/1010.2052} {arXiv:1010.2052 [gr-qc]}
  \BibitemShut {NoStop}%
\bibitem [{\citenamefont {{Jacobson}}\ and\ \citenamefont
  {{Sotiriou}}(2010)}]{Jacobson10}%
  \BibitemOpen
  \bibfield  {author} {\bibinfo {author} {\bibfnamefont {T.}~\bibnamefont
  {{Jacobson}}}\ and\ \bibinfo {author} {\bibfnamefont {T.~P.}\ \bibnamefont
  {{Sotiriou}}},\ }\href {\doibase 10.1103/PhysRevLett.104.021101} {\bibfield
  {journal} {\bibinfo  {journal} {Phys. Rev. Lett.}\ }\textbf {\bibinfo
  {volume} {104}},\ \bibinfo {eid} {021101} (\bibinfo {year} {2010})},\ \Eprint
  {http://arxiv.org/abs/0911.3363} {arXiv:0911.3363 [gr-qc]} \BibitemShut
  {NoStop}%
\bibitem [{\citenamefont {{Harada}}\ and\ \citenamefont
  {{Kimura}}(2011)}]{Harada11b}%
  \BibitemOpen
  \bibfield  {author} {\bibinfo {author} {\bibfnamefont {T.}~\bibnamefont
  {{Harada}}}\ and\ \bibinfo {author} {\bibfnamefont {M.}~\bibnamefont
  {{Kimura}}},\ }\href {\doibase 10.1103/PhysRevD.83.024002} {\bibfield
  {journal} {\bibinfo  {journal} {Phys. Rev. D}\ }\textbf {\bibinfo {volume}
  {83}},\ \bibinfo {eid} {024002} (\bibinfo {year} {2011})},\ \Eprint
  {http://arxiv.org/abs/1010.0962} {arXiv:1010.0962 [gr-qc]} \BibitemShut
  {NoStop}%
\bibitem [{\citenamefont {{Wei}}\ \emph {et~al.}(2010)\citenamefont {{Wei}},
  \citenamefont {{Liu}}, \citenamefont {{Guo}},\ and\ \citenamefont
  {{Fu}}}]{Wei10}%
  \BibitemOpen
  \bibfield  {author} {\bibinfo {author} {\bibfnamefont {S.-W.}\ \bibnamefont
  {{Wei}}}, \bibinfo {author} {\bibfnamefont {Y.-X.}\ \bibnamefont {{Liu}}},
  \bibinfo {author} {\bibfnamefont {H.}~\bibnamefont {{Guo}}}, \ and\ \bibinfo
  {author} {\bibfnamefont {C.-E.}\ \bibnamefont {{Fu}}},\ }\href {\doibase
  10.1103/PhysRevD.82.103005} {\bibfield  {journal} {\bibinfo  {journal} {Phys.
  Rev. D}\ }\textbf {\bibinfo {volume} {82}},\ \bibinfo {eid} {103005}
  (\bibinfo {year} {2010})},\ \Eprint {http://arxiv.org/abs/1006.1056}
  {arXiv:1006.1056 [hep-th]} \BibitemShut {NoStop}%
\bibitem [{\citenamefont {{Zaslavskii}}(2010)}]{Zaslavskii10}%
  \BibitemOpen
  \bibfield  {author} {\bibinfo {author} {\bibfnamefont {O.~B.}\ \bibnamefont
  {{Zaslavskii}}},\ }\href {\doibase 10.1103/PhysRevD.82.083004} {\bibfield
  {journal} {\bibinfo  {journal} {Phys. Rev. D}\ }\textbf {\bibinfo {volume}
  {82}},\ \bibinfo {eid} {083004} (\bibinfo {year} {2010})},\ \Eprint
  {http://arxiv.org/abs/1007.3678} {arXiv:1007.3678 [gr-qc]} \BibitemShut
  {NoStop}%
\bibitem [{\citenamefont {{Zaslavskii}}(2011{\natexlab{a}})}]{Zaslavskii11b}%
  \BibitemOpen
  \bibfield  {author} {\bibinfo {author} {\bibfnamefont {O.~B.}\ \bibnamefont
  {{Zaslavskii}}},\ }\href {\doibase 10.1134/S0021364010210010} {\bibfield
  {journal} {\bibinfo  {journal} {Soviet Journal of Experimental and
  Theoretical Physics Letters}\ }\textbf {\bibinfo {volume} {92}},\ \bibinfo
  {pages} {571} (\bibinfo {year} {2011}{\natexlab{a}})},\ \Eprint
  {http://arxiv.org/abs/1007.4598} {arXiv:1007.4598 [gr-qc]} \BibitemShut
  {NoStop}%
\bibitem [{\citenamefont {{Zaslavskii}}(2011{\natexlab{b}})}]{Zaslavskii11c}%
  \BibitemOpen
  \bibfield  {author} {\bibinfo {author} {\bibfnamefont {O.~B.}\ \bibnamefont
  {{Zaslavskii}}},\ }\href {\doibase 10.1088/0264-9381/28/10/105010} {\bibfield
   {journal} {\bibinfo  {journal} {Class. Quantum Grav.}\ }\textbf {\bibinfo
  {volume} {28}},\ \bibinfo {eid} {105010} (\bibinfo {year}
  {2011}{\natexlab{b}})},\ \Eprint {http://arxiv.org/abs/1011.0167}
  {arXiv:1011.0167 [gr-qc]} \BibitemShut {NoStop}%
\bibitem [{\citenamefont {{Kimura}}\ \emph {et~al.}(2011)\citenamefont
  {{Kimura}}, \citenamefont {{Nakao}},\ and\ \citenamefont
  {{Tagoshi}}}]{Kimura11}%
  \BibitemOpen
  \bibfield  {author} {\bibinfo {author} {\bibfnamefont {M.}~\bibnamefont
  {{Kimura}}}, \bibinfo {author} {\bibfnamefont {K.-I.}\ \bibnamefont
  {{Nakao}}}, \ and\ \bibinfo {author} {\bibfnamefont {H.}~\bibnamefont
  {{Tagoshi}}},\ }\href {\doibase 10.1103/PhysRevD.83.044013} {\bibfield
  {journal} {\bibinfo  {journal} {Phys. Rev. D}\ }\textbf {\bibinfo {volume}
  {83}},\ \bibinfo {eid} {044013} (\bibinfo {year} {2011})},\ \Eprint
  {http://arxiv.org/abs/1010.5438} {arXiv:1010.5438 [gr-qc]} \BibitemShut
  {NoStop}%
\bibitem [{\citenamefont {{Ba{\~n}ados}}\ \emph {et~al.}(2011)\citenamefont
  {{Ba{\~n}ados}}, \citenamefont {{Hassanain}}, \citenamefont {{Silk}},\ and\
  \citenamefont {{West}}}]{Banados11}%
  \BibitemOpen
  \bibfield  {author} {\bibinfo {author} {\bibfnamefont {M.}~\bibnamefont
  {{Ba{\~n}ados}}}, \bibinfo {author} {\bibfnamefont {B.}~\bibnamefont
  {{Hassanain}}}, \bibinfo {author} {\bibfnamefont {J.}~\bibnamefont {{Silk}}},
  \ and\ \bibinfo {author} {\bibfnamefont {S.~M.}\ \bibnamefont {{West}}},\
  }\href {\doibase 10.1103/PhysRevD.83.023004} {\bibfield  {journal} {\bibinfo
  {journal} {Physical Review D}\ }\textbf {\bibinfo {volume} {83}},\ \bibinfo
  {eid} {023004} (\bibinfo {year} {2011})},\ \Eprint
  {http://arxiv.org/abs/1010.2724} {arXiv:1010.2724 [astro-ph.CO]} \BibitemShut
  {NoStop}%
\bibitem [{\citenamefont {{Frolov}}(2012)}]{Frolov12}%
  \BibitemOpen
  \bibfield  {author} {\bibinfo {author} {\bibfnamefont {V.~P.}\ \bibnamefont
  {{Frolov}}},\ }\href {\doibase 10.1103/PhysRevD.85.024020} {\bibfield
  {journal} {\bibinfo  {journal} {Phys. Rev. D}\ }\textbf {\bibinfo {volume}
  {85}},\ \bibinfo {eid} {024020} (\bibinfo {year} {2012})},\ \Eprint
  {http://arxiv.org/abs/1110.6274} {arXiv:1110.6274 [gr-qc]} \BibitemShut
  {NoStop}%
\bibitem [{\citenamefont {{Abdujabbarov}}\ \emph {et~al.}(2013)\citenamefont
  {{Abdujabbarov}}, \citenamefont {{Tursunov}}, \citenamefont {{Ahmedov}},\
  and\ \citenamefont {{Kuvatov}}}]{Abdujabbarov13a}%
  \BibitemOpen
  \bibfield  {author} {\bibinfo {author} {\bibfnamefont {A.~A.}\ \bibnamefont
  {{Abdujabbarov}}}, \bibinfo {author} {\bibfnamefont {A.~A.}\ \bibnamefont
  {{Tursunov}}}, \bibinfo {author} {\bibfnamefont {B.~J.}\ \bibnamefont
  {{Ahmedov}}}, \ and\ \bibinfo {author} {\bibfnamefont {A.}~\bibnamefont
  {{Kuvatov}}},\ }\href {\doibase 10.1007/s10509-012-1251-y} {\bibfield
  {journal} {\bibinfo  {journal} {Astrophys Space Sci}\ }\textbf {\bibinfo
  {volume} {343}},\ \bibinfo {pages} {173} (\bibinfo {year} {2013})},\ \Eprint
  {http://arxiv.org/abs/1209.2680} {arXiv:1209.2680 [gr-qc]} \BibitemShut
  {NoStop}%
\bibitem [{\citenamefont {{Liu}}\ \emph {et~al.}(2011)\citenamefont {{Liu}},
  \citenamefont {{Chen}}, \citenamefont {{Ding}},\ and\ \citenamefont
  {{Jing}}}]{Liu11}%
  \BibitemOpen
  \bibfield  {author} {\bibinfo {author} {\bibfnamefont {C.}~\bibnamefont
  {{Liu}}}, \bibinfo {author} {\bibfnamefont {S.}~\bibnamefont {{Chen}}},
  \bibinfo {author} {\bibfnamefont {C.}~\bibnamefont {{Ding}}}, \ and\ \bibinfo
  {author} {\bibfnamefont {J.}~\bibnamefont {{Jing}}},\ }\href {\doibase
  10.1016/j.physletb.2011.05.070} {\bibfield  {journal} {\bibinfo  {journal}
  {Phys. Lett. B}\ }\textbf {\bibinfo {volume} {701}},\ \bibinfo {pages} {285}
  (\bibinfo {year} {2011})},\ \Eprint {http://arxiv.org/abs/1012.5126}
  {arXiv:1012.5126 [gr-qc]} \BibitemShut {NoStop}%
\bibitem [{\citenamefont {{Atamurotov}}\ \emph {et~al.}(2013)\citenamefont
  {{Atamurotov}}, \citenamefont {{Ahmedov}},\ and\ \citenamefont
  {{Shaymatov}}}]{Atamurotov13a}%
  \BibitemOpen
  \bibfield  {author} {\bibinfo {author} {\bibfnamefont {F.}~\bibnamefont
  {{Atamurotov}}}, \bibinfo {author} {\bibfnamefont {B.}~\bibnamefont
  {{Ahmedov}}}, \ and\ \bibinfo {author} {\bibfnamefont {S.}~\bibnamefont
  {{Shaymatov}}},\ }\href {\doibase 10.1007/s10509-013-1527-x} {\bibfield
  {journal} {\bibinfo  {journal} {Astrophys. Space Sci.}\ }\textbf {\bibinfo
  {volume} {347}},\ \bibinfo {pages} {277} (\bibinfo {year}
  {2013})}\BibitemShut {NoStop}%
\bibitem [{\citenamefont {{Stuchl{\'{\i}}k}}\ \emph {et~al.}(2011)\citenamefont
  {{Stuchl{\'{\i}}k}}, \citenamefont {{Hled{\'{\i}}k}},\ and\ \citenamefont
  {{Truparov{\'a}}}}]{Stuchlik11a}%
  \BibitemOpen
  \bibfield  {author} {\bibinfo {author} {\bibfnamefont {Z.}~\bibnamefont
  {{Stuchl{\'{\i}}k}}}, \bibinfo {author} {\bibfnamefont {S.}~\bibnamefont
  {{Hled{\'{\i}}k}}}, \ and\ \bibinfo {author} {\bibfnamefont {K.}~\bibnamefont
  {{Truparov{\'a}}}},\ }\href {\doibase 10.1088/0264-9381/28/15/155017}
  {\bibfield  {journal} {\bibinfo  {journal} {Class. Quantum Grav.}\ }\textbf
  {\bibinfo {volume} {28}},\ \bibinfo {eid} {155017} (\bibinfo {year}
  {2011})}\BibitemShut {NoStop}%
\bibitem [{\citenamefont {{Stuchl{\'{\i}}k}}\ and\ \citenamefont
  {{Schee}}(2012)}]{Stuchlik12a}%
  \BibitemOpen
  \bibfield  {author} {\bibinfo {author} {\bibfnamefont {Z.}~\bibnamefont
  {{Stuchl{\'{\i}}k}}}\ and\ \bibinfo {author} {\bibfnamefont {J.}~\bibnamefont
  {{Schee}}},\ }\href {\doibase 10.1088/0264-9381/29/6/065002} {\bibfield
  {journal} {\bibinfo  {journal} {Class. Quantum Grav.}\ }\textbf {\bibinfo
  {volume} {29}},\ \bibinfo {eid} {065002} (\bibinfo {year}
  {2012})}\BibitemShut {NoStop}%
\bibitem [{\citenamefont {{Igata}}\ \emph {et~al.}(2012)\citenamefont
  {{Igata}}, \citenamefont {{Harada}},\ and\ \citenamefont
  {{Kimura}}}]{Igata12}%
  \BibitemOpen
  \bibfield  {author} {\bibinfo {author} {\bibfnamefont {T.}~\bibnamefont
  {{Igata}}}, \bibinfo {author} {\bibfnamefont {T.}~\bibnamefont {{Harada}}}, \
  and\ \bibinfo {author} {\bibfnamefont {M.}~\bibnamefont {{Kimura}}},\ }\href
  {\doibase 10.1103/PhysRevD.85.104028} {\bibfield  {journal} {\bibinfo
  {journal} {Phys. Rev. D}\ }\textbf {\bibinfo {volume} {85}},\ \bibinfo {eid}
  {104028} (\bibinfo {year} {2012})},\ \Eprint {http://arxiv.org/abs/1202.4859}
  {arXiv:1202.4859 [gr-qc]} \BibitemShut {NoStop}%
\bibitem [{\citenamefont {{Shaymatov}}\ \emph {et~al.}(2013)\citenamefont
  {{Shaymatov}}, \citenamefont {{Ahmedov}},\ and\ \citenamefont
  {{Abdujabbarov}}}]{Shaymatov13}%
  \BibitemOpen
  \bibfield  {author} {\bibinfo {author} {\bibfnamefont {S.~R.}\ \bibnamefont
  {{Shaymatov}}}, \bibinfo {author} {\bibfnamefont {B.~J.}\ \bibnamefont
  {{Ahmedov}}}, \ and\ \bibinfo {author} {\bibfnamefont {A.~A.}\ \bibnamefont
  {{Abdujabbarov}}},\ }\href {\doibase 10.1103/PhysRevD.88.024016} {\bibfield
  {journal} {\bibinfo  {journal} {Phys. Rev. D}\ }\textbf {\bibinfo {volume}
  {88}},\ \bibinfo {eid} {024016} (\bibinfo {year} {2013})}\BibitemShut
  {NoStop}%
\bibitem [{\citenamefont {{Tursunov}}\ \emph {et~al.}(2013)\citenamefont
  {{Tursunov}}, \citenamefont {{Kolo{\v s}}}, \citenamefont {{Abdujabbarov}},
  \citenamefont {{Ahmedov}},\ and\ \citenamefont
  {{Stuchl{\'{\i}}k}}}]{Tursunov13}%
  \BibitemOpen
  \bibfield  {author} {\bibinfo {author} {\bibfnamefont {A.}~\bibnamefont
  {{Tursunov}}}, \bibinfo {author} {\bibfnamefont {M.}~\bibnamefont {{Kolo{\v
  s}}}}, \bibinfo {author} {\bibfnamefont {A.}~\bibnamefont {{Abdujabbarov}}},
  \bibinfo {author} {\bibfnamefont {B.}~\bibnamefont {{Ahmedov}}}, \ and\
  \bibinfo {author} {\bibfnamefont {Z.}~\bibnamefont {{Stuchl{\'{\i}}k}}},\
  }\href {\doibase 10.1103/PhysRevD.88.124001} {\bibfield  {journal} {\bibinfo
  {journal} {Phys. Rev. D}\ }\textbf {\bibinfo {volume} {88}},\ \bibinfo {eid}
  {124001} (\bibinfo {year} {2013})}\BibitemShut {NoStop}%
\bibitem [{\citenamefont {{Shaymatov}}\ \emph {et~al.}(2018)\citenamefont
  {{Shaymatov}}, \citenamefont {{Ahmedov}}, \citenamefont {{Stuchl{\'\i}k}},\
  and\ \citenamefont {{Abdujabbarov}}}]{Shaymatov18a}%
  \BibitemOpen
  \bibfield  {author} {\bibinfo {author} {\bibfnamefont {S.}~\bibnamefont
  {{Shaymatov}}}, \bibinfo {author} {\bibfnamefont {B.}~\bibnamefont
  {{Ahmedov}}}, \bibinfo {author} {\bibfnamefont {Z.}~\bibnamefont
  {{Stuchl{\'\i}k}}}, \ and\ \bibinfo {author} {\bibfnamefont {A.}~\bibnamefont
  {{Abdujabbarov}}},\ }\href {\doibase 10.1142/S0218271818500888} {\bibfield
  {journal} {\bibinfo  {journal} {International Journal of Modern Physics D}\
  }\textbf {\bibinfo {volume} {27}},\ \bibinfo {eid} {1850088} (\bibinfo {year}
  {2018})}\BibitemShut {NoStop}%
\bibitem [{\citenamefont {{Atamurotov}}\ \emph {et~al.}(2021)\citenamefont
  {{Atamurotov}}, \citenamefont {{Shaymatov}}, \citenamefont {{Sheoran}},\ and\
  \citenamefont {{Siwach}}}]{Atamurotov21JCAP}%
  \BibitemOpen
  \bibfield  {author} {\bibinfo {author} {\bibfnamefont {F.}~\bibnamefont
  {{Atamurotov}}}, \bibinfo {author} {\bibfnamefont {S.}~\bibnamefont
  {{Shaymatov}}}, \bibinfo {author} {\bibfnamefont {P.}~\bibnamefont
  {{Sheoran}}}, \ and\ \bibinfo {author} {\bibfnamefont {S.}~\bibnamefont
  {{Siwach}}},\ }\href {\doibase 10.1088/1475-7516/2021/08/045} {\bibfield
  {journal} {\bibinfo  {journal} {JCAP}\ }\textbf {\bibinfo {volume} {2021}},\
  \bibinfo {eid} {045} (\bibinfo {year} {2021})},\ \Eprint
  {http://arxiv.org/abs/2105.02214} {arXiv:2105.02214 [gr-qc]} \BibitemShut
  {NoStop}%
\bibitem [{\citenamefont {{Stuchl{\'{\i}}k}}\ \emph {et~al.}(2014)\citenamefont
  {{Stuchl{\'{\i}}k}}, \citenamefont {{Schee}},\ and\ \citenamefont
  {{Abdujabbarov}}}]{Stuchlik14a}%
  \BibitemOpen
  \bibfield  {author} {\bibinfo {author} {\bibfnamefont {Z.}~\bibnamefont
  {{Stuchl{\'{\i}}k}}}, \bibinfo {author} {\bibfnamefont {J.}~\bibnamefont
  {{Schee}}}, \ and\ \bibinfo {author} {\bibfnamefont {A.}~\bibnamefont
  {{Abdujabbarov}}},\ }\href {\doibase 10.1103/PhysRevD.89.104048} {\bibfield
  {journal} {\bibinfo  {journal} {Phys. Rev. D}\ }\textbf {\bibinfo {volume}
  {89}},\ \bibinfo {eid} {104048} (\bibinfo {year} {2014})}\BibitemShut
  {NoStop}%
\bibitem [{\citenamefont {{Patil}}\ and\ \citenamefont
  {{Joshi}}(2010)}]{Patil10}%
  \BibitemOpen
  \bibfield  {author} {\bibinfo {author} {\bibfnamefont {M.}~\bibnamefont
  {{Patil}}}\ and\ \bibinfo {author} {\bibfnamefont {P.~S.}\ \bibnamefont
  {{Joshi}}},\ }\href {\doibase 10.1103/PhysRevD.82.104049} {\bibfield
  {journal} {\bibinfo  {journal} {Phys. Rev. D}\ }\textbf {\bibinfo {volume}
  {82}},\ \bibinfo {eid} {104049} (\bibinfo {year} {2010})},\ \Eprint
  {http://arxiv.org/abs/1011.5550} {arXiv:1011.5550 [gr-qc]} \BibitemShut
  {NoStop}%
\bibitem [{\citenamefont {{Patil}}\ and\ \citenamefont
  {{Joshi}}(2011)}]{Patil11}%
  \BibitemOpen
  \bibfield  {author} {\bibinfo {author} {\bibfnamefont {M.}~\bibnamefont
  {{Patil}}}\ and\ \bibinfo {author} {\bibfnamefont {P.}~\bibnamefont
  {{Joshi}}},\ }\href {\doibase 10.1088/0264-9381/28/23/235012} {\bibfield
  {journal} {\bibinfo  {journal} {Class. Quantum Grav.}\ }\textbf {\bibinfo
  {volume} {28}},\ \bibinfo {eid} {235012} (\bibinfo {year} {2011})},\ \Eprint
  {http://arxiv.org/abs/1103.1082} {arXiv:1103.1082 [gr-qc]} \BibitemShut
  {NoStop}%
\bibitem [{\citenamefont {{Patil}}\ \emph {et~al.}(2011)\citenamefont
  {{Patil}}, \citenamefont {{Joshi}},\ and\ \citenamefont
  {{Malafarina}}}]{Patil11b}%
  \BibitemOpen
  \bibfield  {author} {\bibinfo {author} {\bibfnamefont {M.}~\bibnamefont
  {{Patil}}}, \bibinfo {author} {\bibfnamefont {P.~S.}\ \bibnamefont
  {{Joshi}}}, \ and\ \bibinfo {author} {\bibfnamefont {D.}~\bibnamefont
  {{Malafarina}}},\ }\href {\doibase 10.1103/PhysRevD.83.064007} {\bibfield
  {journal} {\bibinfo  {journal} {Phys. Rev. D}\ }\textbf {\bibinfo {volume}
  {83}},\ \bibinfo {eid} {064007} (\bibinfo {year} {2011})},\ \Eprint
  {http://arxiv.org/abs/1102.2030} {arXiv:1102.2030 [gr-qc]} \BibitemShut
  {NoStop}%
\bibitem [{\citenamefont {{Stuchl{\'{\i}}k}}\ and\ \citenamefont
  {{Schee}}(2015)}]{Stuchlik15}%
  \BibitemOpen
  \bibfield  {author} {\bibinfo {author} {\bibfnamefont {Z.}~\bibnamefont
  {{Stuchl{\'{\i}}k}}}\ and\ \bibinfo {author} {\bibfnamefont {J.}~\bibnamefont
  {{Schee}}},\ }\href {\doibase 10.1142/S0218271815500200} {\bibfield
  {journal} {\bibinfo  {journal} {Int. J. Mod. Phys. D}\ }\textbf {\bibinfo
  {volume} {24}},\ \bibinfo {eid} {1550020} (\bibinfo {year} {2015})},\ \Eprint
  {http://arxiv.org/abs/1501.00015} {arXiv:1501.00015 [astro-ph.HE]}
  \BibitemShut {NoStop}%
\bibitem [{\citenamefont {{Abdujabbarov}}\ \emph {et~al.}(2011)\citenamefont
  {{Abdujabbarov}}, \citenamefont {{Ahmedov}}, \citenamefont {{Shaymatov}},\
  and\ \citenamefont {{Rakhmatov}}}]{Abdujabbarov11}%
  \BibitemOpen
  \bibfield  {author} {\bibinfo {author} {\bibfnamefont {A.~A.}\ \bibnamefont
  {{Abdujabbarov}}}, \bibinfo {author} {\bibfnamefont {B.~J.}\ \bibnamefont
  {{Ahmedov}}}, \bibinfo {author} {\bibfnamefont {S.~R.}\ \bibnamefont
  {{Shaymatov}}}, \ and\ \bibinfo {author} {\bibfnamefont {A.~S.}\ \bibnamefont
  {{Rakhmatov}}},\ }\href {\doibase 10.1007/s10509-011-0740-8} {\bibfield
  {journal} {\bibinfo  {journal} {Astrophys Space Sci}\ }\textbf {\bibinfo
  {volume} {334}},\ \bibinfo {pages} {237} (\bibinfo {year} {2011})},\ \Eprint
  {http://arxiv.org/abs/1105.1910} {arXiv:1105.1910 [astro-ph.SR]} \BibitemShut
  {NoStop}%
\bibitem [{\citenamefont {{Okabayashi}}\ and\ \citenamefont
  {{Maeda}}(2020)}]{Okabayashi20}%
  \BibitemOpen
  \bibfield  {author} {\bibinfo {author} {\bibfnamefont {K.}~\bibnamefont
  {{Okabayashi}}}\ and\ \bibinfo {author} {\bibfnamefont {K.-i.}\ \bibnamefont
  {{Maeda}}},\ }\href {\doibase 10.1093/ptep/ptz143} {\bibfield  {journal}
  {\bibinfo  {journal} {Prog. Theor. Exp. Phys.}\ }\textbf {\bibinfo {volume}
  {2020}},\ \bibinfo {eid} {013E01} (\bibinfo {year} {2020})},\ \Eprint
  {http://arxiv.org/abs/1907.07126} {arXiv:1907.07126 [gr-qc]} \BibitemShut
  {NoStop}%
\bibitem [{\citenamefont {{Blandford}}\ and\ \citenamefont
  {{Znajek}}(1977)}]{Blandford1977}%
  \BibitemOpen
  \bibfield  {author} {\bibinfo {author} {\bibfnamefont {R.~D.}\ \bibnamefont
  {{Blandford}}}\ and\ \bibinfo {author} {\bibfnamefont {R.~L.}\ \bibnamefont
  {{Znajek}}},\ }\href@noop {} {\bibfield  {journal} {\bibinfo  {journal} {Mon.
  Not. Roy. Astron. Soc.}\ }\textbf {\bibinfo {volume} {179}},\ \bibinfo
  {pages} {433} (\bibinfo {year} {1977})}\BibitemShut {NoStop}%
\bibitem [{\citenamefont {{Wagh}}\ and\ \citenamefont
  {{Dadhich}}(1989)}]{Wagh89}%
  \BibitemOpen
  \bibfield  {author} {\bibinfo {author} {\bibfnamefont {S.~M.}\ \bibnamefont
  {{Wagh}}}\ and\ \bibinfo {author} {\bibfnamefont {N.}~\bibnamefont
  {{Dadhich}}},\ }\href {\doibase 10.1016/0370-1573(89)90156-7} {\bibfield
  {journal} {\bibinfo  {journal} {Phys. Rep.}\ }\textbf {\bibinfo {volume}
  {183}},\ \bibinfo {pages} {137} (\bibinfo {year} {1989})}\BibitemShut
  {NoStop}%
\bibitem [{\citenamefont {{Morozova}}\ \emph {et~al.}(2014)\citenamefont
  {{Morozova}}, \citenamefont {{Rezzolla}},\ and\ \citenamefont
  {{Ahmedov}}}]{Morozova14}%
  \BibitemOpen
  \bibfield  {author} {\bibinfo {author} {\bibfnamefont {V.~S.}\ \bibnamefont
  {{Morozova}}}, \bibinfo {author} {\bibfnamefont {L.}~\bibnamefont
  {{Rezzolla}}}, \ and\ \bibinfo {author} {\bibfnamefont {B.~J.}\ \bibnamefont
  {{Ahmedov}}},\ }\href {\doibase 10.1103/PhysRevD.89.104030} {\bibfield
  {journal} {\bibinfo  {journal} {Phys. Rev. D}\ }\textbf {\bibinfo {volume}
  {89}},\ \bibinfo {eid} {104030} (\bibinfo {year} {2014})},\ \Eprint
  {http://arxiv.org/abs/1310.3575} {arXiv:1310.3575 [gr-qc]} \BibitemShut
  {NoStop}%
\bibitem [{\citenamefont {{Alic}}\ \emph {et~al.}(2012)\citenamefont {{Alic}},
  \citenamefont {{Moesta}}, \citenamefont {{Rezzolla}}, \citenamefont
  {{Zanotti}},\ and\ \citenamefont {{Jaramillo}}}]{Alic12ApJ}%
  \BibitemOpen
  \bibfield  {author} {\bibinfo {author} {\bibfnamefont {D.}~\bibnamefont
  {{Alic}}}, \bibinfo {author} {\bibfnamefont {P.}~\bibnamefont {{Moesta}}},
  \bibinfo {author} {\bibfnamefont {L.}~\bibnamefont {{Rezzolla}}}, \bibinfo
  {author} {\bibfnamefont {O.}~\bibnamefont {{Zanotti}}}, \ and\ \bibinfo
  {author} {\bibfnamefont {J.~L.}\ \bibnamefont {{Jaramillo}}},\ }\href
  {\doibase 10.1088/0004-637X/754/1/36} {\bibfield  {journal} {\bibinfo
  {journal} {Astrophys. J.}\ }\textbf {\bibinfo {volume} {754}},\ \bibinfo
  {eid} {36} (\bibinfo {year} {2012})},\ \Eprint
  {http://arxiv.org/abs/1204.2226} {arXiv:1204.2226 [gr-qc]} \BibitemShut
  {NoStop}%
\bibitem [{\citenamefont {{Moesta}}\ \emph {et~al.}(2012)\citenamefont
  {{Moesta}}, \citenamefont {{Alic}}, \citenamefont {{Rezzolla}}, \citenamefont
  {{Zanotti}},\ and\ \citenamefont {{Palenzuela}}}]{Moesta12ApJ}%
  \BibitemOpen
  \bibfield  {author} {\bibinfo {author} {\bibfnamefont {P.}~\bibnamefont
  {{Moesta}}}, \bibinfo {author} {\bibfnamefont {D.}~\bibnamefont {{Alic}}},
  \bibinfo {author} {\bibfnamefont {L.}~\bibnamefont {{Rezzolla}}}, \bibinfo
  {author} {\bibfnamefont {O.}~\bibnamefont {{Zanotti}}}, \ and\ \bibinfo
  {author} {\bibfnamefont {C.}~\bibnamefont {{Palenzuela}}},\ }\href {\doibase
  10.1088/2041-8205/749/2/L32} {\bibfield  {journal} {\bibinfo  {journal}
  {Astrophys. J.}\ }\textbf {\bibinfo {volume} {749}},\ \bibinfo {eid} {L32}
  (\bibinfo {year} {2012})},\ \Eprint {http://arxiv.org/abs/1109.1177}
  {arXiv:1109.1177 [gr-qc]} \BibitemShut {NoStop}%
\bibitem [{\citenamefont {{McKinney}}\ and\ \citenamefont
  {{Narayan}}(2007)}]{McKinney07}%
  \BibitemOpen
  \bibfield  {author} {\bibinfo {author} {\bibfnamefont {J.~C.}\ \bibnamefont
  {{McKinney}}}\ and\ \bibinfo {author} {\bibfnamefont {R.}~\bibnamefont
  {{Narayan}}},\ }\href {\doibase 10.1111/j.1365-2966.2006.11220.x} {\bibfield
  {journal} {\bibinfo  {journal} {Mon. Not. Roy. Astron. Soc.}\ }\textbf
  {\bibinfo {volume} {375}},\ \bibinfo {pages} {531} (\bibinfo {year}
  {2007})},\ \Eprint {http://arxiv.org/abs/astro-ph/0607576}
  {arXiv:astro-ph/0607576 [astro-ph]} \BibitemShut {NoStop}%
\bibitem [{\citenamefont {Ginzburg V.~L.}(1964)}]{Ginzburg1964}%
  \BibitemOpen
  \bibfield  {author} {\bibinfo {author} {\bibfnamefont {O.~L.~M.}\
  \bibnamefont {Ginzburg V.~L.}},\ }\href@noop {} {\bibfield  {journal}
  {\bibinfo  {journal} {Zh. Eksp. Teor. Fiz.}\ }\textbf {\bibinfo {volume}
  {47}},\ \bibinfo {pages} {1030} (\bibinfo {year} {1964})}\BibitemShut
  {NoStop}%
\bibitem [{\citenamefont {{Anderson}}\ and\ \citenamefont
  {{Cohen}}(1970)}]{Anderson70}%
  \BibitemOpen
  \bibfield  {author} {\bibinfo {author} {\bibfnamefont {J.~L.}\ \bibnamefont
  {{Anderson}}}\ and\ \bibinfo {author} {\bibfnamefont {J.~M.}\ \bibnamefont
  {{Cohen}}},\ }\href {\doibase 10.1007/BF00649960} {\bibfield  {journal}
  {\bibinfo  {journal} {Astrophys. Space Sci.}\ }\textbf {\bibinfo {volume}
  {9}},\ \bibinfo {pages} {146} (\bibinfo {year} {1970})}\BibitemShut {NoStop}%
\bibitem [{\citenamefont {{Morozova}}\ \emph {et~al.}(2010)\citenamefont
  {{Morozova}}, \citenamefont {{Ahmedov}},\ and\ \citenamefont
  {{Zanotti}}}]{Morozova10}%
  \BibitemOpen
  \bibfield  {author} {\bibinfo {author} {\bibfnamefont {V.~S.}\ \bibnamefont
  {{Morozova}}}, \bibinfo {author} {\bibfnamefont {B.~J.}\ \bibnamefont
  {{Ahmedov}}}, \ and\ \bibinfo {author} {\bibfnamefont {O.}~\bibnamefont
  {{Zanotti}}},\ }\href {\doibase 10.1111/j.1365-2966.2010.17131.x} {\bibfield
  {journal} {\bibinfo  {journal} {Mon. Not. R. Astron. Soc.}\ }\textbf
  {\bibinfo {volume} {408}},\ \bibinfo {pages} {490} (\bibinfo {year}
  {2010})},\ \Eprint {http://arxiv.org/abs/1004.1739} {arXiv:1004.1739
  [astro-ph.HE]} \BibitemShut {NoStop}%
\bibitem [{\citenamefont {{Morozova}}\ \emph {et~al.}(2012)\citenamefont
  {{Morozova}}, \citenamefont {{Ahmedov}},\ and\ \citenamefont
  {{Zanotti}}}]{Morozova12}%
  \BibitemOpen
  \bibfield  {author} {\bibinfo {author} {\bibfnamefont {V.~S.}\ \bibnamefont
  {{Morozova}}}, \bibinfo {author} {\bibfnamefont {B.~J.}\ \bibnamefont
  {{Ahmedov}}}, \ and\ \bibinfo {author} {\bibfnamefont {O.}~\bibnamefont
  {{Zanotti}}},\ }\href {\doibase 10.1111/j.1365-2966.2011.19866.x} {\bibfield
  {journal} {\bibinfo  {journal} {Mon. Not. R. Astron. Soc.}\ }\textbf
  {\bibinfo {volume} {419}},\ \bibinfo {pages} {2147} (\bibinfo {year}
  {2012})},\ \Eprint {http://arxiv.org/abs/1107.3327} {arXiv:1107.3327
  [astro-ph.HE]} \BibitemShut {NoStop}%
\bibitem [{\citenamefont {{Rezzolla}}\ \emph {et~al.}(2001)\citenamefont
  {{Rezzolla}}, \citenamefont {{Ahmedov}},\ and\ \citenamefont
  {{Miller}}}]{Rezzolla01}%
  \BibitemOpen
  \bibfield  {author} {\bibinfo {author} {\bibfnamefont {L.}~\bibnamefont
  {{Rezzolla}}}, \bibinfo {author} {\bibfnamefont {B.~J.}\ \bibnamefont
  {{Ahmedov}}}, \ and\ \bibinfo {author} {\bibfnamefont {J.~C.}\ \bibnamefont
  {{Miller}}},\ }\href {\doibase 10.1046/j.1365-8711.2001.04161.x} {\bibfield
  {journal} {\bibinfo  {journal} {Mon. Not. R. Astron. Soc.}\ }\textbf
  {\bibinfo {volume} {322}},\ \bibinfo {pages} {723} (\bibinfo {year}
  {2001})},\ \Eprint {http://arxiv.org/abs/astro-ph/0011316}
  {arXiv:astro-ph/0011316 [astro-ph]} \BibitemShut {NoStop}%
\bibitem [{\citenamefont {{de Felice}}\ and\ \citenamefont
  {{Sorge}}(2003)}]{deFelice03}%
  \BibitemOpen
  \bibfield  {author} {\bibinfo {author} {\bibfnamefont {F.}~\bibnamefont {{de
  Felice}}}\ and\ \bibinfo {author} {\bibfnamefont {F.}~\bibnamefont
  {{Sorge}}},\ }\href@noop {} {\bibfield  {journal} {\bibinfo  {journal}
  {Class. Quantum Grav.}\ }\textbf {\bibinfo {volume} {20}},\ \bibinfo {pages}
  {469} (\bibinfo {year} {2003})}\BibitemShut {NoStop}%
\bibitem [{\citenamefont {{de Felice}}\ \emph {et~al.}(2004)\citenamefont {{de
  Felice}}, \citenamefont {{Sorge}},\ and\ \citenamefont
  {{Zilio}}}]{deFelice04}%
  \BibitemOpen
  \bibfield  {author} {\bibinfo {author} {\bibfnamefont {F.}~\bibnamefont {{de
  Felice}}}, \bibinfo {author} {\bibfnamefont {F.}~\bibnamefont {{Sorge}}}, \
  and\ \bibinfo {author} {\bibfnamefont {S.}~\bibnamefont {{Zilio}}},\ }\href
  {\doibase 10.1088/0264-9381/21/4/016} {\bibfield  {journal} {\bibinfo
  {journal} {Class. Quantum Grav.}\ }\textbf {\bibinfo {volume} {21}},\
  \bibinfo {pages} {961} (\bibinfo {year} {2004})}\BibitemShut {NoStop}%
\bibitem [{\citenamefont {{Frolov}}\ and\ \citenamefont
  {{Shoom}}(2010)}]{Frolov10}%
  \BibitemOpen
  \bibfield  {author} {\bibinfo {author} {\bibfnamefont {V.~P.}\ \bibnamefont
  {{Frolov}}}\ and\ \bibinfo {author} {\bibfnamefont {A.~A.}\ \bibnamefont
  {{Shoom}}},\ }\href {\doibase 10.1103/PhysRevD.82.084034} {\bibfield
  {journal} {\bibinfo  {journal} {Phys. Rev. D}\ }\textbf {\bibinfo {volume}
  {82}},\ \bibinfo {eid} {084034} (\bibinfo {year} {2010})},\ \Eprint
  {http://arxiv.org/abs/1008.2985} {arXiv:1008.2985 [gr-qc]} \BibitemShut
  {NoStop}%
\bibitem [{\citenamefont {{Aliev}}\ and\ \citenamefont
  {{{\"O}zdemir}}(2002)}]{Aliev02}%
  \BibitemOpen
  \bibfield  {author} {\bibinfo {author} {\bibfnamefont {A.~N.}\ \bibnamefont
  {{Aliev}}}\ and\ \bibinfo {author} {\bibfnamefont {N.}~\bibnamefont
  {{{\"O}zdemir}}},\ }\href {\doibase 10.1046/j.1365-8711.2002.05727.x}
  {\bibfield  {journal} {\bibinfo  {journal} {Mon. Not. R. Astron. Soc.}\
  }\textbf {\bibinfo {volume} {336}},\ \bibinfo {pages} {241} (\bibinfo {year}
  {2002})},\ \Eprint {http://arxiv.org/abs/gr-qc/0208025} {gr-qc/0208025}
  \BibitemShut {NoStop}%
\bibitem [{\citenamefont {{Abdujabbarov}}\ and\ \citenamefont
  {{Ahmedov}}(2010)}]{Abdujabbarov10}%
  \BibitemOpen
  \bibfield  {author} {\bibinfo {author} {\bibfnamefont {A.}~\bibnamefont
  {{Abdujabbarov}}}\ and\ \bibinfo {author} {\bibfnamefont {B.}~\bibnamefont
  {{Ahmedov}}},\ }\href {\doibase 10.1103/PhysRevD.81.044022} {\bibfield
  {journal} {\bibinfo  {journal} {Phys. Rev. D}\ }\textbf {\bibinfo {volume}
  {81}},\ \bibinfo {eid} {044022} (\bibinfo {year} {2010})},\ \Eprint
  {http://arxiv.org/abs/0905.2730} {arXiv:0905.2730 [gr-qc]} \BibitemShut
  {NoStop}%
\bibitem [{\citenamefont {{Shaymatov}}\ \emph {et~al.}(2014)\citenamefont
  {{Shaymatov}}, \citenamefont {{Atamurotov}},\ and\ \citenamefont
  {{Ahmedov}}}]{Shaymatov14}%
  \BibitemOpen
  \bibfield  {author} {\bibinfo {author} {\bibfnamefont {S.}~\bibnamefont
  {{Shaymatov}}}, \bibinfo {author} {\bibfnamefont {F.}~\bibnamefont
  {{Atamurotov}}}, \ and\ \bibinfo {author} {\bibfnamefont {B.}~\bibnamefont
  {{Ahmedov}}},\ }\href {\doibase 10.1007/s10509-013-1752-3} {\bibfield
  {journal} {\bibinfo  {journal} {Astrophys Space Sci}\ }\textbf {\bibinfo
  {volume} {350}},\ \bibinfo {pages} {413} (\bibinfo {year}
  {2014})}\BibitemShut {NoStop}%
\bibitem [{\citenamefont {{Kolo{\v s}}}\ \emph {et~al.}(2015)\citenamefont
  {{Kolo{\v s}}}, \citenamefont {{Stuchl{\'{\i}}k}},\ and\ \citenamefont
  {{Tursunov}}}]{Kolos15}%
  \BibitemOpen
  \bibfield  {author} {\bibinfo {author} {\bibfnamefont {M.}~\bibnamefont
  {{Kolo{\v s}}}}, \bibinfo {author} {\bibfnamefont {Z.}~\bibnamefont
  {{Stuchl{\'{\i}}k}}}, \ and\ \bibinfo {author} {\bibfnamefont
  {A.}~\bibnamefont {{Tursunov}}},\ }\href {\doibase
  10.1088/0264-9381/32/16/165009} {\bibfield  {journal} {\bibinfo  {journal}
  {Class. Quantum Grav.}\ }\textbf {\bibinfo {volume} {32}},\ \bibinfo {eid}
  {165009} (\bibinfo {year} {2015})},\ \Eprint
  {http://arxiv.org/abs/1506.06799} {arXiv:1506.06799 [gr-qc]} \BibitemShut
  {NoStop}%
\bibitem [{\citenamefont {{Jamil}}\ \emph {et~al.}(2015)\citenamefont
  {{Jamil}}, \citenamefont {{Hussain}},\ and\ \citenamefont
  {{Majeed}}}]{Jamil15}%
  \BibitemOpen
  \bibfield  {author} {\bibinfo {author} {\bibfnamefont {M.}~\bibnamefont
  {{Jamil}}}, \bibinfo {author} {\bibfnamefont {S.}~\bibnamefont {{Hussain}}},
  \ and\ \bibinfo {author} {\bibfnamefont {B.}~\bibnamefont {{Majeed}}},\
  }\href {\doibase 10.1140/epjc/s10052-014-3230-7} {\bibfield  {journal}
  {\bibinfo  {journal} {Eur. Phys. J. C}\ }\textbf {\bibinfo {volume} {75}},\
  \bibinfo {eid} {24} (\bibinfo {year} {2015})},\ \Eprint
  {http://arxiv.org/abs/1404.7123} {arXiv:1404.7123 [gr-qc]} \BibitemShut
  {NoStop}%
\bibitem [{\citenamefont {{Stuchl{\'{\i}}k}}\ and\ \citenamefont {{Kolo{\v
  s}}}(2016)}]{Stuchlik16}%
  \BibitemOpen
  \bibfield  {author} {\bibinfo {author} {\bibfnamefont {Z.}~\bibnamefont
  {{Stuchl{\'{\i}}k}}}\ and\ \bibinfo {author} {\bibfnamefont {M.}~\bibnamefont
  {{Kolo{\v s}}}},\ }\href {\doibase 10.1140/epjc/s10052-015-3862-2} {\bibfield
   {journal} {\bibinfo  {journal} {Eur. Phys. J. C}\ }\textbf {\bibinfo
  {volume} {76}},\ \bibinfo {eid} {32} (\bibinfo {year} {2016})},\ \Eprint
  {http://arxiv.org/abs/1511.02936} {arXiv:1511.02936 [gr-qc]} \BibitemShut
  {NoStop}%
\bibitem [{\citenamefont {{Tursunov}}\ \emph {et~al.}(2016)\citenamefont
  {{Tursunov}}, \citenamefont {{Stuchl{\'{\i}}k}},\ and\ \citenamefont
  {{Kolo{\v s}}}}]{Tursunov16}%
  \BibitemOpen
  \bibfield  {author} {\bibinfo {author} {\bibfnamefont {A.}~\bibnamefont
  {{Tursunov}}}, \bibinfo {author} {\bibfnamefont {Z.}~\bibnamefont
  {{Stuchl{\'{\i}}k}}}, \ and\ \bibinfo {author} {\bibfnamefont
  {M.}~\bibnamefont {{Kolo{\v s}}}},\ }\href {\doibase
  10.1103/PhysRevD.93.084012} {\bibfield  {journal} {\bibinfo  {journal} {Phys.
  Rev. D}\ }\textbf {\bibinfo {volume} {93}},\ \bibinfo {eid} {084012}
  (\bibinfo {year} {2016})},\ \Eprint {http://arxiv.org/abs/1603.07264}
  {arXiv:1603.07264 [gr-qc]} \BibitemShut {NoStop}%
\bibitem [{\citenamefont {{Hussain}}\ and\ \citenamefont
  {{Jamil}}(2015)}]{Hussain15}%
  \BibitemOpen
  \bibfield  {author} {\bibinfo {author} {\bibfnamefont {S.}~\bibnamefont
  {{Hussain}}}\ and\ \bibinfo {author} {\bibfnamefont {M.}~\bibnamefont
  {{Jamil}}},\ }\href {\doibase 10.1103/PhysRevD.92.043008} {\bibfield
  {journal} {\bibinfo  {journal} {Phys. Rev. D}\ }\textbf {\bibinfo {volume}
  {92}},\ \bibinfo {eid} {043008} (\bibinfo {year} {2015})},\ \Eprint
  {http://arxiv.org/abs/1508.02123} {arXiv:1508.02123 [gr-qc]} \BibitemShut
  {NoStop}%
\bibitem [{\citenamefont {{Piotrovich}}\ \emph {et~al.}(2010)\citenamefont
  {{Piotrovich}}, \citenamefont {{Silant'ev}}, \citenamefont {{Gnedin}},\ and\
  \citenamefont {{Natsvlishvili}}}]{Piotrovich10}%
  \BibitemOpen
  \bibfield  {author} {\bibinfo {author} {\bibfnamefont {M.~Y.}\ \bibnamefont
  {{Piotrovich}}}, \bibinfo {author} {\bibfnamefont {N.~A.}\ \bibnamefont
  {{Silant'ev}}}, \bibinfo {author} {\bibfnamefont {Y.~N.}\ \bibnamefont
  {{Gnedin}}}, \ and\ \bibinfo {author} {\bibfnamefont {T.~M.}\ \bibnamefont
  {{Natsvlishvili}}},\ }\href@noop {} {\bibfield  {journal} {\bibinfo
  {journal} {ArXiv e-prints}\ } (\bibinfo {year} {2010})},\ \Eprint
  {http://arxiv.org/abs/1002.4948} {arXiv:1002.4948 [astro-ph.CO]} \BibitemShut
  {NoStop}%
\bibitem [{\citenamefont {{Baczko}}\ \emph {et~al.}(2016)\citenamefont
  {{Baczko}}, \citenamefont {{Schulz}},\ and\ \citenamefont {{et
  al.}}}]{Baczko16}%
  \BibitemOpen
  \bibfield  {author} {\bibinfo {author} {\bibfnamefont {A.-K.}\ \bibnamefont
  {{Baczko}}}, \bibinfo {author} {\bibfnamefont {R.}~\bibnamefont {{Schulz}}},
  \ and\ \bibinfo {author} {\bibnamefont {{et al.}}},\ }\href {\doibase
  10.1051/0004-6361/201527951} {\bibfield  {journal} {\bibinfo  {journal}
  {Astron. Astrophys.}\ }\textbf {\bibinfo {volume} {593}},\ \bibinfo {eid}
  {A47} (\bibinfo {year} {2016})},\ \Eprint {http://arxiv.org/abs/1605.07100}
  {arXiv:1605.07100} \BibitemShut {NoStop}%
\bibitem [{\citenamefont {{Dallilar}}\ and\ \citenamefont
  {et~al.}(2017)}]{Dallilar2018}%
  \BibitemOpen
  \bibfield  {author} {\bibinfo {author} {\bibfnamefont {Y.}~\bibnamefont
  {{Dallilar}}}\ and\ \bibinfo {author} {\bibnamefont {et~al.}},\ }\href
  {\doibase 10.1126/science.aan0249} {\bibfield  {journal} {\bibinfo  {journal}
  {Science}\ }\textbf {\bibinfo {volume} {358}},\ \bibinfo {pages} {1299}
  (\bibinfo {year} {2017})}\BibitemShut {NoStop}%
\bibitem [{\citenamefont {Misner}\ \emph {et~al.}(1973)\citenamefont {Misner},
  \citenamefont {Thorne},\ and\ \citenamefont {Wheeler}}]{Misner73}%
  \BibitemOpen
  \bibfield  {author} {\bibinfo {author} {\bibfnamefont {C.~W.}\ \bibnamefont
  {Misner}}, \bibinfo {author} {\bibfnamefont {K.~S.}\ \bibnamefont {Thorne}},
  \ and\ \bibinfo {author} {\bibfnamefont {J.~A.}\ \bibnamefont {Wheeler}},\
  }\href@noop {} {\emph {\bibinfo {title} {Gravitation}}}\ (\bibinfo
  {publisher} {W. H. Freeman},\ \bibinfo {address} {San Francisco},\ \bibinfo
  {year} {1973})\BibitemShut {NoStop}%
\bibitem [{\citenamefont {{Boshkayev}}\ \emph {et~al.}(2020)\citenamefont
  {{Boshkayev}}, \citenamefont {{Idrissov}}, \citenamefont {{Luongo}},\ and\
  \citenamefont {{Malafarina}}}]{Boshkayev20}%
  \BibitemOpen
  \bibfield  {author} {\bibinfo {author} {\bibfnamefont {K.}~\bibnamefont
  {{Boshkayev}}}, \bibinfo {author} {\bibfnamefont {A.}~\bibnamefont
  {{Idrissov}}}, \bibinfo {author} {\bibfnamefont {O.}~\bibnamefont
  {{Luongo}}}, \ and\ \bibinfo {author} {\bibfnamefont {D.}~\bibnamefont
  {{Malafarina}}},\ }\href {\doibase 10.1093/mnras/staa1564} {\bibfield
  {journal} {\bibinfo  {journal} {Mon. Not. R. Astron. Soc.}\ }\textbf
  {\bibinfo {volume} {496}},\ \bibinfo {pages} {1115} (\bibinfo {year}
  {2020})},\ \Eprint {http://arxiv.org/abs/2006.01269} {arXiv:2006.01269}
  \BibitemShut {NoStop}%
\bibitem [{\citenamefont {{Bardeen}}\ \emph {et~al.}(1972)\citenamefont
  {{Bardeen}}, \citenamefont {{Press}},\ and\ \citenamefont
  {{Teukolsky}}}]{Bardeen72}%
  \BibitemOpen
  \bibfield  {author} {\bibinfo {author} {\bibfnamefont {J.~M.}\ \bibnamefont
  {{Bardeen}}}, \bibinfo {author} {\bibfnamefont {W.~H.}\ \bibnamefont
  {{Press}}}, \ and\ \bibinfo {author} {\bibfnamefont {S.~A.}\ \bibnamefont
  {{Teukolsky}}},\ }\href {\doibase 10.1086/151796} {\bibfield  {journal}
  {\bibinfo  {journal} {Astrophys. J.}\ }\textbf {\bibinfo {volume} {178}},\
  \bibinfo {pages} {347} (\bibinfo {year} {1972})}\BibitemShut {NoStop}%
\bibitem [{\citenamefont {{Bambi}}(2017)}]{Bambi17-BHs}%
  \BibitemOpen
  \bibfield  {author} {\bibinfo {author} {\bibfnamefont {C.}~\bibnamefont
  {{Bambi}}},\ }\href {\doibase 10.1103/RevModPhys.89.025001} {\bibfield
  {journal} {\bibinfo  {journal} {Rev. Mod. Phys.}\ }\textbf {\bibinfo {volume}
  {89}},\ \bibinfo {eid} {025001} (\bibinfo {year} {2017})},\ \Eprint
  {http://arxiv.org/abs/1509.03884} {arXiv:1509.03884 [gr-qc]} \BibitemShut
  {NoStop}%
\bibitem [{\citenamefont {{Walton}}\ \emph {et~al.}(2013)\citenamefont
  {{Walton}}, \citenamefont {{Nardini}}, \citenamefont {{Fabian}},
  \citenamefont {{Gallo}},\ and\ \citenamefont {{Reis}}}]{Walton13}%
  \BibitemOpen
  \bibfield  {author} {\bibinfo {author} {\bibfnamefont {D.~J.}\ \bibnamefont
  {{Walton}}}, \bibinfo {author} {\bibfnamefont {E.}~\bibnamefont {{Nardini}}},
  \bibinfo {author} {\bibfnamefont {A.~C.}\ \bibnamefont {{Fabian}}}, \bibinfo
  {author} {\bibfnamefont {L.~C.}\ \bibnamefont {{Gallo}}}, \ and\ \bibinfo
  {author} {\bibfnamefont {R.~C.}\ \bibnamefont {{Reis}}},\ }\href {\doibase
  10.1093/mnras/sts227} {\bibfield  {journal} {\bibinfo  {journal} {Mon. Not.
  R. Astron. Soc.}\ }\textbf {\bibinfo {volume} {428}},\ \bibinfo {pages}
  {2901} (\bibinfo {year} {2013})},\ \Eprint {http://arxiv.org/abs/1210.4593}
  {arXiv:1210.4593 [astro-ph.HE]} \BibitemShut {NoStop}%
\bibitem [{\citenamefont {{Patrick}}\ \emph
  {et~al.}(2011{\natexlab{a}})\citenamefont {{Patrick}}, \citenamefont
  {{Reeves}}, \citenamefont {{Porquet}}, \citenamefont {{Markowitz}},
  \citenamefont {{Lobban}},\ and\ \citenamefont
  {{Terashima}}}]{Patrick11b-Seyfert}%
  \BibitemOpen
  \bibfield  {author} {\bibinfo {author} {\bibfnamefont {A.~R.}\ \bibnamefont
  {{Patrick}}}, \bibinfo {author} {\bibfnamefont {J.~N.}\ \bibnamefont
  {{Reeves}}}, \bibinfo {author} {\bibfnamefont {D.}~\bibnamefont {{Porquet}}},
  \bibinfo {author} {\bibfnamefont {A.~G.}\ \bibnamefont {{Markowitz}}},
  \bibinfo {author} {\bibfnamefont {A.~P.}\ \bibnamefont {{Lobban}}}, \ and\
  \bibinfo {author} {\bibfnamefont {Y.}~\bibnamefont {{Terashima}}},\ }\href
  {\doibase 10.1111/j.1365-2966.2010.17852.x} {\bibfield  {journal} {\bibinfo
  {journal} {Mon. Not. R. Astron. Soc.}\ }\textbf {\bibinfo {volume} {411}},\
  \bibinfo {pages} {2353} (\bibinfo {year} {2011}{\natexlab{a}})},\ \Eprint
  {http://arxiv.org/abs/1010.2080} {arXiv:1010.2080 [astro-ph.HE]} \BibitemShut
  {NoStop}%
\bibitem [{\citenamefont {{Patrick}}\ \emph
  {et~al.}(2011{\natexlab{b}})\citenamefont {{Patrick}}, \citenamefont
  {{Reeves}}, \citenamefont {{Lobban}}, \citenamefont {{Porquet}},\ and\
  \citenamefont {{Markowitz}}}]{Patrick11a-Seyfert}%
  \BibitemOpen
  \bibfield  {author} {\bibinfo {author} {\bibfnamefont {A.~R.}\ \bibnamefont
  {{Patrick}}}, \bibinfo {author} {\bibfnamefont {J.~N.}\ \bibnamefont
  {{Reeves}}}, \bibinfo {author} {\bibfnamefont {A.~P.}\ \bibnamefont
  {{Lobban}}}, \bibinfo {author} {\bibfnamefont {D.}~\bibnamefont {{Porquet}}},
  \ and\ \bibinfo {author} {\bibfnamefont {A.~G.}\ \bibnamefont
  {{Markowitz}}},\ }\href {\doibase 10.1111/j.1365-2966.2011.19224.x}
  {\bibfield  {journal} {\bibinfo  {journal} {Mon. Not. R. Astron. Soc.}\
  }\textbf {\bibinfo {volume} {416}},\ \bibinfo {pages} {2725} (\bibinfo {year}
  {2011}{\natexlab{b}})},\ \Eprint {http://arxiv.org/abs/1106.2135}
  {arXiv:1106.2135 [astro-ph.HE]} \BibitemShut {NoStop}%
\bibitem [{\citenamefont {{Tan}}\ \emph {et~al.}(2012)\citenamefont {{Tan}},
  \citenamefont {{Wang}}, \citenamefont {{Shu}},\ and\ \citenamefont
  {{Zhou}}}]{Tan12}%
  \BibitemOpen
  \bibfield  {author} {\bibinfo {author} {\bibfnamefont {Y.}~\bibnamefont
  {{Tan}}}, \bibinfo {author} {\bibfnamefont {J.~X.}\ \bibnamefont {{Wang}}},
  \bibinfo {author} {\bibfnamefont {X.~W.}\ \bibnamefont {{Shu}}}, \ and\
  \bibinfo {author} {\bibfnamefont {Y.}~\bibnamefont {{Zhou}}},\ }\href
  {\doibase 10.1088/2041-8205/747/1/L11} {\bibfield  {journal} {\bibinfo
  {journal} {Astrophys. J.}\ }\textbf {\bibinfo {volume} {747}},\ \bibinfo
  {eid} {L11} (\bibinfo {year} {2012})},\ \Eprint
  {http://arxiv.org/abs/1202.0400} {arXiv:1202.0400 [astro-ph.HE]} \BibitemShut
  {NoStop}%
\bibitem [{\citenamefont {{Gallo}}\ \emph {et~al.}(2005)\citenamefont
  {{Gallo}}, \citenamefont {{Fabian}}, \citenamefont {{Boller}},\ and\
  \citenamefont {{Pietsch}}}]{Gallo05}%
  \BibitemOpen
  \bibfield  {author} {\bibinfo {author} {\bibfnamefont {L.~C.}\ \bibnamefont
  {{Gallo}}}, \bibinfo {author} {\bibfnamefont {A.~C.}\ \bibnamefont
  {{Fabian}}}, \bibinfo {author} {\bibfnamefont {T.}~\bibnamefont {{Boller}}},
  \ and\ \bibinfo {author} {\bibfnamefont {W.}~\bibnamefont {{Pietsch}}},\
  }\href {\doibase 10.1111/j.1365-2966.2005.09418.x} {\bibfield  {journal}
  {\bibinfo  {journal} {Mon. Not. R. Astron. Soc.}\ }\textbf {\bibinfo {volume}
  {363}},\ \bibinfo {pages} {64} (\bibinfo {year} {2005})},\ \Eprint
  {http://arxiv.org/abs/astro-ph/0508229} {astro-ph/0508229} \BibitemShut
  {NoStop}%
\bibitem [{\citenamefont {{Gallo}}\ \emph {et~al.}(2011)\citenamefont
  {{Gallo}}, \citenamefont {{Miniutti}}, \citenamefont {{Miller}},
  \citenamefont {{Brenneman}}, \citenamefont {{Fabian}}, \citenamefont
  {{Guainazzi}},\ and\ \citenamefont {{Reynolds}}}]{Gallo11}%
  \BibitemOpen
  \bibfield  {author} {\bibinfo {author} {\bibfnamefont {L.~C.}\ \bibnamefont
  {{Gallo}}}, \bibinfo {author} {\bibfnamefont {G.}~\bibnamefont {{Miniutti}}},
  \bibinfo {author} {\bibfnamefont {J.~M.}\ \bibnamefont {{Miller}}}, \bibinfo
  {author} {\bibfnamefont {L.~W.}\ \bibnamefont {{Brenneman}}}, \bibinfo
  {author} {\bibfnamefont {A.~C.}\ \bibnamefont {{Fabian}}}, \bibinfo {author}
  {\bibfnamefont {M.}~\bibnamefont {{Guainazzi}}}, \ and\ \bibinfo {author}
  {\bibfnamefont {C.~S.}\ \bibnamefont {{Reynolds}}},\ }\href {\doibase
  10.1111/j.1365-2966.2010.17705.x} {\bibfield  {journal} {\bibinfo  {journal}
  {Mon. Not. R. Astron. Soc.}\ }\textbf {\bibinfo {volume} {411}},\ \bibinfo
  {pages} {607} (\bibinfo {year} {2011})},\ \Eprint
  {http://arxiv.org/abs/1009.2987} {arXiv:1009.2987 [astro-ph.HE]} \BibitemShut
  {NoStop}%
\bibitem [{\citenamefont {{Berti}}\ \emph {et~al.}(2009)\citenamefont
  {{Berti}}, \citenamefont {{Cardoso}}, \citenamefont {{Gualtieri}},
  \citenamefont {{Pretorius}},\ and\ \citenamefont {{Sperhake}}}]{Berti09}%
  \BibitemOpen
  \bibfield  {author} {\bibinfo {author} {\bibfnamefont {E.}~\bibnamefont
  {{Berti}}}, \bibinfo {author} {\bibfnamefont {V.}~\bibnamefont {{Cardoso}}},
  \bibinfo {author} {\bibfnamefont {L.}~\bibnamefont {{Gualtieri}}}, \bibinfo
  {author} {\bibfnamefont {F.}~\bibnamefont {{Pretorius}}}, \ and\ \bibinfo
  {author} {\bibfnamefont {U.}~\bibnamefont {{Sperhake}}},\ }\href {\doibase
  10.1103/PhysRevLett.103.239001} {\bibfield  {journal} {\bibinfo  {journal}
  {Phys. Rev. Lett.}\ }\textbf {\bibinfo {volume} {103}},\ \bibinfo {eid}
  {239001} (\bibinfo {year} {2009})},\ \Eprint {http://arxiv.org/abs/0911.2243}
  {arXiv:0911.2243 [gr-qc]} \BibitemShut {NoStop}%
\bibitem [{\citenamefont {{Abramowicz}}\ and\ \citenamefont
  {{Fragile}}(2013)}]{Abramowicz13}%
  \BibitemOpen
  \bibfield  {author} {\bibinfo {author} {\bibfnamefont {M.~A.}\ \bibnamefont
  {{Abramowicz}}}\ and\ \bibinfo {author} {\bibfnamefont {P.~C.}\ \bibnamefont
  {{Fragile}}},\ }\href {\doibase 10.12942/lrr-2013-1} {\bibfield  {journal}
  {\bibinfo  {journal} {Living Rev. Relativ.}\ }\textbf {\bibinfo {volume}
  {16}},\ \bibinfo {eid} {1} (\bibinfo {year} {2013})},\ \Eprint
  {http://arxiv.org/abs/1104.5499} {arXiv:1104.5499 [astro-ph.HE]} \BibitemShut
  {NoStop}%
\end{thebibliography}%

\end{document}